\PassOptionsToPackage{svgnames}{xcolor}
\documentclass[a4paper,12pt]{article}

\usepackage{a4wide}

\usepackage{xcolor}
\definecolor{marron}{RGB}{60,30,10}
\definecolor{darkblue}{RGB}{0,0,80}
\definecolor{lightblue}{RGB}{80,80,80}
\definecolor{darkgreen}{RGB}{0,80,0}
\definecolor{darkgray}{RGB}{35,31,32}
\definecolor{Lundblue}{RGB}{0,52,119}
\definecolor{Lundgold}{RGB}{154,91,11}

\usepackage{fancyhdr}
\usepackage{fourier-orns}
\newcommand{\ornamento}{\vspace{2em}\noindent \textcolor{Lundgold}{\hrulefill~ \raisebox{-2.5pt}[10pt][10pt]{\leafright \decofourleft \decothreeleft  \aldineright \decotwo \floweroneleft \decoone   \floweroneright \decotwo \aldineleft\decothreeright \decofourright \leafleft} ~  \hrulefill \\ \vspace{2em}}}

\newcommand{\colorrule}[3]{\textcolor{#1}{\rule{#2}{#3}}}


\newcommand{\estcab}[1]{\itshape\textcolor{Lundgold}{\nouppercase #1}}
\usepackage[utf8]{inputenc}

\usepackage{axodraw2}

\usepackage[compat=1.0.0]{tikz-feynman}
\usepackage[object=vectorian]{pgfornament}

\usepackage{graphicx}
\usepackage{float}

\usepackage{amssymb}

\usepackage{eso-pic}
\usepackage{rotating}

\usepackage{amsmath}
\usepackage{latexsym}
\newcommand\numberthis{\addtocounter{equation}{1}\tag{\theequation}}

\usepackage{mathtools}
\usepackage{fancyvrb}

\usepackage{physics}

\usepackage{caption}
\usepackage{subcaption}

\usepackage{threeparttable}

\usepackage{array}

\newcolumntype{C}{>{$}c<{$}}
\newcolumntype{L}{>{$}l<{$}}

\usepackage{hyperref}
\hypersetup{
    colorlinks=true,
    linkcolor=blue,
    filecolor=magenta,      
    urlcolor=cyan,
}

\renewcommand{\theequation}{\thesection.\arabic{equation}}

\usepackage[acronym,toc]{glossaries}
\makeglossaries
\newacronym{gr}{GR}{General Relativity} 
\newacronym{sm}{SM}{Standard Model}
\newacronym{qft}{QFT}{Quantum Field Theory}
\newacronym{eft}{EFT}{Effective Field Theory}
\newacronym{ym}{YM}{Yang-Mills gauge theory}
\newacronym{cm}{CM}{Center-of-Mass frame}
\newacronym{emtt}{EMT}{Energy-Momentum Tensor}
\newacronym{gct}{GCT}{General Coordinate Transformations}

\newcommand{\sectionlinetwo}[2]{%
  \nointerlineskip \vspace{.5\baselineskip}\hspace{\fill}
  {\color{#1}
    \resizebox{0.5\linewidth}{2ex}
    {{%
    {\begin{tikzpicture}
    \node  (C) at (0,0) {};
    \node (D) at (9,0) {};
    \path (C) to [ornament=#2] (D);
    \end{tikzpicture}}}}}%
    \hspace{\fill}
    \par\nointerlineskip \vspace{.5\baselineskip}
  }

\begin{document}

\begin{titlepage}
\AddToShipoutPicture*{\put(35,680){\rotatebox{0}{\scalebox{1}{
        \includegraphics[height=4cm]{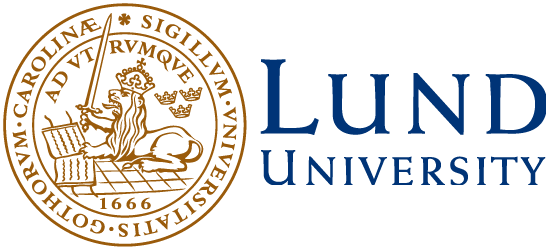}
      }}}}
\begin{flushright} 
LU TP 18-08\\
August 2018\\
\end{flushright}
\vfill
\begin{center}
\colorrule{Lundgold}{\textwidth}{1.6pt}\\[\baselineskip]
{\Large\bf\textcolor{Lundblue}{SIMPLIFYING QUANTUM GRAVITY \\[3mm] CALCULATIONS}  \\[3mm]}
\colorrule{Lundgold}{\textwidth}{1.6pt}\\[\baselineskip]
\sectionlinetwo{Lundgold}{88}\vspace{3mm}
{ \textcolor{Lundblue}{With examples of scalar-graviton and graviton-graviton scattering} \\[2mm]}
\sectionlinetwo{Lundgold}{88}\vspace{20mm}
{\bf Safi Rafie-Zinedine}
\\[10mm]
{Department of Astronomy and Theoretical Physics, Lund University}
\\[1cm]
{Master thesis supervised by Prof. Johan Bijnens}
\vfill
\end{center}
\vfill
\AddToShipoutPicture*{\put(135,0){\rotatebox{0}{\scalebox{1}{ \includegraphics[width=1\textwidth,height=10cm]{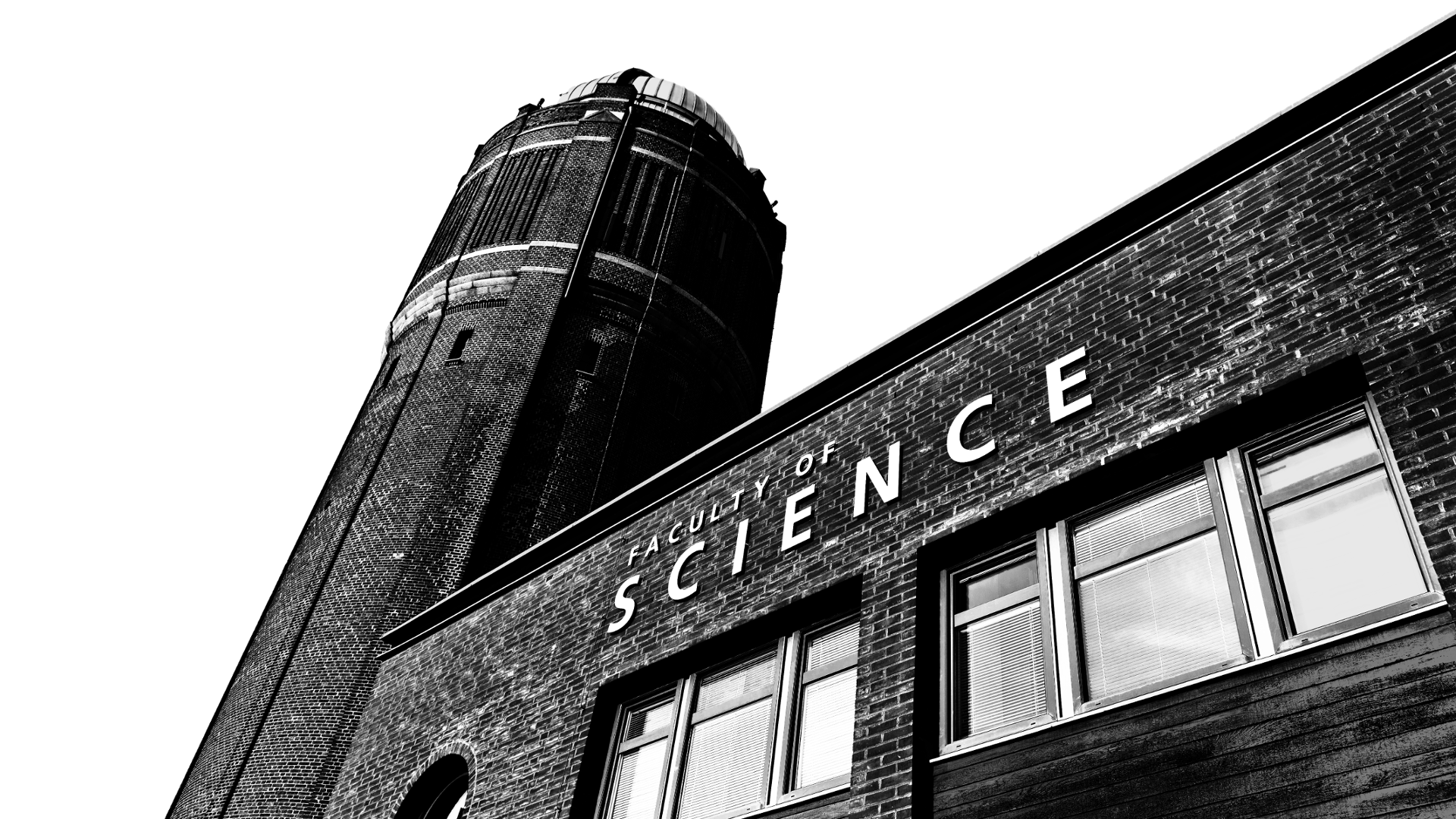}
      }}}}
\vfill
\end{titlepage}
\thispagestyle{empty} 
\phantom{p}
\vfill

{\color{darkgreen}
\begin{tikzpicture}[every node/.style={inner sep=0pt}]
  \node[text width=13cm,align=center](Text){%
    \textcolor{black}{\bf{\emph{\Large{ This work is dedicated to the memory \hspace*{18mm}   \\[6mm] \hspace*{40mm} of my friend Rabeeh Zinedine.}}}} 
} ;
\node[shift={(-1cm,2cm)},anchor=north west](CNW)
at (Text.north west) {};
\node[shift={(1cm,2cm)},anchor=north east](CNE)
at (Text.north east) {};
\node[shift={(-1cm,-2cm)},anchor=south west](CSW)
at (Text.south west) {};
\node[shift={(1cm,-2cm)},anchor=south east](CSE)
at (Text.south east) {};
\pgfornamenthline{CNW}{CNE}{north}{87}
\pgfornamenthline{CSW}{CSE}{south}{87}
\end{tikzpicture}}

\vfill
\clearpage
\thispagestyle{empty} 

\phantom{p}

\section*{\centering \Huge{Abstract}}
\addcontentsline{toc}{section}{Abstract}
The Einstein-Hilbert Lagrangian for gravity is non-renormalizable at loop level.
However, it can be treated in the effective field theory framework which means
that gravity as an effective theory can be renormalized when a proper expansion
of the effective Lagrangian is made. At the same time, the Feynman rules for
gravity are very complicated, although the resulting amplitudes do not have the
same complications. Therefore, in this thesis we want to simplify the Feynman
rules as much as possible by using the most general parameterized gauge condition,
adding all possible parameterized total derivative terms and redefining the
gravitational, ghosts and scalar fields in a general parameterization way.
By choosing the parameters in a specific way, we obtain simplified Feynman
rules, especially the triple and quadruple graviton vertices are simplified. In addition, we verify
our simplified rules by calculating the amplitudes of scalar-graviton and
graviton-graviton scattering at tree level using the simplified and standard Feynman
rules. Finally, we show the utility of these simplified rules by calculating
some one-loop diagrams for scalar-graviton scattering and comparing to the
standard Feynman rules.


\newpage
\thispagestyle{empty}
\phantom{p}
\section*{\centering \LARGE{Popular Science Description}}
In physics, the story of gravity is still incomplete. It began under an apple
tree when Newton started his journey to discover his laws about the gravitational force, but these laws were
not enough to describe all the gravitational phenomena. To describe gravity in
a more accurate way, Einstein came with the theory of general relativity. This
theory treats the gravitational force as a consequence of the curvature of
spacetime. This should be compared with the other forces in nature (the electromagnetic, weak
and strong force) which are described by the standard model of particle physics. This model
treats the forces as a consequence of exchanging particles which are called
quanta. There have been many attempts in the last century to study gravity as a quantized theory, quantum gravity, where the exchanged particles are called gravitons. Quantum gravity
is still not fully understood because of many obstacles, one of them being
its complicated calculations.


The purpose of this thesis is to address this latter problem of
complicated calculations, following the belief that nature should be described
in a beautiful and simple mathematical way. Moreover, a simplified form with
fewer terms that contribute to gravitational effects can lead to a deeper understanding of
gravity. To treat this problem, we want to find mathematical tools that can
simplify the math of the theory without changing the information that it
contains. Fortunately, in quantum physics such tools exist as field redefinition which
means that we can redefine the gravitons in order to find a simpler expression that
can describe exchanging these particles. As a result of applying these mathematical
tools, we successfully simplify the math that describes the gravitational
interaction between particles. In particular, we show that the interaction
between three gravitons can be reduced from 40 to just 4 terms, and the
interaction between four gravitons can be reduced from 113 to 12 terms. Finally,
we verify our simplification by comparing the results for physical processes using the standard approach and using our simplified approach.


\vspace{5mm}
\section*{\centering \LARGE{Acknowledgments}}
\thispagestyle{empty}
\addcontentsline{toc}{section}{Acknowledgments}
\hspace*{\parindent}\ignorespaces Firstly, I would like to express my sincere gratitude to my supervisor
Prof. Johan Bijnens for his support, patience, motivation and immense knowledge.
Besides my supervisor, I would like to thank Prof. Johan Rathsman for his
insightful comments and guidance throughout writing this thesis. Finally, I must express my very great appreciation to my family and everyone who supported and contributed to this thesis. I am also thankful to my friend Jordi Ferré for his valuable comments.

\vfill

\newpage
\tableofcontents

\newpage

\section{Introduction}
\setcounter{equation}{0}
\label{sec:introduction}
In physics, there have been many attempts to study the gravitational field as a
quantized field in order to unify the gravitational force, which is described by General Relativity
(\acrshort{gr}), with the other forces in nature, which are described by the Standard Model
(\acrshort{sm}) using the Quantum Field Theory (\acrshort{qft}) framework
\cite{Peskin,GR}. These attempts have met various obstacles,
either the lack of experimental abilities to explore sufficiently high energy or the lack of needed mathematical tools that
allow the study of gravity as a quantized theory, quantum gravity. The reason
behind the latter problem is that the Einstein-Hilbert Lagrangian of quantum gravity diverges at loop
level so we consider it as a non-renormalizable theory
\cite{GRasQFTDonoghue,HigherSpin}. Thus, such a quantum theory for gravity has not been solved yet, and a full unified theory has not been found yet.

Even so, it is still possible to construct and solve a renormalizable effective theory for
gravity order by order by using the Effective Field Theory (\acrshort{eft}) framework \cite{GRasQFTDonoghue,EFT}. In this
framework, we can study gravity at a particular loop level by expanding the
Lagrangian in the energy expansion up to the relevant terms for this loop level.
To determine which terms are relevant, we use Weinberg’s power counting
theorem \cite{GRasQFTDonoghue}. These new terms in the expansion contain new
parameters which can absorb the divergences from the loop diagrams, and at the
same time can be measured experimentally. Therefore, gravity can be renormalized
at loop level when we make the proper expansion of the Lagrangian. 

At the same time, the Einstein–Hilbert Lagrangian is very complicated and leads to equally
complicated Feynman rules as given in \cite{GhostVertex}, which we will call the
standard Feynman rules. These rules give lengthy complicated calculations, but
they also lead to scattering amplitudes that are simple in general \cite{TreeLevelResults,DonoghueTreeLevel}. Therefore, there is
a strong indication that manipulating this Lagrangian can lead to a simpler form
which still gives the same scattering amplitudes. In addition, when the Lagrangian contains
fewer terms, it is possible to understand the math in the theory better.
Consequently, there has been some efforts to simplify these rules as in \cite{SameWork}, where a parameterized metric field was used.

As starting point for the simplification approach in this thesis, we derive the
Lagrangian for gravity, the Einstein–Hilbert Lagrangian, by using an analogy
with the Yang-Mills (\acrshort{ym}) gauge theory. Then we construct the most
general effective Lagrangian by using Effective Field Theory (\acrshort{eft}).
After that, we set out to simplify the Feynman rules as much as possible by manipulating the Lagrangian for
gravity using the three freedoms \cite{Peskin,GRasQFTDonoghue,EquivalenceTheorem}: choosing a gauge, adding total derivative
terms to the Lagrangian and re-parameterization of the fields (gravitational,
scalar, ghost fields). In other words, we choose the most general parameterized gauge condition, and we
add a parameterization of all possible total derivative terms. Then, we redefine
the gravitational, ghosts and scalar fields using a general parameterization. As a result of parameterizing the previous freedoms, we efficiently reduce the problem of simplifying Feynman rules as much as possible to solve a system of linear equations for eight sets of parameters.


In addition, for comparison and verification purposes, we perform the
calculations in two approaches. In the standard calculations, we use
the de Donder gauge to obtain the standard Feynman rules as shown in App.~\ref{Ap:FeynmanRules}, which agree with those in \cite{GhostVertex}. In the simplified calculations, we use the three freedoms that we mentioned before, and then choose the parameters in order to obtain Feynman rules as simple as possible, especially the triple and quadruple graviton vertices.

To check our simplified rules, we compare the resulting amplitudes of
scalar-graviton and graviton-graviton scattering at tree level using the
standard rules and using the simplified ones \cite{TreeLevelResults,DonoghueTreeLevel,HelicityAmplitudes}. Furthermore, to show the utility of
these simplified rules, we compare the standard and simplified
calculations of some one-loop diagrams for scalar-graviton scattering, where we
use the dimensional regularization scheme with the Passarino-Veltman method to
calculate the loop integrals \cite{MasslessTadpole,PassarinoVeltman}. Finally,
since we are interested in the calculations up to one-loop level for scalar-graviton scattering, we only simplified the lowest order vertices.

In this thesis, we start in Sec.~\ref{TheoreticalBackground} by discussing
the theoretical background for deriving the Lagrangian for gravity and show the
tools needed for manipulating the Lagrangian and calculating scattering
processes. Then in Sec.~\ref{se:FeynmanRules} we calculate the Feynman rules in
the standard and simplified way. After that, in Sec.~\ref{se:TreeLevel} we use the Feynman
rules to calculate the amplitudes of scalar-graviton and graviton-graviton
scattering at tree level in both ways. Moreover, in
Sec.~\ref{se:OneLoopCorrection} we show the usefulness of our Feynman rules by
comparing calculations of some one-loop diagrams for scalar-graviton scattering
using both approaches. Finally, the conclusions of our work is in Sec.~\ref{se:conclusion}.

Before we start, it is important to mention that we follow the conventions in
\cite{GRasQFTDonoghue} throughout this thesis, such as the metric signature
$(+,-,-,-)$ and the natural units $c=\hbar=1$. In addition, since the calculations of Feynman rules
and the amplitudes are extremely lengthy to do by hand, we use the FORM program \cite{FORM,FORMGitHub} to perform them.

\section{Theoretical Background \label{TheoreticalBackground}}
\setcounter{equation}{0}
In this section, we start by discussing our motivation behind choosing spin-2
for the graviton \cite{Peskin,GR,GRasQFTDonoghue,HigherSpin}. Then, we give the derivation of the
Einstein–Hilbert Lagrangian using an analogy with \acrshort{ym} theory
\cite{GR,GRasQFTDonoghue}. After that, using the \acrshort{eft} framework, we build the most general effective
Lagrangian for gravity which is needed to study gravity at one-loop level
\cite{GRasQFTDonoghue,EFT}. In addition, we discuss the three freedoms that we
use to simplify Feynman rules: choosing the gauge \cite{Peskin,GRasQFTDonoghue}, adding
total derivative terms \cite{Peskin}, and field redefinitions
\cite{EquivalenceTheorem}.

\subsection{Spin of Graviton \label{se:GR}}
\begin{figure}[H]
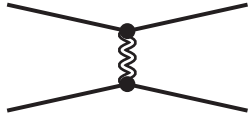

	\centering
  \noindent
  \begin{axopicture}(90,50)
    \SetWidth{1.5}
    \Line(0,10)(45,20)
    \Line(45,20)(90,10)
    \Line(0,50)(45,40)
    \Line(45,40)(90,50)
    \SetWidth{1.}
    \DoublePhoton(45,20)(45,40){2}{3.5}{2}
    \Vertex(45,20){3}
    \Vertex(45,40){3}
  \end{axopicture}
  \caption{Scalar-Scalar scattering at tree level.}
	\label{fig:spin0}
\end{figure}

As in this thesis we want to quantize the gravitational field, let us start by
discussing the possible spins for its quanta, which are called gravitons \cite{GRasQFTDonoghue}. Since the graviton is a boson, its spin has to be an integer
number. In this case, there are three possibilities: spin-0, spin-1 and spin-2 while higher spins are not consistent with \acrshort{qft} for
fundamental particles with interactions \cite{HigherSpin}. Firstly, if we start
with a spin-0 particle as the Higgs boson, the scalar-scalar scattering via a spin-0
graviton, as shown in Fig.~\ref{fig:spin0}, leads to a Newtonian
gravitational potential with the bare mass as the source
\cite{Peskin,GRasQFTDonoghue}. However, we know from \acrshort{gr} that the bare
mass of an object is not the only source of the gravitational field \cite{GR}.

\noindent Secondly, with a spin-1 particle as the photon, the scalar-scalar
scattering leads to an attractive or repulsive potential, so spin–1
particles are not appropriate to describe the gravitational force \cite{GRasQFTDonoghue}.
Finally, in the case of a spin-2 particle, the scalar-scalar scattering leads to a
gravitational potential with the energy-momentum tensor $\mathcal{T}_{\mu\nu}$ as the source
for gravity, which is consistent with \acrshort{gr} \cite{GR,GRasQFTDonoghue}.

From the above discussion, it is clear that the graviton should be a spin-2 boson in order to be
consistent with \acrshort{gr} and \acrshort{qft}, and later on we will see more
indications that the graviton should be a spin-2 particle.

\subsection{Analogy with Yang-Mills Theory \label{se:YangMill}}
The main idea of a Yang-Mills (\acrshort{ym}) theory is to search for an
appropriate global symmetry which is relevant to the force that we study. Then,
we convert this symmetry to a local or gauge symmetry, where this conversion is
called gauging the symmetry. After that, to preserve the local gauge
invariance, we need to insert a new field which will be the field of the gauge boson for this force.

For example, consider the following, matter Lagrangian for a massive real scalar field $\phi$:
\begin{equation}
  \mathcal{L}_{\text{Matter}}  = \frac{1}{2} \eta^{ab} \partial_a\phi \partial_b\phi - \frac{1}{2} m^2 \phi^2 \, , \label{eq:flatL}
\end{equation}
where $\eta^{ab}$ is the Minkowski metric, and $m$ is the mass. This Lagrangian is invariant under the global translational symmetry
\begin{equation}
  y^a  \;\;\rightarrow\;\;  y^{\prime a} = y^a + d^a \, . \label{eq:GlobalSymmetry}
\end{equation}
Then, we convert this global translational symmetry to a local translational symmetry, where the latter is called the General Coordinate Transformations (\acrshort{gct}),
\begin{equation}
    \label{eq:GCT}
 x^{\mu} \;\;\rightarrow\;\;  x^{\prime \mu} = x^{\mu} + d^{\mu}(x) \, ,
\end{equation}
where Lorentz indices $ a,b,\cdots $ in flat space-time have been replaced by
world indices $ \mu,\nu,\cdots $ in curved space-time. In addition, we need to
replace the Minkowski metric in flat space $ \eta_{a b}$ by the metric field in
curved space $ g_{\mu\nu}$ in order to make the Lagrangian Eq.~(\ref{eq:flatL}) invariant under \acrshort{gct} Eq.~(\ref{eq:GCT}). Thus, the Lagrangian becomes

\begin{equation}
    \label{eq:MLagrangian}
    \mathcal{L}_{\text{Matter}}  = \frac{1}{2} g^{\mu\nu} \partial_{\mu}\phi \partial_{\nu}\phi - \frac{1}{2} m^2 \phi^2  \, ,
\end{equation}
which contains a new field, the metric field $ g^{\mu\nu} $, that represents
the spin-2 gauge boson for gravity (i.e., the graviton).

As with other gauge theories, we need to introduce a kinetic term
for the gravitational field $g^{\mu\nu}$. We search for a quantity similar to the
\acrshort{ym} field strengh tensor $F_{\mu\nu}$, which we recall is related to the
commutator of covariant derivatives $  [D_{\mu},D_{\nu}] $. For gravity, with a
vector field $ V^{\beta} $, this commutator can be written as \cite{GRasQFTDonoghue}
\begin{align}
	[D_{\mu},D_{\nu}] V^{\beta} & = R_{\mu\nu\alpha}^{\;\;\;\;\;\;\beta} V^{\alpha} \label{eq:Covariant}  \, , 
\end{align}                                
where $D_{\nu}V_{\beta}  =\partial_{\nu} V_{\beta} -
\Gamma_{\beta\nu}^{\gamma}V_{\gamma}$ is the covariant derivative, $R_{\mu\nu\alpha}^{\;\;\;\;\;\;\beta}$ is the Riemann tensor given by
\begin{align}
  R_{\mu\nu\alpha}^{\;\;\;\;\;\;\beta}& = \partial_{\mu} \Gamma_{\nu\alpha}^{\;\;\;\;\beta} - \partial_{\nu} \Gamma_{\mu\alpha}^{\;\;\;\;\beta} + \Gamma_{\mu\sigma}^{\;\;\;\;\beta} \Gamma_{\nu\alpha}^{\;\;\;\;\sigma} - \Gamma_{\nu\sigma}^{\;\;\;\;\beta} \Gamma_{\mu\alpha}^{\;\;\;\;\sigma} \, , \label{eq:Riemann} 
\end{align}                                
and $\Gamma_{\nu\alpha}^{\;\;\;\;\beta}$ is the Christoffel symbol given by
\begin{align}
  \Gamma_{\nu\alpha}^{\;\;\;\;\beta}&=\frac{1}{2}g^{\beta\rho}(\partial_{\nu} g_{\rho\alpha} + \partial_{\alpha} g_{\rho\nu}-\partial_{\rho} g_{\nu\alpha}  ) \, .
\end{align}
In addition, from the Riemann tensor we can also get the Ricci tensor $R_{\nu\alpha}$ and the scalar curvature $R$ as
\begin{align}
  R_{\nu\alpha}&=R_{\mu\nu\alpha}^{\;\;\;\;\;\;\mu} \label{eq:Ricci} \, ,\\[1mm]
  R & =g^{\nu\alpha} R_{\nu\alpha}=g^{\nu\alpha} R_{\mu\nu\alpha}^{\;\;\;\;\;\;\mu} \label{eq:ScalarCurvature} \, .
\end{align}
We also use the standard expansion of the weak gravitational field around the
Minkowski metric $\eta_{\mu\nu}$, that is used to raise or lower indices, given by
\begin{equation}
  \label{eq:WeakFieldExpansion}
  g_{\mu\nu}=   \eta_{\mu\nu} +  \kappa \; h_{\mu \nu} \, ,
\end{equation}
where $h_{\mu\nu}$ is the canonical quantized gravitational field, and $\kappa$ is the Newtonian strength of gravitational interactions.

In this situation, we have three quantities: $R_{\mu\nu\alpha\beta}$, $R_{\nu\alpha}$ and $R$ that are related to the commutator of covariant derivatives.
One of them can be chosen as a field strengh tensor for the gravitational field
to build the kinetic term in analogy with $\mathcal{L}_{\text{(Kin, YM)}}=-\frac{1}{4}F^{\mu\nu}F_{\mu\nu}$. To do this, we use that the \acrshort{ym} field strengh tensor is antisymmetric,
$F_{\mu\nu}=-F_{\nu\mu}$, and has one partial derivative $F_{\mu\nu}\sim\partial A$, while the Riemann tensor
$R_{\mu\nu\alpha\beta}=R_{\alpha\beta\mu\nu}$ and the Ricci tensor
$R_{\nu\alpha}=R_{\alpha\nu}$ do not have the same symmetry property and have
two partial derivatives $R_{\mu\nu\alpha\beta},R_{\mu\nu}\sim\partial\partial
h$. In addition, the kinetic term should be Lorentz invariant and \acrshort{gct}
invariant. So, the simplest combinations for the kinetic term are
$R\sim\partial\partial h$ as well as $R_{\mu\nu\alpha\beta}R^{\mu\nu\alpha\beta},R_{\mu\nu}R^{\mu\nu}\sim\partial\partial\partial\partial
h$. It follows that $R$ is the most relevant one of them for weak-field since it
has only two partial derivatives while the other have four. Thus, we can write
the kinetic term for the gravitational field in terms of the scalar curvature as
the Einstein-Hilbert Lagrangian for gravity \cite{GRasQFTDonoghue}
\begin{equation}
  \label{eq:GLagrangian}
	\mathcal{L}_{\text{Gravity}} = - \frac{2}{\kappa^2} R   \, .
\end{equation}

In addition, when we move from flat to curved space, there is a correction of
the measure $d^4y$ in the action $ \mathcal{S}$ as
\begin{align}
  d^4y=\sqrt{-\det(g_{\mu\nu})}\, d^4x=\sqrt{-g}\, d^4x \, ,
\end{align}
where $\sqrt{-g}$ is the square root of the determinant of the metric tensor
which is given by
\begin{align}
    \sqrt{-g} & = \sqrt{-\det(g_{\mu\nu})} = \Big (-\det(\eta_{\mu\lambda})\det(\delta^{\lambda}_{\nu} + \kappa h^{\lambda}_{\;\;\nu} + \cdots) \Big )^{1/2} =\Big (e^{\tr(\ln(\delta^{\lambda}_{\nu} + \kappa h^{\lambda}_{\;\;\nu} + \cdots))} \Big )^{1/2} \nonumber\\[2mm]
    & = \sum_{i=0}^{\infty} \frac{1}{i!} \Big ( \frac{1}{2} \sum_{j=1}^{\infty} \frac{(-1)^{j+1}}{j} (\kappa h^{\lambda}_{\;\;\lambda} + \cdots  )^j     \Big )^i \, , \label{eq:determinant}
  \end{align}
and we have used that the metric tensor can be written as $g_{\mu\nu} =\eta_{\mu\lambda}
(\delta^{\lambda}_{\nu} + \kappa h^{\lambda}_{\;\;\nu} + \cdots)$. In this
thesis, we take the expansion in $i$ and $j$ up to four which is relevant to the lowest
order vertices as we will see later.

Summarizing, the actions of matter and gravity become
\begin{align}
  \mathcal{S}_{\text{Matter}}  &= \int d^4x \sqrt{-g}  \Big (  \frac{1}{2} g^{\mu\nu} \partial_{\mu}\phi \partial_{\nu}\phi - \frac{1}{2} m^2 \phi^2\Big)   \, ,   \label{eq:MAction}\\[2mm]
	\mathcal{S}_{\text{Gravity}} &= \int d^4x \sqrt{-g}  \Big ( - \frac{2}{\kappa^2} R \Big )  \, . \label{eq:GAction}
\end{align}
From the matter action, the variation with respect to $g^{\mu\nu}$ is
\begin{align}
 \frac{2}{\sqrt{-g}} \frac{\delta \mathcal{S}_{\text{Matter}}}{\delta
  g^{\mu\nu}} = \mathcal{T}_{\mu\nu}  \, ,
\end{align}
where $\mathcal{T}_{\mu\nu} =  \partial_{\mu} \phi \partial_{\nu} \phi -
\frac{1}{2} g_{\mu\nu}  ( g^{\alpha\beta} \partial_{\alpha} \phi
\partial_{\beta} \phi - m^2 \phi^2 )$ is the Energy-Momentum Tensor (\acrshort{emtt}). Thus,
the \acrshort{emtt} is the conserved current that follows from \acrshort{gct}
which is consistent with \acrshort{gr}, where \acrshort{emtt} is the source for gravity. This is an indication that we are dealing with the correct symmetry for gravity. Moreover, the total action is
\begin{equation}
	\mathcal{S}_{\text{Total}} = \int dx^4 \sqrt{-g} \Big (- \frac{2}{\kappa^2} R + \frac{1}{2} g^{\mu\nu} \partial_{\mu}\phi \partial_{\nu}\phi - \frac{1}{2} m^2 \phi^2 \Big )  \, . \label{eq:TotalAction}
\end{equation}
From this action, the equation of motion for $g^{\mu\nu}$ will be
\begin{align}
  \delta \mathcal{S} & = \int dx^4 \sqrt{-g} \Big ( - \frac{2}{\kappa^2} R_{\mu\nu} + \frac{2}{\kappa^2} \frac{1}{2} g_{\mu\nu} R  + \frac{1}{2} \mathcal{T}_{\mu\nu}  \Big ) \delta g^{\mu\nu} = 0  \, , \\[2mm]
  & \qquad \qquad \Rightarrow  \qquad R_{\mu\nu} - \frac{1}{2} g_{\mu\nu} R = \frac{\kappa^2}{4} \mathcal{T}_{\mu\nu}  \, , \label{eq:Einstein}                    
\end{align}
This equation can be recognized from \acrshort{gr} as Einstein's
field equation with $ \frac{\kappa^2}{4}  = 8 \pi G$, where $G$ is the
gravitational constant \cite{GR}. Reaching Einstein's equation is another indication that we are gauging the correct global symmetry for the gravitational force.

However, when we calculate scattering at loop level with this Lagrangian
Eq.~(\ref{eq:GLagrangian}), we get UV divergences. Therefore, we need to
construct gravity as an effective field theory in order to study this theory at loop level.

\subsection{Effective Field Theory \label{se:EFT}}
\acrlong{eft} (EFT) is a way to study physics in a particular range of energy while neglecting the
physics at higher energy. This can be done if the contributions from high energy are
 small when the theory is studied in the low energy range. For example, we can
 study the hydrogen atom using the Schrödinger equation, neglecting the quark and gluon
interactions inside the nucleus. Mathematically, it means that we need to expand
and organize the Lagrangian according to the dimension of the energy operators,
where in the case of gravity, the energy operators are just partial derivatives. This
expansion is called the energy expansion of the Lagrangian, and it separates the terms
which are relevant at high energy, from the terms which are relevant at
low energy. In other words, the effective Lagrangian can be written in the energy expansion as
\begin{align}
  \mathcal{L}_{\textsf{eff}}  = \mathcal{L}_0 + \mathcal{L}_1 + \mathcal{L}_2 + \mathcal{L}_3 + \cdots  \, ,
\end{align}
where the term $ \mathcal{L}_0 $  does not contain any energy operator \(\mathcal{O}(E^0) \), and it is just a constant. The term $\mathcal{L}_1 $ contains energy operators of dimension one \(\mathcal{O}(E^1)\) and so on.

In the case of gravity, the quantities that can be used to construct the
effective Lagrangian are: $R_{\mu\nu\alpha\beta}$, $R_{\nu\alpha}$ and $R$, which as
already mentioned are related to the commutator of the covariant
derivative Eq.~(\ref{eq:Covariant}). However, each of these quantities ($R_{\mu\nu\alpha\beta},
R_{\nu\alpha},R\sim\partial\partial h$) contain two partial derivatives, which
as already mentioned are energy operators of the gravitational field. Therefore, only even energy dimensions are possible in the energy expansion
\begin{align}
\mathcal{L}_{\textsf{eff}}  = \mathcal{L}_0 + \mathcal{L}_2 + \mathcal{L}_4 + \mathcal{L}_6 + \cdots  \, .
\end{align}

At the same time, since in this thesis we are only interested in studying gravity up to
one-loop order, it is useful to use Weinberg’s power counting theorem \cite{GRasQFTDonoghue}. This
theorem can tell us how many terms in the energy expansion should be taken into
account in order that the theory can be renormalized at one-loop. According to this theorem, the energy
dimension $\mathcal{D}$ of a diagram with \(N_L\) loops and \(N_n\) vertices
arising from the effective Lagrangian terms that contain \(n\) derivatives is given by
\begin{align}
  \label{eq:PowerCounting}
  \mathcal{D} = 2 + \sum_{n} N_n (n-2) + 2 N_L  \, .
\end{align}

In our case, we want to calculate the energy dimension \(\mathcal{D}\) for
gravity with $n=2$ up to one-loop order $N_L=1$. This gives
$\mathcal{D}=4$, which means that gravity can be renormalized at one-loop, if
we take into account terms up to energy dimension four \(\mathcal{O}(E^4)\)
in the energy expansion of the Lagrangian.

Let us discuss the possible terms that can be inserted into the effective
Lagrangian for gravity in more detail. First, the terms should be \acrshort{gct}
invariant and Lorentz invariant. Then we need to organize the terms according to
the energy dimension as follows:
\begin{itemize}
\item $\mathcal{L}_0$: This is only a constant such as the Cosmological constant $\Lambda$.
\item $\mathcal{L}_2$: The only possible term is $R\sim\partial\partial$.
\item $\mathcal{L}_4$: There are three possible combinations: $R^2$,
  $\, R_{\mu\nu} R^{\mu\nu} $, $\, R_{\mu\nu\alpha\beta} R^{\mu\nu\alpha\beta} \sim  \partial\partial\partial\partial$.
\item $\mathcal{L}_6$: There are four possible combinations: $R^3$,
  $\, R\, R_{\mu\nu} R^{\mu\nu}$, $\, R\, R_{\mu\nu\alpha\beta}
  R^{\mu\nu\alpha\beta} $, $\, R^{\mu\nu} R^{\alpha\beta} R_{\mu\nu\alpha\beta}\sim  \partial\partial\partial\partial\partial\partial $.
\end{itemize}
Thus, the most general effective Lagrangian in the energy expansion up to energy
dimension four is given by
\begin{align}
	\mathcal{L}_{\text{eff}} & = \mathcal{L}_0 + \mathcal{L}_2 + \mathcal{L}_4 + \cdots \nonumber   \\
                             & = - \Lambda - \frac{2}{\kappa^2} R + c_1 R^2 + c_2 R_{\mu\nu} R^{\mu\nu} + c_3 R_{\mu\nu\alpha\beta} R^{\mu\nu\alpha\beta} + \cdots  
                               \label{eq:EffectiveLagrangian} \, ,
\end{align}
where \(c_i\) are numerical coefficients of the higher order corrections, which
are called counter terms. In addition, \(\Lambda \) is experimentally very small so we will
neglect it as the following.

When we use the \(\kappa\) term ($-\frac{2}{\kappa^2}R$) for one-loop
calculations, we get UV divergences which are absorbed by $c_i$. The finite parts
of $c_i$ have to be determined experimentally. Thus, the theory can be renormalized at one-loop. In
addition, when we calculate scalar-graviton scattering up to one-loop order from the
above Lagrangian, the $\kappa$ term contributes to one-loop diagrams while the $\kappa$ and $c_i$ terms contribute to tree level diagrams.

\begin{figure}[H]
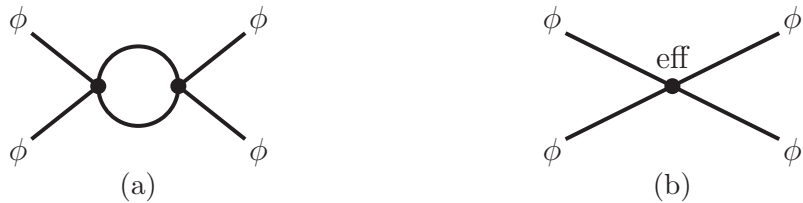

  \vspace{3mm}
    \centering
    \begin{subfigure}[b]{0.4\linewidth}
      \centering
      \begin{axopicture}(90,50)
        \SetWidth{1.5}
        \Line(5,5)(30,25)
        \Line(5,45)(30,25)
        \Text(0,0){$\phi$}
        \Text(90,0){$\phi$}
        \Line(60,25)(85,5)
        \Line(60,25)(85,45)
        \Text(0,50){$\phi$}
        \Text(90,50){$\phi$}
        \Arc(45,25)(15,0,180)
        \Arc[clockwise](45,25)(15,0,180)
        \Vertex(30,25){3}
        \Vertex(60,25){3}
      \end{axopicture}
      \caption{}
      \label{fig:loop EFT}
    \end{subfigure}%
    ~~~
    \begin{subfigure}[b]{0.4\linewidth}
      \centering
      \begin{axopicture}(90,50)
        \SetWidth{1.5}
        \Line(5,5)(45,25)
        \Line(45,25)(85,5)
        \Line(5,45)(45,25)
        \Line(45,25)(85,45)
        \Text(0,0){$\phi$}
        \Text(90,0){$\phi$}
        \Text(0,50){$\phi$}
        \Text(90,50){$\phi$}
        \Vertex(45,25){3}
        \SetWidth{1.}
        \Text(45,32)[b]{eff}
      \end{axopicture}
      \caption{}
      \label{fig:vertex EFT}
    \end{subfigure}
    \caption{(a) one-loop diagram at high energy in a
      full theory. (b) its corresponding vertex diagram in its \acrshort{eft} at
      low energy when locality is restored.}
  \end{figure}

Finally, let us discuss locality of the theory \cite{GRasQFTDonoghue}. \acrshort{eft} is local
if the higher order operators in the energy expansion are taken into account.
However, if we neglect these higher order operators, the theory becomes
non-local at high energy whereas locality will be restored at low energy. For
example, in the full theory, the one-loop diagram as shown in
Fig.~\ref{fig:loop EFT} is local at large energy. On the other hand, in its
corresponding \acrshort{eft} and after neglecting the higher order operators,
this diagram becomes non-local. However, locality will be restored at low energy where this loop will be
reduced to a vertex as shown in Fig.~\ref{fig:vertex EFT}. In addition, using \acrshort{eft} the diagram becomes easier to calculate by reducing this loop to a vertex.

\subsection{Lagrangians for Fixing Gauge and Ghosts \label{se:LFGandLGH}}
Since the Lagrangian for gravity has more degrees of freedom than its gauge
boson, we need to fix the gauge and introduce ghosts in order to get rid of
the extra degrees of freedom. The gauge boson in our case is a spin-2 massless graviton, which
has a transverse, traceless, symmetric polarization tensor, with two degrees of freedom.

Another way to see why we need to fix the gauge and introduce ghosts is by
considering the following \acrshort{gct}, where $\xi^{\mu}(x)$ is an infinitesimal translation,
\begin{align}
	x^{\mu} \; \rightarrow x'^{\mu} = x^{\mu} - \xi^{\mu}(x) \, .
\end{align}
In this case, the metric $g_{\mu\nu}$  transforms as
\begin{align}
  & g_{\mu\nu} \; \rightarrow g'_{\mu\nu}(x')  = g_{\alpha\beta}(x) \Big (\frac{\partial x^{\alpha}}{\partial x'^{\mu}} \Big ) \Big (\frac{\partial x^{\beta}}{\partial x'^{\nu}} \Big ) = g_{\alpha\beta}(x) \Big (\delta_{\mu}^{\alpha}+\partial_{\mu} \xi^{\alpha} (x) \Big ) \Big (\delta_{\nu}^{\beta}+\partial_{\nu} \xi^{\beta} (x) \Big ) \,.
\end{align}
We also consider the weak gravitational field expansion, Eq.~(\ref{eq:WeakFieldExpansion}), so the transformation of the gravitational field $h_{\mu\nu}$, the gauge transformation, is
\begin{align}
    h_{\mu\nu} \; \rightarrow h'_{\mu\nu} & = h_{\mu\nu} + \partial_{\mu} \xi_{\nu}(x) + \partial_{\nu} \xi_{\mu}(x) + h_{\mu\sigma} \partial_{\nu} \xi^{\sigma}(x) + h_{\nu\sigma} \partial_{\mu} \xi^{\sigma}(x) +  \xi^{\sigma}(x)  \partial_{\sigma} h_{\mu\nu} \label{eq:GaugeTransformation} \, .
\end{align}
However, since the Lagrangian for gravity is invariant under this
transformation, we get a redundancy in the description of the physical
system. This redundancy can also be seen from the path integral formulation of
the generating functional
\begin{align}
	\mathcal{Z} = \int D[h] \; \exp( i \, S(h)  ) = \int D[h] \; \exp( i \,\int d^4x  \mathcal{L} (h)  )
  \label{eq:GeneratingFunctional} \, ,
\end{align}
where the measure \(\int D[h]\) is performed over all configurations of
\(h\), including those configurations that are equivalent under the gauge
transformation, Eq.~(\ref{eq:GaugeTransformation}).

Because of this, we need to choose a gauge condition to build two Lagrangians
which together can remove this redundancy. The first Lagrangian is to
fix the gauge, and the second Lagrangian is to correct the first one depending on
the choice of the gauge condition. In other words, we need to insert a
Lagrangian for fixing the gauge $\mathcal{L}_{\text{FG}}$ and a Lagrangian for ghosts $\mathcal{L}_{\text{GH}}$ into the
Lagrangian for gravity in order to remove this
redundancy. In the path integral formalism, this means that using
$\mathcal{L}_{\text{FG}}$ and $\mathcal{L}_{\text{GH}}$  will correct the measure \(\int D[h]\) to be
performed over the correct configurations of $h$. Next, we will follow the
Faddeev-Popov method to derive the Lagrangian for fixing the gauge
$\mathcal{L}_{\text{FG}}$ and the Lagrangian for ghosts $\mathcal{L}_{\text{GH}}$. 

The Faddeev-Popov method depends on multiplying the generating functional by two
identities and then performing the integral in order to get a new generating
functional with two new terms, where one of them is related to $\mathcal{L}_{\text{FG}}$ and the other is
related to $\mathcal{L}_{\text{GH}}$.

To do that, we use the following identity, always taking into account the gauge
transformation Eq.~(\ref{eq:GaugeTransformation}),
\begin{align}
    \label{eq:FirstIdentity}
    1 = & \int D[\xi] \; \delta \Big (  \mathcal{C}_{\mu} (h) - F_{\mu}(x) \Big ) \; \Delta(h) \, ,
\end{align}
where $\mathcal{C}_{\mu}(h) = F_{\mu}(x)$ is the gauge condition, and $\Delta(h)
  = \det \Big ( \frac{\partial \mathcal{C}_{\alpha}(h)}{\partial \xi_{\beta}} \Big )
  $ is Faddeev-Popov determinant. \\
Since the generating functional Eq.~(\ref{eq:GeneratingFunctional}) does not depend
on $\xi$, it is possible to insert the above identity into it as
\begin{equation}
    \label{eq:IndependentFx}
  \mathcal{Z} = \; \int D[h] \; D[\xi] \; \delta \Big (  \mathcal{C}_{\mu} (h) - F_{\mu}(x) \Big ) \; \Delta(h) \; \exp(i S) \, .
\end{equation}
We also use the following identity
\begin{align}
        1 = &  N \; \int D[F] \; \exp \Big ( - \frac{i}{2\epsilon} \int d^4 x  F_{\nu}(x)F^{\nu}(x) \Big ) \, , \label{eq:SecondIdentity}
\end{align}
where $N$ is a normalization constant, and $\epsilon$ is a parameter.\\
Again, since the generating functional Eq.~(\ref{eq:IndependentFx})  does not depend on $F(x)$, it is also possible to
insert the above identity into it as
\begin{align}
  \mathcal{Z} = N \;  \int D[F] \; D[h] \; D[\xi] \; \delta \Big (  \mathcal{C}_{\mu} (h) - F_{\mu}(x) \Big ) \; \Delta(h) \; \exp \Big ( i S - \frac{i}{2\epsilon} \int d^4 x  F_{\nu}(x)F^{\nu}(x) \Big ) \, .
\end{align}
Then, performing the integral over \(\xi, F(x) \) yields 
\begin{align}
	&   \mathcal{Z} = N' \; \int D[h] \; \Delta(h) \; \exp \Big ( i S - \frac{i}{2\epsilon} \int d^4 x \; \mathcal{C}_{\nu}(h) \mathcal{C}^{\nu}(h) \Big )   \, ,
\end{align}
where $N'$ is a new normalization constant. However, the
Faddeev-Popov determinant $\Delta(h)$ can be written in terms of an artificial vector fermion field \(\chi_{\beta}\), the ghost field, and an anti-fermion field $\bar{\chi}_{\alpha}$, the anti-ghost field, as
	\begin{align}
    & \qquad \qquad \Delta(h)  = \det \Big ( \frac{\partial \mathcal{C}_{\alpha}(h) }{\partial \xi_{\beta}} \Big ) = \int D[\bar{\chi}] D[\chi]  \exp \Big ( i \int d^4 x \; \bar{\chi}^{\alpha} \; \frac{\partial \mathcal{C}_{\alpha}(h) }{\partial \xi_{\beta}} \; \chi^{\beta} \Big ) \, .
	\end{align}
Thus, the generating functional becomes
\begin{align*}
	&   \mathcal{Z} = N'  \int D[h] \,D[\bar{\chi}] \, D[\chi] \, \exp \Bigg ( i S + i \int d^4 x \Big ( \bar{\chi}^{\alpha} \; \frac{\partial \mathcal{C}_{\alpha}(h) }{\partial \xi_{\beta}} \; \chi^{\beta} \Big )  - i \int d^4 x \Big ( \frac{1}{2\epsilon} \mathcal{C}_{\nu}(h) \mathcal{C}^{\nu}(h)\Big )  \Bigg )  \, ,
\end{align*}
where we have two new terms:
\begin{itemize}
\item Lagrangian for ghost fields:
\begin{align}
    \label{eq:LagrangianGhosts}
    \mathcal{L}_{\text{ghost}}(\bar{\chi},\chi,h) = \bar{\chi}^{\alpha} \; \frac{\partial \mathcal{C}_{\alpha}(h) }{\partial \xi_{\beta}} \; \chi^{\beta} \, .
\end{align}
\item Lagrangian for fixing gauge:
\begin{equation}
  \label{eq:LagrangianFixing}
  \mathcal{L}_{\text{FG}}(h) = \frac{1}{2\epsilon} \mathcal{C}_{\nu}(h) \mathcal{C}^{\nu}(h) \, ,
\end{equation}
where $\epsilon$ is a parameter that can be chosen differently for different
gauges. However, Feynman–'t Hooft gauge, $\epsilon=1$ will be used throughout the thesis.
\end{itemize}
Finally, we calculate $\mathcal{L}_{\text{FG}}$ and $\mathcal{L}_{\text{ghost}}$ in two approaches. In the standard approach, we use the de Donder (harmonic) gauge condition
\begin{equation}
  \label{eq:GaugeDonder}
  \mathcal{C}_{\mu}(h) = \;\, \kappa \Big [ \partial^{\nu} h_{\mu\nu} - \frac{1}{2} \partial_{\mu} h_{\nu}^{\;\;\nu} \Big ] \, .
\end{equation}  
In the simplified approach, we use a general parameterized gauge condition which
contains all possible combinations for ($\partial h$, $\partial hh$, $\partial
hhh$) as given below
\begin{align*}
  \mathcal{C}_{\mu}(h)  =  & \;\, \kappa \Big [
                      b_1 \partial^{\nu} h_{\mu\nu}
                     +b_2 \partial_{\mu} h_{ \nu}^{\;\;\nu}
                     \Big ]
\\ &  +\kappa^2 \Big [
                   b_3 \partial_{\mu} h_{ \nu}^{\;\;\nu} h_{\alpha}^{\;\;\alpha}
                  +b_4 \partial_{\mu} h^{ \nu \alpha} h_{\nu \alpha}
                  +b_5 \partial^{\nu} h_{ \mu \nu} h_{\alpha}^{\;\;\alpha}
                  +b_6 \partial_{\nu} h_{ \mu \alpha} h^{\nu \alpha}
  \\ &  \qquad\quad               +b_7 \partial_{\nu} h^{ \nu \alpha} h_{\mu \alpha}
                  +b_8 \partial^{\nu} h_{ \alpha}^{\;\;\alpha} h_{\mu \nu}
                  \Big ]
\\ &   +\kappa^3 \Big [
                 b_9 \partial_{\mu} h_{ \nu}^{\;\;\nu} h_{\alpha}^{\;\;\alpha} h_{\beta}^{\;\;\beta}
                + b_{10} \partial_{\mu} h_{ \nu}^{\;\;\nu} h^{\alpha \beta} h_{\alpha \beta}
                + b_{11} \partial_{\mu} h^{ \nu \alpha} h_{\nu \alpha} h_{\beta}^{\;\;\beta}
                + b_{12} \partial_{\mu} h^{ \nu \alpha} h_{\alpha}^{\;\;\beta} h_{\beta \nu}
  \\ & \qquad\quad                + b_{13} \partial^{\nu} h_{ \mu \nu} h_{\alpha}^{\;\;\alpha} h_{\beta}^{\;\;\beta}
                + b_{14} \partial^{\nu} h_{ \mu \nu} h^{\alpha \beta} h_{\alpha \beta}
                + b_{15} \partial_{\nu} h_{ \mu \alpha} h^{\nu \alpha} h_{\beta}^{\;\;\beta}
                + b_{16} \partial^{\nu} h_{ \mu \alpha} h^{\alpha \beta} h_{\beta \nu}
 \\ & \qquad\quad                + b_{17} \partial_{\nu} h^{ \nu \alpha} h_{\mu \alpha} h_{\beta}^{\;\;\beta}
                + b_{18} \partial^{\nu} h^{ \alpha \beta} h_{\mu \alpha} h_{\nu \beta}
                + b_{19} \partial^{\nu} h_{ \nu \alpha} h_{\mu \beta} h^{\alpha \beta}
                + b_{20} \partial_{\alpha} h_{ \nu}^{\;\;\nu} h_{\mu \beta} h^{\alpha \beta}
 \\ & \qquad\quad                + b_{21} \partial^{\nu} h_{ \alpha}^{\;\;\alpha} h_{\mu \nu} h_{\beta}^{\;\;\beta}
                + b_{22} \partial^{\nu} h^{ \alpha \beta} h_{\mu \nu} h_{\alpha \beta}
      \Big ] + \cdots \numberthis \label{eq:Gauge} \, ,
\end{align*}
where $b_i$ are the parameters that will be chosen later to
simplify the Feynman rules as much as possible. In addition, note that we consider only up to
three powers of $h$ which are relevant to the lowest order vertices, that we need to
calculate scalar-graviton scattering to one-loop as will be shown later.

\subsection{Total Derivative Lagrangians \label{se:TD}}
In general, a transformation of the fields is a symmetry transformation if the Lagrangian
changes by a total derivative \cite{Peskin}. This means that adding total derivative
terms to the Lagrangian does not change the physics that it contains. We can see this from the principle of
least action. Adding a total derivative term to the Lagrangian
\begin{align}
 \mathcal{L} \qquad\rightarrow\qquad \tilde{\mathcal{L}} = \mathcal{L} + \partial_{\mu} F^{\mu} (h) \, ,
\end{align}
the variation of the action remains zero. Explicitly,
\begin{align}
  \delta \tilde{S} = \delta \int d^4x \tilde{\mathcal{L}} = \delta \int d^4x \big( \mathcal{L} + \partial_{\mu} F^{\mu} (h) \big ) = \delta \int d^4x \mathcal{L} + \delta \int d^4x \partial_{\mu} F^{\mu} (h) = 0 \, ,
\end{align}
where \( \delta S = \delta \int d^4x \mathcal{L} = 0\), and the infinitesimal
variation of the total derivative part vanishes by the assumption that $F$
vanishes at the boundary of integration.

In other words, adding total derivative terms to the Lagrangian is equivalent to
doing integration by parts. For example, 
\begin{align}
  \int d^4 x\, \phi\; \partial_{\mu}\partial^{\mu}\phi & = \int d^4 x\, \partial_{\mu} (\phi\; \partial^{\mu}\phi) - \int d^4 x\, \partial_{\mu}\phi\; \partial^{\mu}\phi  = \phi\,\partial^{\mu}\phi\Big\rvert_{\delta V} - \int d^4 x\, \partial_{\mu}\phi\; \partial^{\mu}\phi \nonumber  \\
  & = - \int d^4 x\, \partial_{\mu}\phi\; \partial^{\mu}\phi \, .
\end{align}

In addition, the freedom of adding total derivative terms can also be seen from applying momentum
conservation in a vertex in momentum space. For example, if we consider the term
$\phi \,\phi \;  \partial^{\mu}\partial_{\mu}\phi$ and then add the total
derivative $- \partial^{\mu} (\phi \,\phi \; \partial_{\mu}\phi)$, we get
\begin{align}
  & \phi \,\phi \;  \partial^{\mu}\partial_{\mu}\phi = \phi \,\phi \;  \partial^{\mu}\partial_{\mu}\phi - \partial^{\mu} (\phi \,\phi \; \partial_{\mu}\phi)  = - 2 \phi \; \partial^{\mu} \phi \;   \partial_{\mu} \phi \, , \nonumber\\[3mm]
  &  \Rightarrow \qquad\qquad \phi \,\phi \;  \partial^{\mu}\partial_{\mu}\phi =- 2 \phi \; \partial^{\mu} \phi \;   \partial_{\mu} \phi \label{eq:MomentumSpace} \, .
\end{align}
These partial derivatives give momenta in momentum space. So, assuming that all
the momenta in this vertex are ingoing, equation~(\ref{eq:MomentumSpace}) can be represented in momentum space as
\begin{align}
  2(p_1^2+p_2^2+p_3^2) & = -4 (p_1\cdot p_2+p_1\cdot p_3+p_2\cdot p_3) \, .
\end{align}
This result also agrees with momentum conservation in this vertex
{\setlength{\belowdisplayskip}{0pt}
\begin{align*}
p_1+p_2+p_3=0\, , \,\; \Rightarrow \;\, (p_1+p_2+p_3)^2=0 \, , \;\, \Rightarrow \;\, p_1^2+p_2^2+p_3^2 = -2(p_1\cdot p_2+p_1\cdot p_3+p_2\cdot p_3) \, .
\end{align*}}

In this thesis, we use this freedom by adding a parameterization of all possible total derivative
terms which are relevant to the lowest order vertices. This means that we use
the expansion up to four $h$ for the gravitational field,

\begin{align*}
  \mathcal{L}_{\text{TD}}(h) = \frac{1}{\kappa^{2}} \partial^{\mu} \Bigg [ &
       \kappa \Big [
             a_1  \partial_{\mu} h_{ \nu}^{\;\;\nu}
            +a_2  \partial^{\nu} h_{ \mu \nu}
            \Big ]
      +\kappa^2 \Big [
               a_3  \partial_{\mu} h_{ \alpha}^{\;\;\alpha} h_{\beta}^{\;\;\beta}
              +a_4  \partial_{\mu} h^{ \alpha \nu} h_{\alpha \nu}
              +a_5  \partial^{\alpha} h_{ \mu \alpha} h_{\nu}^{\;\;\nu}
         \\ & +a_6  \partial_{\alpha} h_{ \mu \nu} h^{\alpha \nu}
              +a_7  h_{\mu \nu} \partial_{\alpha} h^{ \alpha \nu}
              +a_8  h_{\mu \alpha} \partial^{\alpha} h_{ \nu}^{\;\;\nu}
                \Big ]
       +\kappa^3 \Big [
             a_9  \partial_{\mu} h_{ \nu}^{\;\;\nu} h_{\alpha}^{\;\;\alpha} h_{\beta}^{\;\;\beta}
      \\ &  +a_{10}  \partial_{\mu} h_{ \nu}^{\;\;\nu} h^{\alpha \beta} h_{\alpha \beta}
            +a_{11}  \partial_{\mu} h_{ \nu \alpha} h^{\nu \alpha} h_{\beta}^{\;\;\beta}
            +a_{12}  \partial_{\mu} h^{ \nu \alpha} h_{\nu}^{\;\; \beta} h_{\alpha \beta}
            +a_{13}  \partial^{\nu} h_{ \mu \nu} h_{\alpha}^{\;\;\alpha} h_{\beta}^{\;\;\beta}
     \\ &   +a_{14}  \partial^{\nu} h_{ \mu \nu} h^{\alpha \beta} h_{\alpha \beta}
            +a_{15}  \partial_{\nu} h_{ \mu \alpha} h^{\nu \alpha} h_{\beta}^{\;\;\beta}
            +a_{16}  \partial^{\nu} h_{ \mu \alpha} h_{\nu \beta} h^{\alpha \beta}
            +a_{17}  \partial_{\nu} h^{ \nu \alpha} h_{\mu \alpha} h_{\beta}^{\;\;\beta}
     \\ &   +a_{18}  \partial_{\nu} h^{ \nu \alpha} h_{\mu \beta} h_{\alpha}^{\;\; \beta}
            +a_{19}  \partial^{\nu} h_{ \alpha}^{\;\;\alpha} h_{\mu \nu} h_{\beta}^{\;\;\beta}
            +a_{20}  \partial^{\nu} h^{ \alpha \beta} h_{\mu \nu} h_{\alpha \beta}
            +a_{21}  \partial_{\nu} h_{ \alpha}^{\;\;\alpha} h_{\mu \beta} h^{\nu \beta}
   \\  &   +a_{22}  \partial^{\nu} h^{ \alpha \beta} h_{\mu \alpha} h_{\nu \beta}
                \Big ]
        +\kappa^4 \Big [
            a_{23}  \partial_{\mu} h_{ \alpha}^{\;\;\alpha} h_{\beta}^{\;\;\beta} h_{\gamma}^{\;\;\gamma} h_{\delta}^{\;\;\delta}
            +a_{24}  \partial_{\mu} h_{ \alpha}^{\;\;\alpha} h_{\beta}^{\;\;\beta} h^{\gamma \delta} h_{\gamma \delta}
    \\ &    +a_{25}  \partial_{\mu} h_{ \alpha}^{\;\;\alpha} h^{\beta \gamma} h_{\gamma}^{\;\; \delta} h_{\delta \beta}
            +a_{26}  \partial^{\alpha} h_{ \mu \alpha} h_{\beta}^{\;\;\beta} h_{\gamma}^{\;\;\gamma} h_{\delta}^{\;\;\delta}
            +a_{27}  \partial^{\alpha} h_{ \mu \alpha} h_{\beta}^{\;\;\beta} h^{\gamma \delta} h_{\gamma \delta}
     \\ &   +a_{28}  \partial^{\alpha} h_{ \mu \alpha} h^{\beta \gamma} h_{\gamma}^{\;\; \delta} h_{\delta \beta}
            +a_{29}  \partial_{\mu} h_{ \alpha \beta} h^{\alpha \beta} h_{\gamma}^{\;\;\gamma} h_{\delta}^{\;\;\delta}
            +a_{30}  \partial_{\mu} h_{ \alpha \beta} h^{\alpha \beta} h_{\gamma \delta} h^{\gamma \delta}
     \\ &   +a_{31}  \partial_{\alpha} h_{ \mu \beta} h^{\alpha \beta} h_{\gamma}^{\;\;\gamma} h_{\delta}^{\;\;\delta}
            +a_{32}  \partial_{\alpha} h_{ \mu \beta} h^{\alpha \beta} h^{\gamma \delta} h_{\gamma \delta}
            +a_{33}  \partial^{\beta} h_{ \alpha}^{\;\;\alpha} h_{\mu \beta} h_{\gamma}^{\;\;\gamma} h_{\delta}^{\;\;\delta}
\\  &        +a_{34}  \partial^{\beta} h_{ \alpha}^{\;\;\alpha} h_{\mu \beta} h^{\gamma \delta} h_{\gamma \delta}
            +a_{35}  \partial^{\beta} h^{ \alpha \gamma} h_{\mu \beta} h_{\alpha \gamma} h_{\delta}^{\;\;\delta}
            +a_{36}  \partial^{\beta} h^{ \alpha \gamma} h_{\mu \beta} h_{\gamma}^{\;\; \delta} h_{\delta \alpha}
   \\ &         +a_{37}  \partial_{\alpha} h^{ \beta \alpha} h_{\mu \beta} h_{\gamma}^{\;\;\gamma} h_{\delta}^{\;\;\delta}
         +a_{38}  \partial_{\alpha} h^{ \beta \alpha} h_{\mu \beta} h^{\gamma \delta} h_{\gamma \delta}
            +a_{39}  \partial^{\alpha} h^{ \beta \gamma} h_{\mu \beta} h_{\alpha \gamma} h_{\delta}^{\;\;\delta}
     \\ &       +a_{40}  \partial^{\alpha} h^{ \beta \gamma} h_{\mu \beta} h_{\gamma}^{\;\; \delta} h_{\delta \alpha}
        +a_{41}  \partial_{\gamma} h_{ \alpha}^{\;\;\alpha} h_{\mu \beta} h^{\beta \gamma} h_{\delta}^{\;\;\delta}
            +a_{42}  \partial_{\gamma} h^{ \alpha \delta} h_{\mu \beta} h^{\beta \gamma} h_{\alpha \delta}
     \\ &       +a_{43}  \partial^{\alpha} h_{ \gamma \alpha} h_{\mu \beta} h^{\beta \gamma} h_{\delta}^{\;\;\delta}
            +a_{44}  \partial_{\alpha} h_{ \gamma \delta} h_{\mu \beta} h^{\beta \gamma} h^{\alpha \delta}
            +a_{45}  \partial_{\alpha} h^{ \delta \alpha} h_{\mu \beta} h^{\beta \gamma} h_{\gamma \delta}
    \\ &        +a_{46}  \partial^{\delta} h_{ \alpha}^{\;\;\alpha} h_{\mu \beta} h^{\beta \gamma} h_{\gamma \delta}
            +a_{47}  \partial^{\alpha} h_{ \mu \beta} h_{\alpha \gamma} h^{\beta \gamma} h_{\delta}^{\;\;\delta}
            +a_{48}  \partial_{\alpha} h_{ \mu \beta} h^{\alpha \gamma} h^{\beta \delta} h_{\gamma \delta}
    \\ &        +a_{49}  \partial_{\mu} h^{ \alpha \beta} h_{\alpha}^{\;\; \gamma} h_{\beta \gamma} h_{\delta}^{\;\;\delta}
            +a_{50}  \partial_{\mu} h^{ \alpha \beta} h_{\alpha}^{\;\; \gamma} h_{\beta }^{\;\;\delta} h_{\gamma \delta}         \Big ]
         \Bigg ] + \cdots \numberthis \label{eq:LagrangianTD} \, ,
\end{align*}
up to three $h$ for the scalar field
\begin{align*}
  \mathcal{L}_{\text{TD}}(\phi,h) =  \partial^{\mu} \Bigg [
&         d_1  \phi \partial_{\mu} \phi
      +\kappa \Big [
          d_2  \phi^2\partial_{\mu} h_{ \nu}^{\;\;\nu}
         +d_3  \phi^2 \partial^{\nu} h_{ \nu \mu}
         +d_4  \phi \partial_{\mu} \phi h_{\nu}^{\;\;\nu}
         +d_5  \phi \partial^{\nu} \phi h_{\nu \mu}
                 \Big ]
\\[1mm] &      +\kappa^2 \Big [
          d_6  \phi^2\partial_{\mu} h_{ \nu}^{\;\;\nu} h_{\alpha}^{\;\;\alpha}
         +d_7  \phi^2\partial_{\mu} h^{ \nu \alpha} h_{\nu \alpha}
         +d_8  \phi^2\partial^{\nu} h_{ \mu \nu} h_{\alpha}^{\;\;\alpha}
         +d_9  \phi^2\partial_{\nu} h_{ \mu \alpha} h^{\nu \alpha}
\\[1mm] &         +d_{10}  \phi^2\partial_{\nu} h^{ \nu \alpha} h_{\mu \alpha}
        +d_{11}  \phi^2\partial^{\nu} h_{ \alpha}^{\;\;\alpha} h_{\mu \nu}
        +d_{12}  \phi \partial_{\mu} \phi h_{\nu}^{\;\;\nu} h_{\alpha}^{\;\;\alpha}
        +d_{13}  \phi \partial_{\mu} \phi h^{\nu \alpha} h_{\nu \alpha}
\\[1mm] &        +d_{14}  \phi \partial^{\nu} \phi h_{\mu \nu} h_{\alpha}^{\;\;\alpha}
        +d_{15}  \phi \partial_{\nu} \phi h_{\mu \alpha} h^{\nu \alpha}
       \Big ]
      +\kappa^3 \Big [
        d_{16}  \phi^2\partial_{\mu} h^{ \nu \alpha} h_{\nu \alpha} h_{\beta}^{\;\;\beta}
\\[1mm] &        +d_{17}  \phi^2\partial^{\nu} h_{ \mu \nu} h_{\alpha}^{\;\;\alpha} h_{\beta}^{\;\;\beta}
        +d_{18}  \phi^2\partial_{\nu} h_{ \mu \alpha} h^{\nu \alpha} h_{\beta}^{\;\;\beta}
        +d_{19}  \phi^2\partial_{\nu} h^{ \nu \alpha} h_{\mu \alpha} h_{\beta}^{\;\;\beta}
\\[1mm] &        +d_{20}  \phi^2\partial^{\nu} h_{ \alpha}^{\;\;\alpha} h_{\mu \nu} h_{\beta}^{\;\;\beta}
        +d_{21}  \phi \partial_{\mu} \phi h_{\nu}^{\;\;\nu} h_{\alpha}^{\;\;\alpha} h_{\beta}^{\;\;\beta}
        +d_{22}  \phi \partial_{\mu} \phi h^{\nu \alpha} h_{\nu \alpha} h_{\beta}^{\;\;\beta}
\\[1mm]  &   +d_{23}  \phi \partial^{\nu} \phi h_{\mu \nu} h_{\alpha}^{\;\;\alpha} h_{\beta}^{\;\;\beta}
        +d_{24}  \phi \partial_{\nu} \phi h_{\mu \alpha} h^{\nu \alpha} h_{\beta}^{\;\;\beta}
        +d_{25}  \phi^2\partial_{\mu} h_{ \nu}^{\;\;\nu} h_{\alpha}^{\;\;\alpha} h_{\beta}^{\;\;\beta} 
       \Big  ]
   \Bigg  ] + \cdots \numberthis \label{eq:LagrangianTDScalar} \, ,
\end{align*}
and up to two $h$ for the ghost and antighost fields
\begin{flalign*}
  \mathcal{L}_{\text{TD}} (\chi, \bar{\chi},h) = \partial^{\mu} \Bigg    [
   &      h_1  \bar{\chi}^{\nu} \partial_{\mu} \chi_{ \nu}
      +\kappa \Big [
          h_2  h^{\nu \alpha} \bar{\chi}_{\nu} \partial_{\mu} \chi_{ \alpha}
         +h_3  \bar{\chi}_{\nu} \partial_{\alpha} \chi_{ \mu} h^{\nu \alpha}
         +h_4  \bar{\chi}^{\nu} \partial_{\mu} \chi_{ \nu} h_{\alpha}^{\;\;\alpha}
\\ &         +h_5  \bar{\chi}^{\nu} \partial_{\nu} \chi_{ \mu} h_{\alpha}^{\;\;\alpha}
         +h_6  \bar{\chi}^{\nu} \partial^{\alpha} \chi_{ \alpha} h_{\mu \nu}
         +h_7  \bar{\chi}^{\nu} \partial^{\alpha} \chi_{ \nu} h_{\mu \alpha}
         +h_8  \bar{\chi}_{\nu} \partial^{\nu} \chi^{ \alpha} h_{\mu \alpha}
\\ &         +h_9  \bar{\chi}_{\mu} \partial^{\nu} \chi_{ \nu} h_{\alpha}^{\;\;\alpha}
         +h_{10}  \bar{\chi}_{\mu} \partial^{\nu} \chi^{ \alpha} h_{\nu \alpha}
         +h_{11}  \bar{\chi}^{\nu} \partial^{\alpha} \chi_{ \nu} h_{\mu \alpha}
         +h_{12}  \bar{\chi}^{\nu} \partial^{\alpha} \chi_{ \alpha} h_{\nu \mu}
\\ &         +h_{13}  \bar{\chi}_{\nu} \partial_{\mu} \chi_{ \alpha} h^{\nu \alpha}
         +h_{14}  \bar{\chi}_{\nu} \partial_{\mu} \chi^{\nu} h_{\alpha}^{\;\;\alpha}
         +h_{15}  \bar{\chi}_{\mu} \partial_{\nu} \chi_{ \alpha} h^{\nu \alpha}
         \Big ]
\\ &      +\kappa^2 \Big [
         h_{20}  \bar{\chi}_{\mu} \partial^{\nu} \chi_{ \nu} h_{\alpha}^{\;\;\alpha} h_{\beta}^{\;\;\beta}         
         +h_{21}  \bar{\chi}_{\mu} \partial_{\nu} \chi_{ \alpha} h^{\nu \alpha} h_{\beta}^{\;\;\beta}
         +h_{22}  \bar{\chi}_{\mu} \partial^{\nu} \chi_{ \alpha} h_{\nu \beta} h^{\alpha \beta}
\\ &         +h_{23}  \bar{\chi}_{\nu} \partial_{\mu} \chi^{ \nu} h_{\alpha}^{\;\;\alpha} h_{\beta}^{\;\;\beta}
         +h_{24}  \bar{\chi}_{\nu} \partial_{\mu} \chi^{\nu} h^{\alpha \beta} h_{\alpha \beta}
         +h_{25}  \bar{\chi}_{\nu} \partial_{\mu} \chi_{ \alpha} h^{\nu \alpha} h_{\beta}^{\;\;\beta}
\\ &         +h_{26}  \bar{\chi}^{\nu} \partial_{\mu} \chi_{ \alpha} h_{\nu \beta} h^{\alpha \beta}
         +h_{27}  \bar{\chi}^{\nu} \partial^{\alpha} \chi_{ \nu} h_{\mu \alpha} h_{\beta}^{\;\;\beta}
         +h_{28}  \bar{\chi}^{\nu} \partial_{\alpha} \chi_{ \nu} h_{\mu \beta} h^{\alpha \beta}
\\ &         +h_{29}  \bar{\chi}_{\nu} \partial_{\alpha} \chi_{ \mu} h^{\nu \alpha} h_{\beta}^{\;\;\beta}
         +h_{30}  \bar{\chi}^{\nu} \partial_{\alpha} \chi_{ \mu} h_{\nu \beta} h^{\alpha \beta}
         +h_{31}  \bar{\chi}^{\nu} \partial^{\alpha} \chi_{ \alpha} h_{\nu \mu} h_{\beta}^{\;\;\beta}
\\ &         +h_{32}  \bar{\chi}_{\nu} \partial^{\alpha} \chi_{ \alpha} h^{\nu \beta} h_{\mu \beta}
         +h_{33}  \bar{\chi}^{\nu} \partial_{\alpha} \chi_{ \beta} h_{\nu \mu} h^{\alpha \beta}
         +h_{34}  \bar{\chi}^{\nu} \partial^{\alpha} \chi^{ \beta} h_{\nu \alpha} h_{\mu \beta}
\\ &     +h_{35}  \bar{\chi}_{\nu} \partial^{\alpha} \chi_{ \beta} h^{\nu\beta} h_{\mu \alpha}
         +h_{36}  \bar{\chi}_{\mu} \partial^{\nu} \chi_{ \nu} h^{\alpha \beta} h_{\alpha \beta}
          \Big ]
     \Bigg  ] + \cdots \, , && \numberthis \label{eq:LagrangianTDGhost} 
\end{flalign*}
where $a_i$, $d_i$, $h_i$ are parameters that we choose later to simplify
the Feynman rules. Mainly, we use them to get rid of terms that have
second order derivatives of the fields (e.g., $\partial\partial h$) as we show
later in more detail.

\subsection{Field Redefinition \label{se:FieldRedefinition}}
The field redefinition freedom follows from the equivalence theorem
\cite{EquivalenceTheorem} which states that the S-matrix in quantum field theory
remains unchanged under reparameterization of the field operators. We can
illustrate this theorem by taking a scalar field $\phi$ as an example, where the generating functional is given by
\begin{align}
  \mathcal{Z} = \int D[\phi] \; \exp \Big ( i\; \int d^4x \; \mathcal{L}(\phi,\partial_{\mu} \phi) \Big ) \, .
  \label{eq:ScalarGenerating}
\end{align}
Now, redefining the scalar field \( \phi \rightarrow \tilde{\phi} = \phi +
a_1 \phi^2 + a_2 \phi^3 +  \cdots \), we get
\begin{align}
\mathcal{Z} = \int D[\tilde{\phi}] \; \exp \Big ( i\; \int d^4x \; \mathcal{L}(\tilde{\phi},\partial_{\mu} \tilde{\phi}) \Big ) \, .
\end{align}
This redefinition is allowed as long as the Jacobian of the integral is
essentially one \cite{EquivalenceTheorem}. Similarly, we can also redefine the other fields: $h_{\mu\nu}$, $\chi$, and $\bar{\chi}$.

Now let us explain how the field redefinition can simplify the Lagrangian. Consider the following field redefinition for the gravitational field
\begin{align}
  h_{\mu \nu} \; \rightarrow \; h_{\mu \nu}^{\prime}    = h_{\mu \nu} +\kappa \Big [
  a_{1} h_{\mu \gamma} h_{\nu}^{\;\; \gamma} +a_{2} h_{\mu \nu} h_{\gamma}^{\;\; \gamma} \Big ] + \cdots \ .
\end{align}
Inserting this into $h_{\mu\nu}\,\partial^{\mu} h^{\nu\alpha}
\partial_{\alpha}h_{\beta}^{\;\;\beta}$, which is part of the triple graviton
vertex, as an illustration gives
\begin{align}
  h_{\mu\nu}\, \partial^{\mu} h^{\nu\alpha} \partial_{\alpha}h_{\beta}^{\;\;\beta} \qquad  \,\rightarrow\,\qquad h_{\mu\nu} \, \partial^{\mu} h^{\nu\alpha} \partial_{\alpha}h_{\beta}^{\;\;\beta}\;\; +&\;\;  a_{1}\, \kappa\, h_{\mu \gamma} h_{\nu}^{\;\; \gamma} \, \partial^{\mu} h^{\nu\alpha} \partial_{\alpha}h_{\beta}^{\;\;\beta}\nonumber \\[2mm]
  \;\; +&\;\;  a_{2}\, \kappa\, h_{\mu \nu} h_{\gamma}^{\;\; \gamma} \, \partial^{\mu} h^{\nu\alpha} \partial_{\alpha}h_{\beta}^{\;\;\beta} \;\; + \;\; \cdots \, ,
\end{align}

\begin{figure}[H]
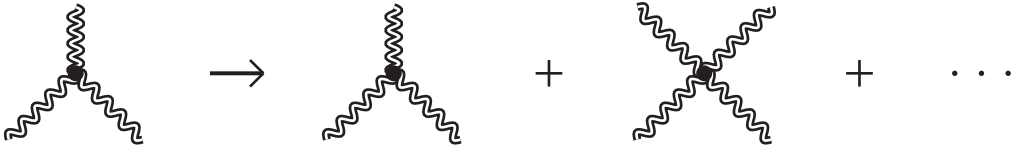

  ~~~
  \begin{axopicture}(50,50)
    \SetWidth{1.}
    \DoublePhoton(25,25)(25,50){2}{5}{2}
    \DoublePhoton(25,25)(0,0){2}{5}{2}
    \DoublePhoton(25,25)(50,0){2}{5}{2}
    \SetWidth{0.8}
    \Vertex(25,25){3}
  \end{axopicture}\hspace{2mm}
  ~~~
  \begin{axopicture}(10,50)(-10,0)
    \SetWidth{1.5}
    \Line(-10,25)(10,25)
    \SetWidth{1.}
    \Line(5,30)(10,25)
    \Line(5,20)(10,25)
  \end{axopicture}\hspace{5mm}
  ~~~
  \begin{axopicture}(50,50)
    \SetWidth{1.}
    \DoublePhoton(25,25)(25,50){2}{5}{2}
    \DoublePhoton(25,25)(0,0){2}{5}{2}
    \DoublePhoton(25,25)(50,0){2}{5}{2}
    \SetWidth{0.8}
    \Vertex(25,25){3}
  \end{axopicture}\hspace{3mm}
  ~~~
  \begin{axopicture}(10,50)
    \SetWidth{1}
    \Line(0,25)(10,25)
    \Line(5,20)(5,30)
  \end{axopicture}\hspace{3mm}
  ~~~
  \begin{axopicture}(50,50)
    \SetWidth{1.}
    \DoublePhoton(25,25)(0,50){2}{5}{2}
    \DoublePhoton(25,25)(50,50){2}{5}{2}
    \DoublePhoton(25,25)(0,0){2}{5}{2}
    \DoublePhoton(25,25)(50,0){2}{5}{2}
    \SetWidth{0.8}
    \Vertex(25,25){3}
  \end{axopicture}\hspace{3mm}
  ~~~
  \begin{axopicture}(10,50)
    \SetWidth{1}
    \Line(0,25)(10,25)
    \Line(5,20)(5,30)
  \end{axopicture}
  ~~~
  \begin{axopicture}(50,50)
    \SetWidth{0.8}
    \Vertex(25,25){1}
    \Vertex(15,25){1}
    \Vertex(35,25){1}
  \end{axopicture}\vspace{2mm}
  \caption{Field redefinition for the triple graviton vertex.}
  \label{fig:FieldRedefinition}
\end{figure}
\noindent Thus, the field redefinition generates an expansion of the triple
graviton vertex as shown in Fig.~\ref{fig:FieldRedefinition}, giving two new
contributions for the quadruple graviton vertex with the two
parameters ($a_1,a_2$). So, by choosing a proper value for these parameters, we
can cancel some of the contributions to the quadruple graviton vertex in the
standard Lagrangian.

For our fields, we use the most general parameterized expansions which are
relevant to the lowest order vertices. This means that we write all possible
parameterized combinations up to four $h$ for the gravitational field
$h_{\mu\nu}$ as
\begin{flalign*}
  h_{\mu \nu}^{\prime} =  h_{\mu \nu}
&           +\kappa \Big [
                             c_{1} h_{\mu \alpha} h_{\nu}^{\;\; \alpha}
                       +c_{2} h_{\mu \nu} h_{\alpha}^{\;\; \alpha}
                       \Big ]
\\ &     +\kappa^2 \Big [
                c_{3} h_{\mu \nu} h_{\alpha}^{\;\; \alpha} h_{\beta}^{\;\; \beta}
               +c_{4} h_{\mu \nu} h^{\alpha \beta} h_{\alpha \beta}
               +c_{5} h_{\mu \alpha} h_{\nu}^{\;\; \alpha} h_{\beta}^{\;\; \beta}
               +c_{6} h_{\mu \alpha} h_{\nu \beta} h^{\alpha \beta}
                      \Big ]
\\ &     +\kappa^3 \Big [
        c_{7 } h_{\mu \nu} h_{\alpha}^{\;\; \alpha} h_{\beta}^{\;\; \beta} h_{\gamma}^{\;\; \gamma}
       +c_{8 } h_{\mu \nu} h_{\alpha}^{\;\; \alpha} h^{\beta \gamma} h_{\beta \gamma}
       +c_{9} h_{\mu \nu} h^{\alpha \beta} h_{\beta}^{\;\; \gamma} h_{\gamma \alpha}
\\ & \qquad\quad       +c_{10} h_{\mu \alpha} h_{\nu}^{\;\; \alpha} h_{\beta}^{\;\; \beta} h_{\gamma}^{\;\; \gamma}
       +c_{11} h_{\mu \alpha} h_{\nu}^{\;\; \alpha} h_{\beta \gamma} h^{\beta \gamma}
       +c_{12} h_{\mu \alpha} h_{\nu \beta} h^{\alpha \beta} h_{\gamma}^{\;\; \gamma}
\\ &   \qquad\quad    +c_{13} h_{\mu \alpha} h_{\nu}^{\;\; \beta} h^{\alpha \gamma} h_{\beta \gamma}
     \Big ] + \cdots \, ,  \numberthis \label{eq:RedefinedField} &&
\end{flalign*}
up to three powers of $h$ for the scalar field $\phi$ as
\begin{flalign*}
  \phi^{\prime} = \phi 
&       +\kappa \Big [
               e_1 h_{\alpha}^{\;\; \alpha} \phi
                   \Big ]
\\ &       +\kappa^2 \Big [
          e_2 h_{\alpha}^{\;\; \alpha} h_{\beta}^{\;\; \beta} \phi
         +e_3 h_{\alpha \beta} h^{\alpha \beta} \phi
                   \Big ]
\\ &       +\kappa^3 \Big [
          e_4 h_{\alpha }^{\;\;\alpha} h_{\beta}^{\;\; \beta} h_{\gamma}^{\;\; \gamma} \phi
         +e_5 h_{\alpha \beta} h^{\alpha \beta} h_{\gamma}^{\;\; \gamma} \phi
         +e_6 h^{\alpha \beta} h_{\beta}^{\;\; \gamma} h_{\gamma \alpha} \phi
     \Big ] + \cdots \, , \numberthis \label{eq:RedefinedFieldScalar} &&
\end{flalign*}
up to two powers of $h$ for the ghost field $\chi_{\mu}$ as 
\begin{flalign*}
  \chi_{\mu}^{\prime} = \chi_{\mu}
&       +\kappa \Big [
               g_{1} h_{\alpha }^{\;\;\alpha} \chi_{\mu}
              +g_{2} h_{\alpha \mu} \chi^{\alpha}
                   \Big ]
\\ &       +\kappa^2 \Big [
          g_{3} h_{\alpha}^{\;\; \alpha} h_{\beta}^{\;\; \beta} \chi_{\mu}
          +g_{4} h^{\alpha \beta} h_{\alpha \beta} \chi_{\mu}
          +g_{5} h_{\alpha }^{\;\;\alpha} h_{\beta \mu} \chi_{\beta}
          +g_{6} h^{\alpha \beta} h_{\alpha \mu} \chi_{\beta}
     \Big ] + \cdots \, , \numberthis \label{eq:RedefinedFieldGhost} &&
\end{flalign*}
and up to two powers of $h$ for the anti-ghost field $\bar{\chi}_{\mu}$ as
\begin{flalign*}
  \bar{\chi}_{\mu}^{\prime} = \bar{\chi}_{\mu}
&       +\kappa \Big [
               f_{1} h_{\alpha}^{\;\; \alpha} \bar{\chi}_{\mu}
              +f_{2} h_{\alpha \mu} \bar{\chi}^{\alpha}
                   \Big ]
\\ &       +\kappa^2 \Big [
          f_{3 }h_{\alpha}^{\;\; \alpha} h_{\beta}^{\;\; \beta} \bar{\chi}_{\mu}
          +f_{4} h^{\alpha \beta} h_{\alpha \beta} \bar{\chi}_{\mu}
          +f_{5} h_{\alpha}^{\;\; \alpha} h_{\beta \mu} \bar{\chi}^{\beta}
          +f_{6} h^{\alpha \beta} h_{\alpha \mu} \bar{\chi}_{\beta}
     \Big ] + \cdots \, , \numberthis \label{eq:RedefinedFieldAntighost} &&
\end{flalign*}
where $c_i$, $e_i$, $g_i$, $f_i$ are parameters that will be chosen later to
simplify the Feynman rules.


\section{Feynman Rules\label{se:FeynmanRules}}
\setcounter{equation}{0}
In this section, we calculate the standard Feynman rules \cite{GhostVertex}. After that, we explain our
strategies to simplify the Feynman rules for gravity, and then we show the
resulting simplified Feynman rules.

To perform these calculations, we use the FORM program which is a symbolic
manipulation system that can manipulate symbolic expressions and do
mathematical operations, then return symbolic results \cite{FORM,FORMGitHub}.
In addition, some short pieces of code that are relevant to our calculations are shown in App.~\ref{Ap:formprogram}.

\subsection{The Standard Calculations \label{se:Standard}}
First, we calculate the Lagrangian for matter up to three powers of $h$ and the Lagrangian
for gravity up to four powers of $h$ from Eqs.~(\ref{eq:MAction}, \ref{eq:GAction})
respectively using the expansion Eq.~(\ref{eq:determinant}) and the definitions (\ref{eq:Riemann}$-$\ref{eq:WeakFieldExpansion}).
Second, we calculate the Lagrangian for fixing the gauge up to four $h$
and the Lagrangian for ghosts up to two $h$ from Eqs.~(\ref{eq:LagrangianFixing},
\ref{eq:LagrangianGhosts}) using the de Donder (harmonic) gauge condition Eq.~(\ref{eq:GaugeDonder}) and the gauge transformation Eq.~(\ref{eq:GaugeTransformation}). Thus, we get the total Lagrangian as
\begin{align} 
  \mathcal{L}_{\text{Total}}(h,\phi,\chi,\bar{\chi}) =&
  \;\; \mathcal{L}_{\text{Gravity}}(h) + \mathcal{L}_{\text{Matter}}(h,\phi) +
  \mathcal{L}_{\text{FG}}(h) + \mathcal{L}_{\text{Ghost}}(\chi,\bar{\chi},h) \, . \label{eq:LagrangianTotal2} 
\end{align}
From this total Lagrangian, we get the standard Feynman rules which are listed in
App.~\ref{Ap:FeynmanRules}. We have also verified that they agree with the Feynman rules in \cite{GhostVertex}.

\subsection{The Simplified Calculations \label{se:Simplified}}
Again, we calculate the Lagrangian for matter up to three powers of $h$ and the Lagrangian
for gravity up to four powers of $h$ from Eqs.~(\ref{eq:MAction}, \ref{eq:GAction})
respectively using the expansion Eq.~(\ref{eq:determinant}) and the definitions
(\ref{eq:Riemann}$-$\ref{eq:WeakFieldExpansion}). Second, we calculate the Lagrangian for fixing gauge up to four $h$ and the
Lagrangian for ghost up to two $h$ from Eqs.~(\ref{eq:LagrangianGhosts}, \ref{eq:LagrangianFixing}) using the general parameterized gauge condition Eq.~(\ref{eq:Gauge}) and the gauge transformation Eq.~(\ref{eq:GaugeTransformation}). Third, we add the total derivative Lagrangians
Eqs.~(\ref{eq:LagrangianTD}$-$\ref{eq:LagrangianTDGhost}). Fourth, we put the
previous Lagrangians together to obtain the total Lagrangian as
\begin{align} 
  \mathcal{L}_{\text{Total}}(h,\phi,\chi,\bar{\chi}) =&
  \;\; \mathcal{L}_{\text{Gravity}}(h) + \mathcal{L}_{\text{Matter}}(h,\phi) +
  \mathcal{L}_{\text{FG}}(h) + \mathcal{L}_{\text{Ghost}}(\chi,\bar{\chi},h) \label{eq:LagrangianTotal3} \\
  & + \mathcal{L}_{\text{TD}}(h) + \mathcal{L}_{\text{TD}}(\phi,h) + \mathcal{L}_{\text{TD}}(\chi,\bar{\chi},h) \, . \nonumber
\end{align}
After that, we redefine all the fields in the general parameterized way as
given in Eqs.~(\ref{eq:RedefinedField}$-$\ref{eq:RedefinedFieldAntighost}). Thus, we get the total Lagrangian Eq.~(\ref{eq:LagrangianTotal3}) with eight sets of parameters:
\begin{table}[H]
  \centering
  \begin{tabular}{L|C|C|C}
  \hline\hline  
                         &     h              &  \phi             &  \chi, \bar{\chi}          \\
  \hline
  \mathcal{L}_{\text{TD}}&\; (a_1, \cdots , a_{50})\; &\; (d_1, \cdots , d_{25}) \; & (h_1, \cdots , h_{36})          \\
  \hline
  \text{Field Redefinition}&(c_1, \cdots , c_{13})&(e_1, \cdots , e_{6})& (f_1, \cdots ,f_{6}), (g_1, \cdots , g_{6})          \\
  \hline
  \mathcal{C}_{\mu}(h)   &(b_1, \cdots , b_{22})   &                  &                              \\
  \hline
  \end{tabular} 
\end{table}
Finding the proper values of these parameters by inspection is what took the
most time in this thesis. In order to find suitable values for the parameters we used the following strategies:
\begin{enumerate}
\item Ensuring the same propagators as in the standard calculations.
\item Minimizing the number of terms as much as possible, especially for the triple and quadruple
  graviton vertices.
\item Cancelling all terms that have second order derivative of our fields as
  $\partial\partial h$. Because when we integrate by parts, the second order
  derivative of our fields as $h \partial\partial h$ gives two terms that have first
  order derivative of our fields as $\partial h \partial h$. Some of these
  terms can cancel terms in the standard Lagrangian.
\item Trying to keep terms that have the same indices for the partial
  derivatives, such as $\partial_{\mu} h_{\nu \alpha} \partial^{\mu} h^{\nu
    \alpha}$ and not $\partial_{\mu} h_{\nu \alpha} \partial^{\nu} h^{\mu
    \alpha}$, in order to get simpler expressions in momentum space.
\end{enumerate}
\noindent Thus, we choose the parameters as given in App.~\ref{Ap:Parameters},
where we list each set of the parameters separately and we also show the parameters that contribute to each Feynman rule. These parameters are the main results of our work.

To illustrate the simplified calculations, let us take an example of the
triple graviton vertex from our work. The FORM code for the total Lagrangian of this vertex is
shown below, where only parameters from the following sets ($c_1, c_2,\cdots, c_{13}$), ($a_1, a_2, \cdots, a_{50}$), ($b_1, b_2, \cdots, b_{22}$) can appear:\vspace{2mm}
\begin{Verbatim}[gobble=2,frame=single,framesep=2mm,label=Result,labelposition=all,numbers=left]
  LagT3 =
   + H(mu,mu)*H(nu,nu)*H(al,al,be,be)*Fact(1 + 4*a9,4)
   + H(mu,mu)*H(nu,nu)*H(al,be,al,be)*Fact( - 1 + 4*a13,4)
   + H(mu,mu)*H(nu,nu,al)*H(al,be,be)*Fact( - 1 - b5 + 2*b3 + a19 + 2*a13,1)
   + H(mu,mu)*H(nu,nu,al)*H(be,al,be)*Fact(1 + 2*b5 + a17,1)
   + H(mu,mu)*H(nu,nu,al,be)*H(al,be)*Fact( - 1 + a11,1)
   + H(mu,mu)*H(nu,al)*H(nu,al,be,be)*Fact( - 1 + a19,1)
   + H(mu,mu)*H(nu,al)*H(nu,be,al,be)*Fact(2 + a17 + a15,2)
   + H(mu,mu)*H(nu,al)*H(al,be,nu,be)*Fact(2 + a17 + a15,2)
   + H(mu,mu)*H(nu,al,al)*H(nu,be,be)*Fact(1 - 4*c2 - 4*b3 + 8*a9,4)
   + H(mu,mu)*H(nu,al,be)^2*Fact( - 3 + 4*c2 + 4*a11,4)
   + H(mu,mu)*H(nu,al,be)*H(al,nu,be)*Fact(1 + 2*a15,2)
   + H(mu,mu,nu)*H(nu,al)*H(al,be,be)*Fact(2 + 2*b8 - b7 + a21 + a17,2)
   + H(mu,mu,nu)*H(nu,al,be)*H(al,be)*Fact(2 + 2*b4 + a20 + 2*a14,2)
   + H(mu,mu,nu)*H(al,nu,be)*H(al,be)*Fact( - 4 + 2*b6 + a22 + a18,2)
   + H(mu,mu,nu,nu)*H(al,be)^2*Fact( - 1 + 2*a10,2)
   + H(mu,mu,nu,al)*H(nu,be)*H(al,be)*Fact(2 + a12,1)
   + H(mu,nu)^2*H(al,be,al,be)*Fact(1 + 2*a14,2)
   + H(mu,nu)*H(mu,nu,al)*H(al,be,be)*Fact(2 - b6 + a21 + a15,2)
   + H(mu,nu)*H(mu,nu,al)*H(be,al,be)*Fact( - 4 + 2*b6 + a22 + a18,4)
   + H(mu,nu)*H(mu,nu,al,be)*H(al,be)*Fact(2 + a20,1)
   + H(mu,nu)*H(mu,al)*H(nu,al,be,be)*Fact(2 + a21,4)
   + H(mu,nu)*H(mu,al)*H(nu,be,al,be)*Fact( - 4 + a18 + a16,1)
   + H(mu,nu)*H(mu,al,nu,be)*H(al,be)*Fact( - 2 + a22,2)
   + H(mu,nu)*H(mu,al,al)*H(nu,be,be)*Fact( - 1 - 2*b8 + 2*a19,2)
   + H(mu,nu)*H(mu,al,al)*H(be,nu,be)*Fact(2 + 2*b8 - b7 + a21 + a17,4)
   + H(mu,nu)*H(mu,al,be)*H(nu,al,be)*Fact(3 + 2*a20,2)
   + H(mu,nu)*H(mu,al,be)*H(al,nu,be)*Fact( - 2 + a22 + a16,1)
   + H(mu,nu)*H(mu,al,be,be)*H(nu,al)*Fact(6 + 3*a21,4)
   + H(mu,nu)*H(nu,mu,al)*H(al,be,be)*Fact(2 - b6 + a21 + a15,2)
   + H(mu,nu)*H(nu,mu,al)*H(be,al,be)*Fact( - 4 + 2*b6 + a22 + a18,4)
   + H(mu,nu)*H(nu,al,mu,be)*H(al,be)*Fact( - 2 + a22,2)
   + H(mu,nu)*H(nu,al,al)*H(be,mu,be)*Fact(2 + 2*b8 - b7 + a21 + a17,4)
   + H(mu,nu)*H(al,mu,nu)*H(al,be,be)*Fact( - 1 + c2 - c1 - b4 + a11 + 2*a10,1)
   + H(mu,nu)*H(al,mu,nu)*H(be,al,be)*Fact(2 + 2*b4 + a20 + 2*a14,2)
   + H(mu,nu)*H(al,mu,al)*H(be,nu,be)*Fact( - 2 + 2*b7 + a18,1)
   + H(mu,nu)*H(al,mu,be)*H(al,nu,be)*Fact(3 + 2*c1 + 2*a12,1)
   + H(mu,nu)*H(al,mu,be)*H(be,nu,al)*Fact( - 1 + a16,1)
   ;
\end{Verbatim}
Here \texttt{Fact(x,y)}$=\frac{x}{y}$, \;\texttt{H(mu,nu)}$=h_{\mu\nu}$,\;
\texttt{H(al,mu,nu)}$=\partial_{\alpha} h_{\mu\nu}$,\;
\texttt{H(al,be,mu,nu)}$=\partial_{\alpha}\partial_{\beta}h_{\mu\nu}\;$, and all
the indices are contracted in the proper way.

Now, if we plug our choice of the parameters as shown in App.~\ref{Ap:Parameters}, we get the following simplified expression for this vertex:\vspace{3mm}
\begin{Verbatim}[gobble=2,frame=single,framesep=2mm,label=Result,labelposition=all,numbers=left]
  Time =       5.66 sec    Generated terms =          4
             LagT3         Terms in output =          4
                           Bytes used      =        444
  
  LagT3 =
       + H(mu,mu)*H(nu,al,al)*H(nu,be,be)*Fact(1,8)
       + H(mu,nu)*H(mu,al,be)*H(nu,al,be)*Fact(-1,2)
       + H(mu,nu)*H(mu,al,be)*H(al,nu,be)*Fact(1,1)
       + H(mu,nu)*H(al,mu,nu)*H(al,be,be)*Fact(-1,4)
      ;
\end{Verbatim}

\noindent As a result, we efficiently reduce the triple graviton vertex from 40 to 4
terms. In the next subsection we will show all the simplified Feynman rules that
we obtain.

\subsection{The Simplified Feynman Rules \label{se:Results}}
Here we present the simplified Feynman rules that we obtain from our choice of
parameters, in App.~\ref{Ap:Parameters}, together with some
comparisons with the standard Feynman rules that are given in
App.~\ref{Ap:FeynmanRules}. As already mentioned, our choice of parameters ensures the same propagators
as in the standard Feynman rules. For completeness, we give below the Lagrangian and the
corresponding propagator in momentum space for the scalar field $\phi(Q)$\\
\noindent\begin{minipage}{.435\textwidth}
  \hfill
  \begin{axopicture}(90,50)
    \SetWidth{0.8}
    \Text(10,35){$\phi(Q)$}
    \Text(80,35){$\phi(Q)$}
    \Line[arrow](35,30)(55,30)
    \SetWidth{1.5}
    \Line(10,20)(80,20)
  \end{axopicture}
  \hspace{19mm}
\end{minipage}
\begin{minipage}{.5\textwidth}
  \hspace{19mm}
		\begin{flalign}
     &  \mathcal{L}_{\phi\phi} =  \frac{1}{2} \partial_{\mu} \phi \partial^{\mu} \phi
     - \frac{1}{2} m^2 \phi^2 \, , \label{eq:OurScalarPropagator} \\[3mm]
    &\text{in momentum space} \nonumber \\[2mm]
    & S_{\{\phi\phi\}}(Q,m) = \frac{i}{Q^2-m^2+i\epsilon} \, , \nonumber &&
		\end{flalign}
\end{minipage}\\
for the ghost field $\chi(Q)$\\
\begin{minipage}{.435\textwidth}
  \hfill
  \begin{axopicture}(90,50)
    \SetWidth{0.8}
    \Text(10,35){$\chi^{\alpha}(Q)$}
    \Text(80,35){$\bar{\chi}^{\beta}(Q)$}
    \Line[arrow](35,30)(55,30)
    \SetWidth{1.5}
    \Line[arrow](10,20)(80,20)
  \end{axopicture}
  \hspace{19mm}
\end{minipage}
\begin{minipage}{.5\textwidth}
  \hspace{19mm}
  \begin{flalign}
    & \mathcal{L}_{\{\bar{\chi}\chi\}}  =  - \eta_{\mu\nu} \partial_{\lambda}
    \bar{\chi}^{\mu} \partial^{\lambda} \chi^{\nu} \, , \label{eq:OurGhostPropagator}\\[3mm]
    &\text{in momentum space} \nonumber \\[2mm]
    & S_{\{\bar{\chi}\chi\}}^{\alpha\beta} (Q)  = - \frac{i}{Q^2}
    \eta^{\alpha\beta} \, , \nonumber &&
		\end{flalign}
\end{minipage}\\
and for the gravitational field $h^{\alpha\beta}(Q)$\\
\noindent\begin{minipage}{.435\textwidth}
  \hfill
  \begin{axopicture}(90,50)
    \SetWidth{0.8}
    \Text(10,35){$h^{\alpha\beta}(Q)$}
    \Text(80,35){$h^{\gamma\delta}(Q)$}
    \SetWidth{1.}
    \DoublePhoton(10,20)(80,20){2}{10}{2}
    \Line[arrow](35,30)(55,30)
  \end{axopicture}
  \hspace{19mm}
\end{minipage}
\begin{minipage}{.5\textwidth}
  \hspace{19mm}
		\begin{flalign}
      & \mathcal{L}_{hh} =  \frac{1}{2} \partial_{\mu} h_{\nu \lambda}
      \partial^{\mu} h^{\nu \lambda} - \frac{1}{4} \partial_{\mu}
      h_{\nu}^{\;\;\nu} \partial^{\mu}
      h_{\lambda}^{\;\;\lambda} \, , \label{eq:OurGravitonPropagator} && \\[3mm]
    &\text{in momentum space} \nonumber \\[2mm]
      &   S_{\{hh\}}^{\alpha\beta\gamma\delta}(Q) = \frac{i}{Q^2} P^{\alpha
        \beta \gamma \delta} \, , \nonumber &&
		\end{flalign}
\end{minipage}\\
where $\displaystyle P^{\alpha\beta \gamma \delta} = \frac{1}{2} (\eta^{\alpha
        \gamma} \eta^{\beta \delta} + \eta^{\alpha \delta} \eta^{\beta \gamma} -
      \eta^{\alpha \beta} \eta^{\gamma \delta})$.\vspace{4mm}

Turning now to the vertices, our choice of parameters successfully simplifies most of them,
comparing with the standard ones as listed in App.~\ref{Ap:FeynmanRules}.
Here we discuss each vertex separately. First, our main effort is to get the
simplest form of the triple graviton vertex
$V_{\gamma\delta\rho\sigma\eta\lambda}^{\{h h h\}} (q_1,q_2)$ since it can
appear in many one-loop diagrams, as shown in Sec.~\ref{se:OneLoopCorrection}, and its standard expression has 40 terms, as shown
in Eq.~(\ref{eq:3Hb}). This can lead to messy calculations when this vertex
appears twice or more in a diagram. For example, a diagram with three triple graviton vertices such as
Fig.~\ref{fig:TripleGravitonVertex}i can give about 64 000 terms. However, using our choice of parameters, we successfully reduce it to just four terms as follows
\noindent\begin{minipage}{.22\textwidth}
	\centering
  \begin{axopicture}(100,100)
    \Text(0,0)[l]{$h_{\rho\sigma}(q_2)$}
    \Text(90,0){$h_{\eta\lambda}(q_3)$}
    \Text(55,90)[l]{$h_{\gamma\delta}(q_1)$}
    \SetWidth{1.}
    \DoublePhoton(50,50)(50,100){2}{10}{2}
    \DoublePhoton(50,50)(10,10){2}{10}{2}
    \DoublePhoton(50,50)(90,10){2}{10}{2}
    \SetWidth{0.8}
    \Line[arrow](40,80)(40,60)
    \Line[arrow](35,45)(20,30)
    \Line[arrow](65,45)(80,30)
    \Vertex(50,50){3}
  \end{axopicture}
	\label{fig:HHH}
  \vspace{3mm}\\
  $ \Big (q_3=q_1-q_2 \Big ) $
\end{minipage}
\noindent\begin{minipage}{.77\textwidth}
  \begin{flalign}
    &\quad \mathcal{L}_{hhh} = \frac{\kappa}{2} \Big (  \frac{1}{4} h_{\mu}^{\;\;\mu}
    \partial_{\nu} h_{\alpha}^{\;\;\alpha}  \partial^{\nu}
    h_{\beta}^{\;\;\beta} -  h^{\mu\nu}   \partial_{\mu} h^{\alpha\beta}
    \partial_{\nu} h_{\alpha\beta} \label{eq:OurTripleGravitonVertex}  && \\[3mm]
    &\qquad\quad\quad\qquad\qquad\qquad\;\;\; + 2\, h^{\mu\nu}   \partial_{\mu} h^{\alpha\beta} \partial_{\alpha}
    h_{\nu\beta} - \frac{1}{2} h^{\mu\nu} \partial_{\alpha} h_{\mu\nu}
    \partial^{\alpha} h_{\beta}^{\;\;\beta} \Big ) \, , \quad\nonumber &&\\[3mm]
    &\quad \text{in momentum space} \nonumber&&\\[3mm]
    &\quad V_{\gamma\delta\rho\sigma\eta\lambda}^{\{h h h\}} (q_1,q_2) = i
    \frac{\kappa}{2} \Big [ \;\; \frac{1}{2} q_1.q_2   (  \eta_{\rho \sigma} \eta_{\gamma \delta} \eta_{\eta \lambda}  -
    \eta_{\rho \sigma} \eta_{\gamma \eta} \eta_{\delta \lambda}  -  \eta_{\rho \eta} \eta_{\sigma \lambda} \eta_{\gamma \delta} )\nonumber\\[3mm]
    &\qquad\qquad\qquad\qquad\quad\;\; + \frac{1}{2}  q_1.q_3   (   \eta_{\rho \sigma} \eta_{\gamma \delta} \eta_{\eta \lambda}  -
    \eta_{\rho \gamma} \eta_{\sigma \delta} \eta_{\eta \lambda}  -  \eta_{\rho \eta} \eta_{\sigma \lambda} \eta_{\gamma \delta} )\nonumber\\[3mm]
    &\qquad\qquad\qquad\qquad\quad\;\; + \frac{1}{2}  q_2.q_3   ( -  \eta_{\rho \sigma} \eta_{\gamma \delta} \eta_{\eta \lambda}  +
    \eta_{\rho \sigma} \eta_{\gamma \eta} \eta_{\delta \lambda}  +  \eta_{\rho
      \gamma} \eta_{\sigma \delta} \eta_{\eta \lambda} ) \nonumber\\[3mm]
    &\qquad\qquad\qquad\qquad\quad\;\;  - 2\, \eta_{\rho \gamma} \eta_{\sigma \delta}  q_{1 \eta} q_{2 \lambda}
    + \eta_{\rho \gamma} \eta_{\sigma \eta}   ( q_{1 \lambda} q_{2 \delta}  - q_{2 \lambda} q_{3 \delta} )\nonumber\\[3mm]
    &\qquad\qquad\qquad\qquad\quad\;\;  + 2\, \eta_{\rho \delta} \eta_{\gamma \eta}   ( q_{1 \sigma} q_{2 \lambda} + q_{1 \lambda} q_{3 \sigma} )
    + \eta_{\rho \eta} \eta_{\sigma \lambda}    q_{2 \gamma} q_{3 \delta} \nonumber\\[3mm]
    &\qquad\qquad\qquad\qquad\quad\;\;   + 2\, \eta_{\rho \eta} \eta_{\gamma \lambda}   ( q_{1 \sigma} q_{3 \delta} - q_{2 \delta} q_{3 \sigma} )
    - \eta_{\gamma \eta} \eta_{\delta \lambda}  q_{1 \rho} q_{3 \sigma}  \Big ] \, . \nonumber  && 
  \end{flalign}
\end{minipage}\vspace{4mm}
Second, in addition to the triple graviton vertex, our main goal is to get the simplest form of the quadruple graviton vertex $V_{\gamma\delta\rho\sigma\eta\lambda\kappa\epsilon}^{\{h h h h\}}
  (q_1,q_2,q_3) $. This vertex has a very complicated standard form, 113 terms, as shown in
  Eq.~(\ref{eq:4Hb}). In our method, the parameters in App.~\ref{Ap:Parameters} successfully reduces the number of terms to just 12 as follows

\begin{minipage}{.22\textwidth}
  \vspace{5mm}
  \centering
  \begin{axopicture}(100,100)
    \Text(0,0)[l]{$h_{\rho\sigma}(q_2)$}
    \Text(100,0)[r]{$h_{\kappa\epsilon}(q_4)$}
    \Text(100,100)[r]{$h_{\eta\lambda}(q_3)$}
    \Text(0,100)[l]{$h_{\gamma\delta}(q_1)$}
    \SetWidth{1.}
    \DoublePhoton(50,50)(10,90){2}{10}{2}
    \DoublePhoton(50,50)(10,10){2}{10}{2}
    \DoublePhoton(50,50)(90,90){2}{10}{2}
    \DoublePhoton(50,50)(90,10){2}{10}{2}
    \SetWidth{0.8}
    \Vertex(50,50){3}
  \end{axopicture}
	\label{fig:HHHH}
\end{minipage}
\noindent\begin{minipage}{.77\textwidth}
		\begin{flalign}
      & \quad \mathcal{L}_{hhhh} =  \frac{\kappa^2}{4} \Big ( - \frac{5}{16} h_{\mu}^{\;\;\mu} h_{\nu}^{\;\; \nu} \partial_{\alpha} h_{\beta}^{\;\; \beta} \partial^{\alpha} h_{\tau}^{\;\; \tau} 
      + \frac{1}{2} h_{\mu}^{\;\;\mu} h^{\nu \alpha} \partial_{\nu} h_{\beta \tau} \partial_{\alpha} h^{\beta \tau} \nonumber&&\\
      &\quad\qquad\quad\qquad\;\; -  h_{\mu}^{\;\;\mu} h^{\nu \alpha} \partial_{\nu} h^{\beta \tau} \partial_{\beta}h_{\alpha \tau}
      +  h_{\mu}^{\;\;\mu} h^{\nu \alpha} \partial_{\beta}h_{\nu \tau} \partial^{\beta}h_{\alpha}^{\;\;\tau} \nonumber&&\\
      &\qquad\quad\quad\qquad\;\; - \frac{1}{8} h_{\mu \nu} h^{\mu \nu} \partial_{\alpha} h_{\beta}^{\;\; \beta} \partial^{\alpha} h_{\tau}^{\;\; \tau}
      +  h^{\mu \nu} \partial_{\mu}h_{\nu \alpha} \partial^{\beta} h^{\alpha \tau} h_{\beta \tau} \nonumber&&\\
      &\qquad\quad\quad\qquad\;\; + \frac{1}{4} h^{\mu \nu} \partial_{\mu} h_{\alpha}^{\;\; \alpha} h_{\nu \beta} \partial^{\beta} h_{\tau}^{\;\; \tau}
      - 2\, h^{\mu \nu} \partial_{\mu} h^{\alpha \beta} h_{\nu \alpha} \partial^{\tau}h_{\beta \tau} \nonumber\\
      &\qquad\quad\quad\qquad\;\; +  h^{\mu \nu} \partial_{\mu}h_{\alpha \beta} h_{\nu \tau} \partial^{\tau} h^{\alpha \beta}
      - 2\, h^{\mu \nu} \partial_{\mu} h_{\alpha \beta} h^{\alpha \tau} \partial_{\tau} h_{\nu}^{\;\;\beta} \nonumber\\
      &\qquad\quad\quad\qquad\;\; +  h^{\mu \nu} h_{\nu \alpha} \partial_{\beta} h_{\mu \tau} \partial^{\beta} h^{\alpha \tau}
      + 2\, h^{\mu \nu} \partial_{\nu} h_{\alpha \beta} h^{\alpha \beta}
      \partial^{\tau} h_{\mu \tau} \Big) \, . \label{eq:OurQuadrupoleGravitonVertex}  &&
		\end{flalign}
\end{minipage}\vspace{4mm}
Third, the scalar-scalar-graviton vertex $V_{\alpha \beta}^{\{\phi\phi h\}}
(p_1,p_2,m)$, which is the only vertex that has the same expression as in the standard rules Eq.~(\ref{eq:2phi1Hb}), is given by \\
\noindent\begin{minipage}{.22\textwidth}
	\centering
  \begin{axopicture}(100,100)
    \Text(0,0)[l]{$\phi(p_1)$}
    \Text(90,0){$\phi(p_2)$}
    \Text(55,90)[l]{$h_{\alpha\beta}(q_1)$}
    \SetWidth{1.}
    \DoublePhoton(50,50)(50,100){2}{10}{2}
    \SetWidth{1.5}
    \Line(50,50)(10,10)
    \Line(50,50)(90,10)
    \SetWidth{0.8}
    \Line[arrow](40,80)(40,60)
    \Line[arrow](20,30)(35,45)
    \Line[arrow](65,45)(80,30)
    \Vertex(50,50){3}
  \end{axopicture}
	\label{fig:Hphiphi}
\end{minipage}
\noindent\begin{minipage}{.77\textwidth}
		\begin{flalign}
      &\quad \mathcal{L}_{\phi\phi h} = \frac{\kappa}{2} \Big ( - \frac{1}{2}
      h_{\mu}^{\;\;\mu} \phi^2 m^2 + \frac{1}{2} h_{\mu}^{\;\;\mu}
      \partial_{\nu} \phi \partial^{\nu} \phi -  h^{\mu\nu} \partial_{\mu} \phi
      \partial_{\nu} \phi \Big ) \, , \label{eq:OurphiphiH} &&\\[4mm]
      &\quad \text{in momentum space} \nonumber&&\\[4mm]
      &\quad V_{\alpha \beta}^{\{\phi\phi h\}} (p_1,p_2,m) = i \frac{\kappa}{2} \Big [ (p_{1\alpha}
      p_{2\beta} +  p_{2\alpha} p_{1\beta}) - \eta_{\alpha \beta} (p_1.p_2 -
      m^2) \Big ] \, . \nonumber  &&\\\nonumber
		\end{flalign}
\end{minipage}\vspace{4mm}
Fourth, as mentioned before, the parameters are chosen in order to get the triple graviton and quadruple
graviton vertices as simple as possible, but at the same time the
scalar-scalar-graviton-graviton vertex $V_{\gamma\delta\rho\sigma}^{\{\phi\phi h
  h\}} (p_1,p_2)$ is surprisingly reduced from six terms, as in
Eq.~(\ref{eq:2phi2Hb}), to just two terms. Moreover, this vertex is now independent of the scalar mass $m$ as follows\\
\noindent\begin{minipage}{.22\textwidth}
	\centering
  \vspace{7mm}
  \begin{axopicture}(100,100)
    \Text(0,0)[l]{$h_{\gamma\delta}(q_1)$}
    \Text(98,0)[r]{$h_{\rho\sigma}(q_2)$}
    \Text(98,100)[r]{$\phi(p_2)$}
    \Text(0,100)[l]{$\phi(p_1)$}
    \SetWidth{1.}
    \Line(50,50)(10,90)
    \DoublePhoton(50,50)(10,10){2}{10}{2}
    \Line(50,50)(90,90)
    \DoublePhoton(50,50)(90,10){2}{10}{2}
    \SetWidth{0.8}
    \Line[arrow](20,30)(35,45)
    \Line[arrow](65,45)(80,30)
    \Line[arrow](20,70)(35,55)
    \Line[arrow](65,55)(80,70)
    \Vertex(50,50){3}
  \end{axopicture}
	\label{fig:HHphiphi}
\end{minipage}
\noindent\begin{minipage}{.77\textwidth}
  \begin{flalign}
    &\quad \mathcal{L}_{\phi\phi h h} = \frac{\kappa^2}{4} \Big ( h^{\mu\nu} h_{\nu}^{\;\;\alpha}
    \partial_{\mu} \phi  \partial_{\alpha} \phi - \frac{1}{2}
    h_{\mu}^{\;\;\mu} h^{\nu\alpha}  \partial_{\nu} \phi \partial_{\alpha}
    \phi \Big ) \, , \quad\quad \label{eq:OurScalarScalarGravitonGravitonVertex} &&\\[2mm]
    &\quad \text{in momentum space} \nonumber&&\\[2mm]
    &\quad V_{\gamma\delta\rho\sigma}^{\{\phi\phi h h\}} (p_1,p_2) = i \frac{\kappa^2}{8}  \Big [
    - \eta_{\gamma\delta}   (  p_{1\rho} p_{2\sigma}  +  p_{1\sigma} p_{2\rho} )
    + \eta_{\gamma\rho}   (  p_{1\delta} p_{2\sigma}  +  p_{1\sigma} p_{2\delta}
    ) \nonumber&&\\[1mm]
    &\qquad\qquad\qquad\qquad\qquad\;\;\,  + \eta_{\gamma\sigma}   (  p_{1\delta} p_{2\rho}  +  p_{1\rho} p_{2\delta} )
    + \eta_{\delta\rho}   (  p_{1\gamma} p_{2\sigma}  +  p_{1\sigma} p_{2\gamma}
    ) \nonumber&&\\[1mm]
    &\qquad\qquad\qquad\qquad\qquad\;\;\, + \eta_{\delta\sigma}   (  p_{1\gamma} p_{2\rho}  +  p_{1\rho} p_{2\gamma} )
    - \eta_{\rho\sigma}   (  p_{1\gamma} p_{2\delta}  +  p_{1\delta} p_{2\gamma} )
    \Big ] \, . \nonumber && 
  \end{flalign}
\end{minipage}\vspace{4mm}
Fifth, the scalar-scalar-graviton-graviton-graviton vertex
$V_{\gamma\delta\rho\sigma\lambda\epsilon}^{\{\phi\phi h h h\}} (p_1,p_2,q_1,q_2)$ is reduced from 10 terms, as in
  Eq.~(\ref{eq:2phi3Hb}), to seven terms in the simplified method, where this
  reduction is due to the choice of the redefinition parameters for scalar field $e_4$, $e_6$
  as shown in Tab.~\ref{table:simplestphi}, as follows\\
\noindent\begin{minipage}{.22\textwidth}
	\centering
  \begin{axopicture}(100,100)
    \Text(0,0)[l]{$\phi(p_1)$}
    \Text(90,0){$\phi(p_2)$}
    \Text(45,100)[r]{$h_{\rho\sigma}(q_2)$}
    \Text(5,55)[b]{$h_{\gamma\delta}(q_1)$}
    \Text(95,55)[b]{$h_{\lambda\epsilon}(q_3)$}
    \SetWidth{1.}
    \DoublePhoton(50,50)(50,100){2}{10}{2}
    \DoublePhoton(50,50)(0,50){2}{10}{2}
    \DoublePhoton(50,50)(100,50){2}{10}{2}
    \SetWidth{1.5}
    \Line(50,50)(10,10)
    \Line(50,50)(90,10)
    \SetWidth{0.8}
    \Vertex(50,50){3}
  \end{axopicture}
	\label{fig:HHHphiphi}
\end{minipage}
\noindent\begin{minipage}{.77\textwidth}
\begin{flalign*}
\quad \mathcal{L}_{\phi\phi h h h} = \; \frac{\kappa^3}{8} \Big [ & - \frac{1}{4} m^2 \phi^2 h_{\mu}^{\;\;\mu} h^{\nu \alpha} h_{\nu \alpha}
       - \frac{1}{16} \phi \; \partial_{\mu} \phi \partial^{\mu} h_{
         \nu}^{\;\;\nu} h_{\alpha}^{\;\;\alpha} h_{\beta}^{\;\;\beta} \\
&       - \frac{1}{2}  \phi \; \partial_{\mu} \phi \partial^{\mu} h^{\nu \alpha} h_{\nu }^{\;\;\beta}
       h_{\alpha \beta} 
       + \frac{1}{4}  \partial_{\mu}\phi \partial^{\mu} \phi h_{\nu}^{\;\;\nu}
       h^{\alpha \beta} h_{\alpha \beta} \\  &
       + \frac{1}{8}  \partial_{\mu} \phi \partial_{\nu} \phi h^{\mu \nu} h_{\alpha}^{\;\;\alpha} h_{\beta}^{\;\;\beta}
       - \frac{1}{2}  \partial_{\mu} \phi \partial^{\nu} \phi h^{\mu \alpha}
       h_{\nu \alpha} h_{\beta}^{\;\;\beta} \\  &
       -  \partial^{\mu} \phi \partial^{\nu} \phi h_{\mu \alpha}
       h_{\nu \beta} h^{\alpha \beta} \Big ] \, . && \numberthis
  \end{flalign*}
\end{minipage}\vspace{5mm}
Sixth, the ghost-ghost-graviton vertex
$V_{\rho\sigma\gamma\delta}^{\{\bar{\chi}\chi h\}} (p_1,p_2)$ also appears in
our one-loop diagrams as shown in Sec.~\ref{se:OneLoopCorrection}. It has 11
terms, compared to 8 in the standard vertex, and it takes the form \\
\noindent\begin{minipage}{.22\textwidth}
  \centering
  \vspace{5mm}
  \begin{axopicture}(100,100)
    \Text(0,0)[l]{$\chi_{\gamma}(p_1)$}
    \Text(90,0){$\chi_{\delta}(p_2)$}
    \Text(55,90)[l]{$h_{\rho\sigma}(q_1)$}
    \SetWidth{1.}
    \DoublePhoton(50,50)(50,100){2}{10}{2}
    \SetWidth{1.5}
    \Line[arrow](10,10)(50,50)
    \Line[arrow](50,50)(90,10)
    \SetWidth{0.8}
    \Vertex(50,50){3}
  \end{axopicture}
	\label{fig:Hghghabr}
\end{minipage}
\noindent\begin{minipage}{.77\textwidth}
  \begin{flalign*}
    \mathcal{L}_{\bar{\chi}\chi h} = \; \frac{\kappa}{2}  \Big ( & -
    \bar{\chi}^{\mu} \chi^{\nu} \partial_{\mu}\partial_{\nu}
    h_{\alpha}^{\;\;\alpha} + \bar{\chi}_{\mu} \partial^{\mu} \chi_{ \nu} \partial_{\alpha} h^{ \nu \alpha}
          + 2 \; \bar{\chi}_{\mu} \chi^{\nu} \partial_{\nu}\partial_{\alpha} h^{ \mu \alpha}
   \\  &  - \frac{1}{2}  \bar{\chi}_{\mu} \partial_{\nu} \chi^{ \nu} \partial^{\mu} h_{ \alpha}^{\;\;\alpha}
          -   \bar{\chi}_{\mu} \partial_{\nu} \chi_{ \alpha} \partial^{\mu} h^{ \nu \alpha}
         +   \bar{\chi}_{\mu} \partial_{\nu} \chi_{ \alpha} \partial^{\alpha} h^{ \mu \nu}
   \\  &  -   \partial^{\mu} \bar{\chi}_{ \mu} \partial_{\nu} \chi_{ \alpha} h^{\nu \alpha}
         -   \partial^{\mu} \bar{\chi}^{ \nu} \partial_{\mu} \chi_{ \nu} h_{\alpha}^{\;\;\alpha}
         -   \partial_{\mu} \bar{\chi}_{ \nu} \partial^{\mu} \chi_{ \alpha} h^{\nu \alpha}
  \\ &  +   \partial_{\mu} \bar{\chi}_{ \nu} \partial_{\alpha} \chi^{\nu} h^{\mu \alpha}
         -   \partial_{\mu} \bar{\chi}_{ \nu} \partial_{\alpha} \chi^{\alpha}
         h^{\mu \nu}  \Big ) \, . && \numberthis
  \end{flalign*}
\end{minipage}\vspace{6mm}
Seventh, the ghost-ghost-graviton-graviton vertex $V_{\gamma\delta\rho\sigma\lambda\epsilon}^{\{\bar{\chi}\chi h h\}}
  (p_1,p_2,q_1)$ is the last vertex that is needed in the calculations of our
  one-loop diagrams, where it appears in Fig.~\ref{fig:TripleGravitonVertex}e. It has 29 terms while it vanishes in the standard rules. It can be written as\\
\noindent\begin{minipage}{.22\textwidth}
	\centering
  \begin{axopicture}(100,100)
    \Text(0,0)[l]{$h_{\gamma\delta}(q_1)$}
    \Text(100,0)[r]{$h_{\rho\sigma}(q_2)$}
    \Text(100,100)[r]{$\chi_{\epsilon}(p_2)$}
    \Text(0,100)[l]{$\chi_{\lambda}(p_1)$}
    \SetWidth{1.}
    \Line[arrow](10,90)(50,50)
    \DoublePhoton(50,50)(10,10){2}{10}{2}
    \Line[arrow](50,50)(90,90)
    \DoublePhoton(50,50)(90,10){2}{10}{2}
    \SetWidth{0.8}
    \Vertex(50,50){3}
  \end{axopicture}
	\label{fig:HHghostghost}
\end{minipage}
\begin{minipage}{.77\textwidth}
\begin{flalign*}  
\mathcal{L}_{\bar{\chi} \chi h h} = 
 \frac{\kappa^2}{8} \Big ( & -  \bar{\chi}^{\mu} \partial^{\nu} \chi_{ \nu} h_{\mu \alpha} \partial_{\beta} h^{ \alpha \beta}
       +   \bar{\chi}_{\mu} \partial^{\mu} \chi^{ \nu} h_{\nu \alpha} \partial_{\beta} h^{ \alpha \beta}
       + 2\,  \bar{\chi}_{\mu} \partial^{\mu} \chi^{ \nu} \partial_{\nu} h^{ \alpha \beta} h_{\alpha \beta}
\\ &   +   \bar{\chi}_{\mu} \partial^{\mu} \chi^{ \nu} \partial^{\alpha} h_{ \nu \alpha} h_{\beta}^{\;\;\beta}
       + 2\,  \bar{\chi}_{\mu} \chi^{\nu} h^{\mu \alpha} \partial_{\nu}\partial_{\alpha} h_{\beta}^{\;\;\beta}
       +  \bar{\chi}^{\mu} \chi^{\nu} \partial_{\mu} h_{ \alpha}^{\;\;\alpha} \partial_{\nu} h_{ \beta}^{\;\;\beta}
\\ &  + 2\,  \bar{\chi}^{\mu} \chi^{\nu} \partial_{\nu} h_{ \mu \alpha} \partial^{\alpha} h_{\beta}^{\;\;\beta}
       + 2\,  \bar{\chi}^{\mu} \chi^{\nu} \partial_{\nu} \partial^{\alpha} h_{\mu \alpha} h_{\beta}^{\;\;\beta}
       +   \bar{\chi}^{\mu} \partial^{\nu} \chi_{ \mu} h_{\nu \alpha} \partial_{\beta} h^{ \alpha \beta}
\\ &   +   \bar{\chi}_{\mu} \partial^{\mu} \chi_{\nu} h^{\nu \alpha} \partial_{\alpha} h_{ \beta}^{\;\;\beta}
       -   \bar{\chi}^{\mu} \partial^{\nu} \chi_{ \nu} \partial_{\alpha} h_{ \mu \beta} h^{\alpha \beta}
       +   \bar{\chi}^{\mu} \partial^{\nu} \chi^{ \alpha} h_{\mu \nu} \partial_{\alpha} h_{ \beta}^{\;\;\beta}
\\ &  +   \bar{\chi}_{\mu} \partial_{\nu} \chi^{ \alpha} h^{\mu \nu} \partial^{\beta} h_{ \alpha \beta}
       -   \bar{\chi}^{\mu} \partial^{\nu} \chi^{ \alpha} \partial_{\mu} h_{ \nu \alpha} h_{\beta}^{\;\;\beta}
       -   \bar{\chi}^{\mu} \partial^{\nu} \chi^{ \alpha} h_{\mu \alpha} \partial_{\nu} h_{ \beta}^{\;\;\beta}
\\ &   +   \bar{\chi}^{\mu} \partial^{\nu} \chi^{ \alpha} h_{\mu \alpha} \partial^{\beta} h_{ \nu \beta}
       -   \bar{\chi}^{\mu} \partial^{\nu} \chi^{ \alpha} \partial_{\mu} h_{ \beta}^{\;\;\beta} h_{\nu \alpha}
       -   \bar{\chi}^{\mu} \partial^{\nu} \chi^{ \alpha} \partial_{\nu} h_{ \mu \alpha} h_{\beta}^{\;\;\beta}
\\ &   - 4\,  \bar{\chi}^{\mu} \partial^{\nu} \chi^{ \alpha} h_{\nu \alpha} \partial^{\beta} h_{ \mu \beta}
       +   \bar{\chi}^{\mu} \partial^{\nu} \chi^{ \alpha} \partial_{\alpha} h_{ \mu \nu} h_{\beta}^{\;\;\beta}
       -   \partial^{\mu} \bar{\chi}_{ \mu} \partial^{\nu} \chi^{ \alpha} h_{\nu \alpha} h_{\beta}^{\;\;\beta}
\\ &  -   \partial^{\mu} \bar{\chi}^{ \nu} \partial_{\mu} \chi_{ \nu} h^{\alpha \beta} h_{\alpha \beta}
       - 2\,  \partial^{\mu} \bar{\chi}^{ \nu} \partial_{\mu} \chi^{ \alpha} h_{\nu \alpha} h_{\beta}^{\;\;\beta}
       +   \partial^{\mu} \bar{\chi}^{ \nu} \partial^{\alpha} \chi_{ \nu} h_{\mu \alpha} h_{\beta}^{\;\;\beta}
\\ &  -   \partial^{\mu} \bar{\chi}^{ \nu} \partial^{\alpha} \chi_{ \alpha} h_{\mu \nu} h_{\beta}^{\;\;\beta}
       -  \partial^{\mu} \bar{\chi}_{ \nu} \partial^{\alpha} \chi_{ \alpha} h_{\mu \beta} h^{\nu \beta}
       - 4\,  \partial^{\mu} \bar{\chi}^{ \nu} \partial^{\alpha} \chi^{ \beta} h_{\mu \nu} h_{\alpha \beta}
\\ & + 2\,  \partial^{\mu} \bar{\chi}^{ \nu} \partial^{\alpha} \chi^{ \beta} h_{\mu \alpha} h_{\nu \beta}
       + 2\,  \partial^{\mu} \bar{\chi}^{ \nu} \partial^{\alpha} \chi^{
         \beta} h_{\mu \beta} h_{\nu \alpha} \Big ) \, . &&  \numberthis \label{eq:HHghostghost}
     \end{flalign*}
\end{minipage}\vspace{5mm}
Our ghost vertices above are more complicated than the standard ghost vertices
because our general parameterized gauge Eq.~(\ref{eq:Gauge}) is more complicated
than the de Donder gauge Eq.~(\ref{eq:GaugeDonder}). However, the ghost vertices
just appear in few diagrams in scalar-graviton scattering as shown in Sec.~\ref{se:OneLoopCorrection}, so they do not affect our calculations so much if they have slightly more complicated form.

Finally, after deriving the simplified Feynman rules, it is time to apply these rules on
scattering processes as shown in the next section.

\section{Tree Level Scattering \label{se:TreeLevel}}
\setcounter{equation}{0}
Now all the ingredients are in place to start the calculations, but before diving into one-loop
diagrams, we calculate the amplitudes of scalar-graviton scattering and
graviton-graviton scattering at tree level. So, we first show how we use the
helicity formalism \cite{HelicityAmplitudes} to express the resulting amplitudes
at tree level. After that, we calculate these amplitudes by using
the simplified Feynman rules, as shown in the previous section, then repeat the
calculations for the same amplitudes but using the standard Feynman rules, as
shown in App.~\ref{Ap:FeynmanRules}. Finally, we compare the results obtained in
the two ways and check against published results where available.

\subsection{Helicity Amplitudes \label{sec:HelicityAmplitude}}
A helicity amplitude is an amplitude $\mathcal{M}$ that is evaluated for fixed helicity
of the external particles, where the helicity of a particle is the projection of its
spin on its momentum. Namely, if we consider the process: $ a_1 + a_2
\rightarrow a_3 + a_4  $, then the total amplitude $\mathcal{M}$ will be decomposed into helicity amplitudes $\mathcal{M}_{(\lambda_1,\lambda_2;\lambda_3,\lambda_4)}$ each
one representing the amplitude of transition from a particular helicity state
$\lambda_1,\lambda_2$  of the incoming particles $a_1,a_2$ to a particular helicity
state $\lambda_3,\lambda_4$ of the outgoing particles $a_3,a_4$.

In our case, the graviton has a polarization tensor $\epsilon_{\mu\nu}^{\pm
  2}(p)$ for helicity $\pm 2$ which can
be written in terms of the polarization vector $\epsilon_{\mu}^{\pm 1}(p)$ for
helicity $\pm 1$ as
$$ \epsilon_{\mu\nu}^{\pm 2}(p)= \epsilon_{\mu}^{\pm 1}(p) \epsilon_{\nu}^{\pm
  1}(p)\, . $$
In addition, the graviton is a spin-2 massless particle of a symmetric gravitational field
$h_{\mu\nu}$. This implies that its polarization tensor $\epsilon_{\mu\nu}$ is
transverse, traceless and symmetric \cite{HelicityAmplitudes}:
\begin{align}
  \label{eq:PolarizationRelations}
  p^{\mu} \epsilon_{\mu\nu}^{\pm 2}(p)&= p^{\nu} \epsilon_{\mu\nu}^{\pm 2}(p)  =0 \, ,\\
  \eta^{\mu\nu} \epsilon_{\mu\nu}^{\pm 2}(p) &=\epsilon_{\;\;\;\nu}^{\pm 2\; \nu }(p)  =0 \, , \\
  \epsilon_{\mu\nu}^{\pm 2}(p)& = \epsilon_{\nu\mu}^{\pm 2}(p) \, ,
\end{align}
where our choice of polarization vectors and four momenta in the
Center-of-Mass frame (\acrshort{cm}) are given in App.~\ref{Ap:Kinematics} together with other useful kinematic relations.

Finally, helicity amplitudes have some useful properties. For example, in
general the results are given by simple expressions and the total amplitude can
be squared directly without having to use the completeness relations. Above all,
only some of the helicity amplitudes are actually independent. The others can be
calculated by using various symmetries \cite{TreeLevelResults}, such as parity
\begin{align}
\mathcal{M}_{(\lambda_3,\lambda_4;\lambda_1,\lambda_2)}  & = (-1)^{m-n} \mathcal{M}_{(-\lambda_3,-\lambda_4;-\lambda_1,-\lambda_2)} \label{eq:Parity} \, ,
\end{align}
time-reversal
\begin{align}
\mathcal{M}_{(\lambda_3,\lambda_4;\lambda_1,\lambda_2)} & = (-1)^{m-n} \mathcal{M}_{(\lambda_1,\lambda_2;\lambda_3,\lambda_4)}     \label{eq:TimeReversal} \, ,
\end{align}
charge conjugation
\begin{align}
\mathcal{M}_{(\lambda_3,\lambda_4;\lambda_1,\lambda_2)} &  = (-1)^{m-n} \mathcal{M}_{(\lambda_4,\lambda_3;\lambda_2,\lambda_1)}  \label{eq:ChargeConjugation} \, ,
\end{align}
and exchanging bosons
\begin{align}
\mathcal{M}_{(\lambda_3,\lambda_4;\lambda_1,\lambda_2)} (s,t,u) & = (-1)^{m-2s_1} \mathcal{M}_{(\lambda_3,\lambda_4;\lambda_2,\lambda_1)} (s,u,t)  \label{eq:Bose} \, ,
\end{align}
where $m =\lambda_1-\lambda_2$, $\; n=\lambda_3-\lambda_4 $ and $s_1,s_2,s_3,s_4$ are the spin of the particles.


\subsection{Scalar-Graviton Scattering \label{se:SG}}
\begin{figure}[!h]
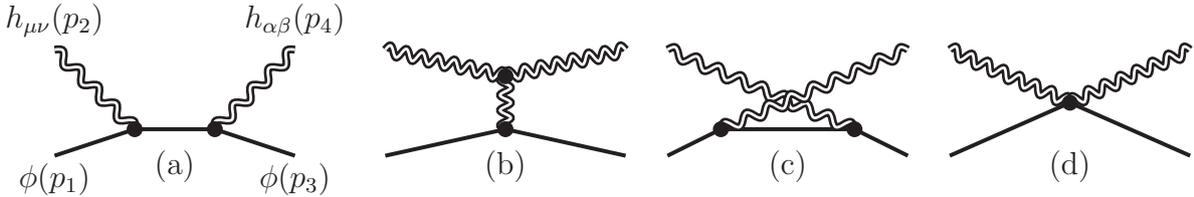

  \vspace{7mm}
	\centering
  \noindent
  \begin{axopicture}(90,50)
    \SetWidth{1.5}
    \Line(0,10)(30,20)
    \Line(30,20)(60,20)
    \Line(60,20)(90,10)
    \SetWidth{1.}
    \DoublePhoton(0,50)(30,20){2}{4.5}{2}
    \DoublePhoton(60,20)(90,50){2}{4.5}{2}
    \Vertex(30,20){3}
    \Vertex(60,20){3}
    \Text(45,0)[b]{(a)}
    \Text(0,-5)[b]{$\phi(p_1)$}
    \Text(90,-5)[b]{$\phi(p_3)$}
    \Text(0,55)[b]{$h_{\mu\nu}(p_2)$}
    \Text(90,55)[b]{$h_{\alpha\beta}(p_4)$}
  \end{axopicture}\hspace{5mm}
  ~~~
  \begin{axopicture}(90,50)
    \SetWidth{1.5}
    \Line(0,10)(45,20)
    \Line(45,20)(90,10)
    \SetWidth{1.}
    \DoublePhoton(0,50)(45,40){2}{7.5}{2}
    \DoublePhoton(45,40)(90,50){2}{7.5}{2}
    \DoublePhoton(45,20)(45,40){2}{3.5}{2}
    \Vertex(45,20){3}
    \Vertex(45,40){3}
    \Text(45,0)[b]{(b)}
  \end{axopicture}
  ~~~
  \begin{axopicture}(90,50)
    \SetWidth{1.5}
    \Line(0,10)(20,20)
    \Line(20,20)(70,20)
    \Line(70,20)(90,10)
    \SetWidth{1.}
    \DoublePhoton(0,50)(70,20){2}{8.5}{2}
    \DoublePhoton(20,20)(90,50){2}{8.5}{2}
    \Vertex(20,20){3}
    \Vertex(70,20){3}
    \Text(45,0)[b]{(c)}
  \end{axopicture}
  ~~~
  \begin{axopicture}(90,50)
    \SetWidth{1.5}
    \Line(0,10)(45,30)
    \Line(45,30)(90,10)
    \SetWidth{1.}
    \DoublePhoton(0,50)(45,30){2}{7.5}{2}
    \DoublePhoton(45,30)(90,50){2}{7.5}{2}
    \Vertex(45,30){3}
    \Text(45,0)[b]{(d)}
  \end{axopicture}
  \caption{Scalar-Graviton scattering at tree level, where (a), (b) and (c)
    represent s, t and u-channels respectively, while (d) is just a simple scalar-scalar-graviton-graviton vertex.}
	\label{fig:TreeLevelScalarGraviton}
\end{figure}

For this process, there are four possible diagrams at tree level as shown in
Fig.~\ref{fig:TreeLevelScalarGraviton}. We use the helicity amplitudes
formalism, as discussed before, to write down the amplitudes. For this
process we have two independent helicity amplitudes $\mathcal{M}_{(0,+2;0,+2)}$,
$\mathcal{M}_{(0,+2;0,-2)}$ and the others can be obtained by applying the
symmetries that are given in the previous subsection.

\noindent The amplitudes of the respective diagrams in
Fig.~\ref{fig:TreeLevelScalarGraviton} can then be written as:
\begin{alignat}{5}
  \mathcal{M}_{\text{(a)}} &=  \epsilon^{\mu\nu}(p_2)  \;\; &&
  V_{\mu\nu}^{\{\phi\phi h\}} (p_1,Q_1,m) \;\; && S(Q_1,m) \;\;
  &&V_{\alpha\beta}^{\{\phi\phi h\}} (p_3,Q_1,m) \;\; && \epsilon^{*
    \alpha\beta}(p_4) \, , \nonumber \\[2mm]
  \mathcal{M}_{\text{(b)}} &=  \epsilon^{\mu\nu}(p_2) \;\; &&
  V_{\gamma\delta}^{\{\phi\phi h\}} (p_1,p_3,m) \;\; &&
  S^{\gamma\delta\rho\sigma}_{\{h h\}}(Q_2) \;\; &&
  V_{\mu\nu\alpha\beta\rho\sigma}^{\{h h h\}} (p_2,p_4) \;\; && \epsilon^{*
    \alpha\beta}(p_4) \, , \nonumber\\[2mm]
  \mathcal{M}_{\text{(c)}} &=  \epsilon^{\mu\nu}(p_2) \;\; &&
  V_{\mu\nu}^{\{\phi\phi h\}} (p_1,Q_3,m) \;\; && S(Q_3,m) \;\; &&
  V_{\alpha\beta}^{\{\phi\phi h\}} (p_3,Q_3,m) \;\; && \epsilon^{*
    \alpha\beta}(p_4) \, , \nonumber\\[2mm]
  \mathcal{M}_{\text{(d)}} &=  \epsilon^{\mu\nu}(p_2) \;\; &&
  V_{\mu\nu\alpha\beta}^{\{\phi\phi h h\}}(p_1,p_3)  \;\; &&\epsilon^{*
    \alpha\beta}(p_4) \, ,  && &&  \label{eq:GeneralAmplitudeScalar}
\end{alignat}
where $Q_1=p_1+p_2$, $Q_2=p_1-p_3$ and $Q_3=p_1-p_4$.\\
Adding these amplitudes together, we get the total amplitude as
\begin{align}
  \mathcal{M}_{\text{(Total)}} &= \mathcal{M}_{\text{(a)}} + \mathcal{M}_{\text{(b)}} + \mathcal{M}_{\text{(c)}} + \mathcal{M}_{\text{(d)}} \, .
\end{align}
In the helicity formalism, this total amplitude can be written as
\begin{align}
  \mathcal{M}_{\text{(Total)}} & = \mathcal{M}_{(0,+2;0,+2)} + \mathcal{M}_{(0,-2;0,-2)} + \mathcal{M}_{(0,+2;0,-2)} + \mathcal{M}_{(0,-2;0,+2)} \, .
\end{align}

\noindent Using the simplified Feynman rules that we derived in the previous
section and applying kinematics in the \acrshort{cm} frame according to the choice
of momenta and polarization vectors as shown in App.~\ref{Ap:Kinematics},
the two independent helicity amplitudes are given by
\begin{align}
  \mathcal{M}_{(0,+2;0,+2)} =& \;\;  \kappa^2  \frac{ k^4 }{(s-m^2)(u-m^2)t} \big (1 + \cos(\theta) \big)^2
                  \Big [  m^4 + 4 k m^2 E + 8 k^2 m^2 + 8 k^3 E + 8 k^4   \Big] \, ,  \nonumber\\
  \mathcal{M}_{(0,+2;0,-2)} =& \;\;  \kappa^2 \frac{ k^4 \, m^4}{(s-m^2)(u-m^2)t} \big (1 - \cos(\theta) \big)^2 \, , \nonumber
\end{align}
where $\theta$ is the scattering angle and $k$ is the momentum of the incoming
particles in the \acrshort{cm} frame.

\noindent However, in the \acrshort{cm} frame, we have:
\begin{align}
   s & = m^2 + 2 k^2 + 2 k E \, , \nonumber\\
  E^2 & = m^2 + k^2 \, . \nonumber
\end{align}
Then, the two independent helicity amplitudes can be written as:

\begin{align}
  \label{eq:ScalarGravitonTreeResults}
  \mathcal{M}_{(0,+2;0,+2)} =&\;\;  \kappa^2 \frac{ k^4 \, s^2 }{(s-m^2)(u-m^2)t} \big (1 + \cos(\theta) \big)^2 \, , \\
  \mathcal{M}_{(0,+2;0,-2)} =& \;\;  \kappa^2 \frac{ k^4 \, m^4}{(s-m^2)(u-m^2)t} \big (1 - \cos(\theta) \big)^2 \nonumber \, .
\end{align}

After obtaining the results with our simplified Feynman rules, we can start
the process of comparison and checking the results. 
First, we have verified that the standard Feynman rules in App.~\ref{Ap:FeynmanRules} give the same results for the independent helicity
amplitudes Eq.~(\ref{eq:ScalarGravitonTreeResults}). In addition, our results
for the scalar-graviton scattering helicity amplitudes also agree with the results of M.
T. Grisaru and P. Van Nieuwenhuizen and C. C. Wu in \cite{TreeLevelResults}.

\subsection{Graviton-Graviton Scattering \label{se:GG}}
\begin{figure}[!h]
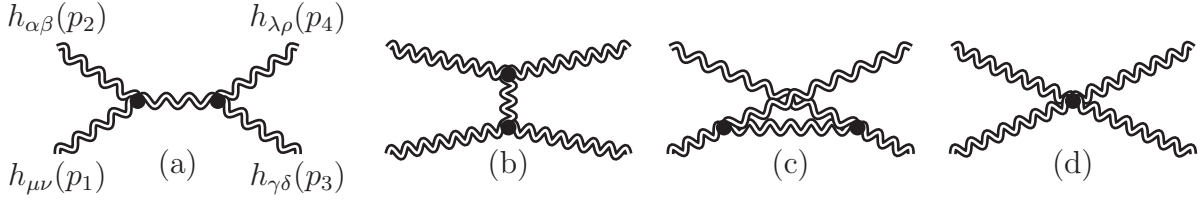

  \vspace{8mm}
	\centering
  \begin{axopicture}(90,50)
    \SetWidth{1.}
    \DoublePhoton(0,50)(30,30){2}{4.5}{2}
    \DoublePhoton(60,30)(90,50){2}{4.5}{2}
    \DoublePhoton(0,10)(30,30){2}{4.5}{2}
    \DoublePhoton(60,30)(90,10){2}{4.5}{2}
    \DoublePhoton(30,30)(60,30){2}{3.5}{2}
    \Vertex(30,30){3}
    \Vertex(60,30){3}
    \Text(45,0)[b]{(a)}
    \Text(0,-5)[b]{$h_{\mu\nu}(p_1)$}
    \Text(90,-5)[b]{$h_{\gamma\delta}(p_3)$}
    \Text(0,55)[b]{$h_{\alpha\beta}(p_2)$}
    \Text(90,55)[b]{$h_{\lambda\rho}(p_4)$}
  \end{axopicture}\hspace{5mm}
  ~~~
  \begin{axopicture}(90,50)
    \SetWidth{1.}
    \DoublePhoton(0,50)(45,40){2}{7.5}{2}
    \DoublePhoton(45,40)(90,50){2}{7.5}{2}
    \DoublePhoton(0,10)(45,20){2}{7.5}{2}
    \DoublePhoton(45,20)(90,10){2}{7.5}{2}
    \DoublePhoton(45,20)(45,40){2}{3.5}{2}
    \Vertex(45,20){3}
    \Vertex(45,40){3}
    \Text(45,0)[b]{(b)}
  \end{axopicture}
  ~~~
  \begin{axopicture}(90,50)
    \SetWidth{1.}
    \DoublePhoton(0,10)(20,20){2}{3.5}{2}
    \DoublePhoton(20,20)(70,20){2}{6.5}{2}
    \DoublePhoton(70,20)(90,10){2}{3.5}{2}
    \DoublePhoton(0,50)(70,20){2}{8.5}{2}
    \DoublePhoton(20,20)(90,50){2}{8.5}{2}
    \Vertex(20,20){3}
    \Vertex(70,20){3}
    \Text(45,0)[b]{(c)}
  \end{axopicture}
  ~~~
  \begin{axopicture}(90,50)
    \SetWidth{1.}
    \DoublePhoton(0,50)(45,30){2}{7.5}{2}
    \DoublePhoton(45,30)(90,50){2}{7.5}{2}
    \DoublePhoton(0,10)(45,30){2}{7.5}{2}
    \DoublePhoton(45,30)(90,10){2}{7.5}{2}
    \Vertex(45,30){3}
    \Text(45,0)[b]{(d)}
  \end{axopicture}
    \caption{Graviton-Graviton scattering at tree level, where (a), (b) and (c)
    represent s, t and u-channels respectively, while (d) is just a simple quadruple graviton vertex.}
	\label{fig:TreeLevelGravitonGraviton}
\end{figure}

For this process, there are also four possible diagrams at tree level as shown in
Fig.~\ref{fig:TreeLevelGravitonGraviton}, and the helicity amplitude formalism
is again used to express the final results. This process has four
independent helicity amplitudes $\mathcal{M}_{(+2,+2;+2,+2)}$,
$\mathcal{M}_{(+2,-2;+2,-2)}$, $\mathcal{M}_{(+2,+2;+2,-2)}$,
$\mathcal{M}_{(+2,+2;-2,-2)}$.

\noindent The amplitudes of the respective diagrams in Fig.~\ref{fig:TreeLevelGravitonGraviton} can then be written as:
\begin{align}
  \mathcal{M}_{\text{(a)}} &=  \epsilon^{\mu\nu}(p_1) \epsilon^{\alpha\beta}(p_2) \;\; V_{\mu\nu\alpha\beta\eta\sigma}^{\{h h h\}} (p_1,p_2) \;\; S^{\eta\sigma\epsilon\kappa}_{\{h h \}}(Q_1) \;\; V_{\gamma\delta\lambda\rho\epsilon\kappa}^{\{h h h\}} (p_3,p_4) \;\; \epsilon^{* \gamma\delta}(p_3) \epsilon^{* \lambda\rho}(p_4) \, , \nonumber \\[5mm]
  \mathcal{M}_{\text{(b)}} &=  \epsilon^{\mu\nu}(p_1) \epsilon^{\alpha\beta}(p_2) \;\; V_{\mu\nu\alpha\beta\eta\sigma}^{\{h h h\}}(p_1,p_3) \;\; S^{\eta\sigma\epsilon\kappa}_{\{h  h\}}(Q_2) \;\; V_{\gamma\delta\lambda\rho\epsilon\kappa}^{\{h h h\}} (p_2,p_4) \;\; \epsilon^{* \gamma\delta}(p_3) \epsilon^{\lambda\rho}(p_4) \, , \nonumber \\[5mm]
  \mathcal{M}_{\text{(c)}} &=  \epsilon^{\mu\nu}(p_1) \epsilon^{\alpha\beta}(p_2) \;\; V_{\mu\nu\alpha\beta\eta\sigma}^{\{h h h\}}(p_1,p_4) \;\; S^{\eta\sigma\epsilon\kappa}_{\{h h\}}(Q_3) \;\; V_{\gamma\delta\lambda\rho\epsilon\kappa}^{\{h h h\}} (p_2,p_3) \;\; \epsilon^{* \gamma\delta}(p_3) \epsilon^{* \lambda\rho}(p_4) \, , \nonumber \\[5mm]
  \mathcal{M}_{\text{(d)}} &=  \epsilon^{\mu\nu}(p_1) \epsilon^{\alpha\beta}(p_2) \;\; V_{\mu\nu\alpha\beta\gamma\delta\lambda\rho}^{\{h h h h\}}(p_1,p_2,p_3) \;\; \epsilon^{* \gamma\delta}(p_3) \epsilon^{* \lambda\rho}(p_4) \, ,   \label{eq:GeneralAmplitudeGraviton}
\end{align}
where $Q_1=p_1+p_2$, $Q_2=p_1-p_3$ and $Q_3=p_1-p_4$.\\
Adding these amplitudes together, we get the total amplitude as
\begin{align}
  \mathcal{M}_{\text{(Total)}} &= \mathcal{M}_{\text{(a)}} + \mathcal{M}_{\text{(b)}} + \mathcal{M}_{\text{(c)}} + \mathcal{M}_{\text{(d)}} \, .
\end{align}
In the helicity formalism, this total amplitude can be written as
\begin{align*}
  \mathcal{M}_{\text{(Total)}}  = & \;\;\;\; \mathcal{M}_{(+2,+2;+2,+2)} + \mathcal{M}_{(-2,-2;-2,-2)} + \mathcal{M}_{(+2,-2;+2,-2)} + \mathcal{M}_{(-2,+2;-2,+2)} + \mathcal{M}_{(+2,-2;-2,+2)} \\ & +  \mathcal{M}_{(-2,+2;+2,-2)} + \mathcal{M}_{(+2,+2;+2,-2)} + \mathcal{M}_{(-2,-2;-2,+2)} + \mathcal{M}_{(+2,-2;+2,+2)} + \mathcal{M}_{(-2,+2;-2,-2)} \\ & + \mathcal{M}_{(+2,+2;-2,+2)} + \mathcal{M}_{(-2,-2;+2,-2)} + \mathcal{M}_{(-2,+2;+2,+2)} + \mathcal{M}_{(+2,-2;-2,-2)} + \mathcal{M}_{(+2,+2;-2,-2)} \\ & +  \mathcal{M}_{(-2,-2;+2,+2)}  \, . 
\end{align*}
Again, using our simplified Feynman rules and applying kinematics in the
\acrshort{cm} frame gives the four independent helicity amplitudes:
\begin{align}
  \mathcal{M}_{(+2,+2;+2,+2)} &=  \kappa^2 \, \frac{1}{4}   \,  \frac{s^3}{ t\, u} \, , \nonumber\\[4mm]
  \mathcal{M}_{(+2,-2;+2,-2)} &=  \kappa^2 \, \frac{1}{4}  \,  \frac{u^3}{ s\, t }  \, , \label{eq:IndependentResultsGravitonGraviton}\\[4mm]
  \mathcal{M}_{(+2,+2;+2,-2)} &= 0 \, , \nonumber\\[4mm]
  \mathcal{M}_{(+2,+2;-2,-2)} &= 0 \, . \nonumber
\end{align}

\noindent Applying the symmetry relations Eqs.~(\ref{eq:Parity}$-$\ref{eq:Bose}) gives all helicity
amplitudes for graviton-graviton scattering:

\begin{align}
  \mathcal{M}_{(+2,+2;+2,+2)} &= \mathcal{M}_{(-2,-2;-2,-2)} =  \kappa^2 \, \frac{1}{4}   \,  \frac{s^3}{ t\, u} \, , \nonumber\\[2mm]
  \mathcal{M}_{(+2,-2;+2,-2)} &= \mathcal{M}_{(-2,+2;-2,+2)} =  \kappa^2 \, \frac{1}{4}   \,  \frac{u^3}{ s\, t } \, , \nonumber\\[2mm]
  \mathcal{M}_{(+2,-2;-2,+2)} &= \mathcal{M}_{(-2,+2;+2,-2)} =  \kappa^2 \, \frac{1}{4}   \,  \frac{t^3}{ s\, u }   \, , \label{eq:AllResultsGravitonGraviton}\\[2mm]
  \mathcal{M}_{(+2,+2;+2,-2)} &= \mathcal{M}_{(-2,-2;-2,+2)} = \mathcal{M}_{(+2,-2;+2,+2)} = \mathcal{M}_{(-2,+2;-2,-2)} = 0  \, , \nonumber\\[2mm]
  \mathcal{M}_{(+2,+2;-2,+2)} &= \mathcal{M}_{(-2,-2;+2,-2)} = \mathcal{M}_{(-2,+2;+2,+2)} = \mathcal{M}_{(+2,-2;-2,-2)} = 0  \, , \nonumber\\[2mm]
  \mathcal{M}_{(+2,+2;-2,-2)} &= \mathcal{M}_{(-2,-2;+2,+2)} = 0 \, . \nonumber
\end{align}

Once more, comparing with the standard Feynman rules, the same results are
reached. Our results for these helicity
amplitudes also agree with the results of J. F.
Donoghue and T. Torma in their paper \cite{DonoghueTreeLevel} and M. T. Grisaru, P. Van Nieuwenhuizen and C. C. Wu in \cite{TreeLevelResults}.

\section{One-Loop Correction \label{se:OneLoopCorrection}}
\setcounter{equation}{0}
While the goal in the previous section was to verify the simplified Feynman
rules, the goal in this section is to show the usefulness of these rules at loop
level. So, we
first show how to treat the loop integrals in the loop calculations by using
dimensional regularization and the Passarino-Veltman method
\cite{Peskin,MasslessTadpole,PassarinoVeltman}. After that, we calculate some
one-loop diagrams for scalar-graviton scattering using the simplified rules,
then repeat the calculations using the standard rules, and finally compare the
results obtained in the two ways.

\subsection{Loop Integral \label{se:loopintegral}}
To calculate our loop integrals, we use dimensional regularization with Passarino-Veltman reduction since it preserves gauge and Lorentz
invariance \cite{Peskin,PassarinoVeltman}.

\subsubsection{Dimensional Regularization}
Dimensional regularization is widely used to regularize loop integrals and
separate out the UV divergences \cite{Peskin}. The main idea is
to change the dimensionality of the loop integral from the dimension where it diverges to a
lower dimension $d$ where the integral converges.

In our case, the loop integrals are in four-dimensional Minkowski space.
So, we move them to dimension $d=4-2\epsilon$, where $\epsilon$ is a parameter and the limit
$\epsilon \rightarrow 0 $ will be taken at the end of the calculation. Then, we
perform the integrals and go back to the original dimension by doing analytic continuation. Moreover, there are some considerations when using
$d \neq 4$ dimensions that have to be taken into account: the metric tensor becomes
$$ g^{\mu \nu}_4 \; \rightarrow \; g^{\mu \nu}_d \, ,\qquad \Rightarrow \qquad g^{\mu\nu} g_{\mu \nu} = \delta^\mu_\mu = d =4-2\epsilon\, , $$
and the measures become
$$\int \frac{d^4 p}{(2\pi)^4} \qquad  \rightarrow \qquad  \int
\frac{(\mu)^{2\epsilon} \; d^d p}{(2\pi)^d}\, ,$$
where $\mu$ is a regulator parameter with dimension $[\mu]=M$.

Briefly, the standard procedure of this regularization is \cite{Peskin}: transfer to
Euclidean space, do the Wick rotation, apply Feynman parameters, shift the
integration variable, perform the integral, go back to Minkowski space. These steps will be explained in more detail and applied in the next section.
\subsubsection{Scalar Integrals}
\begin{figure}[H]
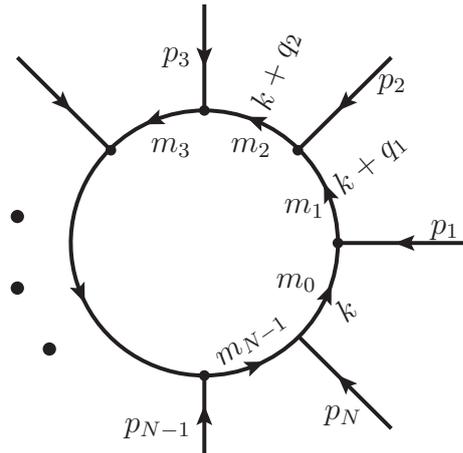

  \vspace*{7mm}
  \centering
  \begin{axopicture}(140,140)
    \SetWidth{1.5}
    \Arc[arrow](70,70)(50,0,45)
    \Arc[arrow](70,70)(50,45,90)
    \Arc[arrow](70,70)(50,90,135)
    \Arc[arrow](70,70)(50,135,270)
    \Arc[arrow](70,70)(50,270,315)
    \Arc[arrow](70,70)(50,315,360)
    \Line[arrow](170,70)(120,70) 
    \Text(112,55)[r]{$ m_0 $}
    \Text(123,45)(40){$ k $}
    \Text(160,75){$p_1$}
    \Text(115,83)[r]{$ m_1 $}
    \Text(132,100)(40){$ k+q_1 $}
    \Line[arrow](70,-10)(70,20) 
    \Text(65,0)[r]{$p_{N-1}$}
    \Text(88,33)(35){$ m_{N-1} $}
    \Line[arrow](70,160)(70,120) 
    \Text(65,140)[r]{$p_3$}
    \Text(65,106)[r]{$ m_3 $}
    \Line[arrow](140,0)(105,35) 
    \Text(115,5)[l]{$p_{N}$}
    \Line[arrow](140,140)(105,105) 
    \Text(135,130)[l]{$p_2$}
    \Text(95,106)[r]{$ m_2 $}
    \Text(97,135)(70){$ k+q_2 $}
    \Line[arrow](0,140)(35,105)
    \Vertex(0,80){2.5}
    \Vertex(0,53){2.5}
    \Vertex(12,30){2.5}
    \Vertex(120,70){2}
    \Vertex(70,120){2}
    \Vertex(70,20){2}
    \Vertex(105,105){2}
    \Vertex(35,105){2}
  \end{axopicture}    
  \caption{Generic diagram of N-point function with N external momenta at
    one-loop.}
  \label{fig:GeneralLoop}
\end{figure}

The general form of a scalar one-loop integral for a N-point function with external momenta
$p_1,\cdots,p_{N-1}$ (with $p_N$ from momentum conservation) as shown in Fig.~\ref{fig:GeneralLoop} is given by
\begin{align}
  \label{eq:GeneralScalarLoopIntegral}
  I^N( p_1,\cdots,p_{N-1}&,m_0,\cdots,m_{N-1}) \sim \\[3mm]
                      &  \int \frac{d^d k}{(2\pi)^d} \frac{1}{(k^2-m_0^2+i\epsilon)((k+q_1)^2-m_1^2+i\epsilon)\, \cdots \, ((k+q_{N-1})^2-m_{N-1}^2+i\epsilon)} \, , \nonumber
\end{align}
where $k$ is the undefined momentum in the loop that will be integrated over, and
$q_1,\cdots,q_{N-1}$ are the internal momenta that are related to the external
momenta by $q_i=\sum_{k=1}^i p_k$ as shown in Fig.~\ref{fig:GeneralLoop},
and $m_0,\cdots,m_{N-1}$ are the masses of the propagators involved in the loop.

In our case of scalar-graviton scattering to one-loop, only the scalar field
propagator has a mass $m$, so the masses $m_0,\cdots,m_{N-1}$ are either zero or
$m$. In the following, we will only need one-loop diagrams up to the box diagrams with four propagators so we limit ourselves to the following scalar integrals:
\begin{align}
  &A_0(m_0) =  \int \frac{d^d k}{(2\pi)^d}  \frac{1}{(k^2-{m}^2_0+i\epsilon)} \nonumber\, ,  \\[6mm]
  &B_0(p_1,m_0,m_1) =  \int \frac{d^d k}{(2\pi)^d}  \frac{1}{(k^2-{m}^2_0+i\epsilon)((k+q_1)^2-{m}^2_1+i\epsilon)} \label{eq:BZero} \, , \\[6mm]
  & C_0(p_1,p_2,m_0,m_1,m_2) =  \int \frac{d^d k}{(2\pi)^d}  \frac{1}{(k^2-{m}^2_0+i\epsilon)((k+q_1)^2-{m}^2_1+i\epsilon)((k+q_2)^2-{m}^2_2+i\epsilon)}\nonumber \, , \\[6mm]
  & D_0(p_1,p_2,p_3,m_0,m_1,m_2,m_3) = \nonumber \\
  & \qquad\quad\;\,\, \int \frac{d^d k}{(2\pi)^d} \frac{1}{(k^2-{m}^2_0+i\epsilon)((k+q_1)^2-{m}^2_1+i\epsilon)((k+q_2)^2-{m}^2_2+i\epsilon)((k+q_3)^2-{m}^2_3+i\epsilon)} \nonumber \, , 
\end{align}
where the loop integrals are denoted with respect to the number of propagators involved in the
loop: ($A_0$) for one propagator, ($B_0$) for two propagators, ($C_0$) for three propagators and ($D_0$) for four propagators.

\begin{figure}[H]
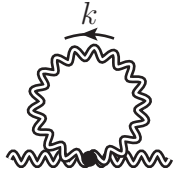

  \centering
  \vspace{3mm}
  \begin{axopicture}(60,60)
    \SetWidth{1.}
    \Arc[arrow](30,30)(27,70,110)
    \Text(30,61)[b]{$k$}
    \DoublePhoton(0,10)(60,10){2}{10}{2}
    \DoublePhotonArc(30,30)(20,0,360){2}{18}{2}
    \Vertex(30,10){3}
  \end{axopicture}
  \caption{One-loop massless tad-pole diagram.}
	\label{fig:Tad-pole}
\end{figure}

Before continuing we note that as a result of the standard procedure of dimensional regularization, all
massless tad-pole diagrams, as illustrated in Fig.~\ref{fig:Tad-pole},
vanish. 
\noindent More specifically these diagrams correspond to the massless scalar integral
$A_0$ given by
\begin{align}
  &A_0(0) =  \int \frac{d^d k}{(2\pi)^d}  \frac{1}{(k^2+i\epsilon)} \, .
\end{align}
As can be seen in App.~\ref{Ap:DimensionalRegularization}, this integral
vanishes for all $d$ since it is proportional to $(M^2)^{d/2-2}$ where in this case $M^2=0$. Thus, the whole integral is zero. The general form of this result can be written as
\begin{equation}
  \label{eq:Tadpole}
  \int \frac{d^{d}k}{(k^2)^{\alpha}} = 0 \qquad\qquad \text{for} \quad \forall\, \alpha, \,  d\, \in \,  \mathbb{C}  \, ,
\end{equation}
which is known as Veltman’s formula \cite{MasslessTadpole}.

\subsubsection{Tensor Integrals}
The general form of a tensor one-loop integral of rank-$M$ for a N-point function with external momenta
$p_1,\cdots,p_{N-1}$ (with $p_N$ from momentum conservation) as shown in Fig.~\ref{fig:GeneralLoop} is given by
\begin{align}
  \label{eq:GeneralLoopIntegral}
  I^N_{\mu_1, \cdots, \mu_M}(& p_1,\cdots,p_{N-1},m_0,\cdots,m_{N-1}) \sim \\
  &  \int \frac{d^d k}{(2\pi)^d} \frac{k_{\mu_1} \cdots k_{\mu_M}}{(k^2-m_0^2+i\epsilon)((k+q_1)^2-m_1^2+i\epsilon) \cdots ((k+q_{N-1})^2-m_{N-1}^2+i\epsilon)} \, , \nonumber
\end{align}
where $\mu_1,\cdots,\mu_M $ are indices that represent the rank of the integral,
and the rest is the same as in Eq.~(\ref{eq:GeneralScalarLoopIntegral}).

In our case of scalar-graviton scattering to one-loop, the maximum number of indices that can appear in the
numerator for $A$ and $D$ integrals is four, for $B$ is five, and $C$ is six. To calculate these tensor integrals, we will follow the Passarino-Veltman method, which will be discussed in the next section.

\subsubsection{Passarino-Veltman Reduction \label{se:PassarinoVeltman}}
The idea of the Passarino-Veltman method \cite{PassarinoVeltman} is to write the tensor integrals in terms
of scalar integrals, Eq.~(\ref{eq:BZero}), with the help of Passarino-Veltman reduction formula which
takes the general form
\begin{equation}
  \label{eq:GeneralReduction}
  k \cdot p_i = \frac{1}{2} [((k+q_i)^2 - m_i^2) - ((k+q_{i-1})^2 - m_{i-1}^2) + m_i^2 - m_{i-1}^2 - q_i^2 + q_{i-1}^2 ] \, .
\end{equation}

To illustrate this idea, let us take an example of a rank-one tensor integral, a
vector integral, for a two-point function which is given by
\begin{align}
  \label{eq:RankOneLoopIntegral}
  I^2_{\mu}(p_1,m_0,m_1)= B_{\mu}(p_1,m_0,m_1) =  \int \frac{d^d k}{(2\pi)^d} \frac{k_{\mu}}{(k^2-m_0^2)((k+q_1)^2-m_1^2)} \, ,
\end{align}
where $q_1=p_1$. Since $p_1$ is the only four vector in this integral which
can carry the index $\mu$ in the result, it follows that this integral can be written
as $p_{1\mu}$ multiplied by a scalar function $B_1(p_1,m_0,m_1)$
\begin{align}
 \int \frac{d^d k}{(2\pi)^d} \frac{k_{\mu}}{(k^2-m_0^2)((k+p_1)^2-m_1^2)} = p_{1\mu} B_1(p_1,m_0,m_1) \, .
  \label{eq:VectorIntegral}
\end{align}
Multiplying by $p_1^{\mu}$ from both sides gives
\begin{align}
  \int \frac{d^d k}{(2\pi)^d} \frac{k \cdot p_1}{(k^2-m_0^2)((k+p_1)^2-m_1^2)} = p_1 \cdot p_{1} \;B_1(p_1,m_0,m_1) \, . \label{eq:ConvertedScalar}
\end{align}
Now we can use the Passarino-Veltman reduction formula Eq.~(\ref{eq:GeneralReduction}) which takes the following form in this example,
\begin{align}
  \label{eq:ReductionRankOne}
  k \cdot p_1  & = \frac{1}{2} [( (k+p_1)^2 - m_1^2) - (k^2 - m_0^2)  + m^2_1 - m^2_0 - p_1^2 ] \, .
\end{align}
Inserting Eq.~(\ref{eq:ReductionRankOne}) into Eq.~(\ref{eq:ConvertedScalar}) gives
\begin{align}
  p_1^2 \, B_1(p_1,m_0,m_1) =  \; \frac{1}{2} \Bigg [ &  \int \frac{d^d k}{(2\pi)^d} \frac{1}{(k^2-m_0^2)} - \int \frac{d^d k}{(2\pi)^d} \frac{1}{((k+p_1)^2-m_1^2)}  \nonumber\\[2mm]
         & + (m_1^2 -m_0^2-p_1^2)  \int \frac{d^d k}{(2\pi)^d}  \frac{1}{(k^2-m_0^2)((k+p_1)^2-m_1^2)} \Bigg ] \nonumber \\[2mm]
  =  \; \frac{1}{2} \Big [ & A_0(m_0) - A_0(m_1) - (- m_1^2 + m_0^2 + p_1^2) B_0(p_1,m_0,m_1) \Big ] \, . \nonumber
\end{align}
From this it follows that
\begin{align}
  B_{\mu}(p_1,m_0,m_1) & =  p_{1\mu} B_1(p_1,m_0,m_1) \label{eq:FinalPassarino} \\[2mm]
 & =  \, \frac{p_{1\mu}}{2 p_1^2} \Big [  A_0(m_0) - A_0(m_1) - (- m_1^2 + m_0^2 + p_1^2) B_0(p_1,m_0,m_1) \Big ] \, . \nonumber
\end{align}
As a result, the tensor integral $B_{\mu}(p_1,m_0,m_1)$ can be written in terms of the
scalar integrals $A_0(m_0)$, $A_0(m_1)$ and $B_0(p_1,m_0,m_1)$. This procedure
can be generalized for any tensor integral, and for completeness we give all
relevant tensor integrals in App.~\ref{Ap:PassarinoVeltman}.


\subsection{Scalar-Graviton Scattering to One-Loop Order}
\begin{figure}[H]
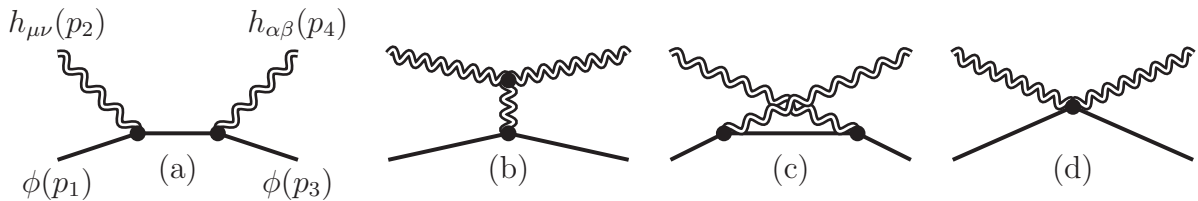

	\centering
  \noindent
  \begin{axopicture}(90,50)
    \SetWidth{1.5}
    \Line(0,10)(30,20)
    \Line(30,20)(60,20)
    \Line(60,20)(90,10)
    \SetWidth{1.}
    \DoublePhoton(0,50)(30,20){2}{4.5}{2}
    \DoublePhoton(60,20)(90,50){2}{4.5}{2}
    \Vertex(30,20){3}
    \Vertex(60,20){3}
    \Text(45,0)[b]{(a)}
    \Text(0,-5)[b]{$\phi(p_1)$}
    \Text(90,-5)[b]{$\phi(p_3)$}
    \Text(0,55)[b]{$h_{\mu\nu}(p_2)$}
    \Text(90,55)[b]{$h_{\alpha\beta}(p_4)$}
  \end{axopicture}\hspace{5mm}
  ~~~
  \begin{axopicture}(90,50)
    \SetWidth{1.5}
    \Line(0,10)(45,20)
    \Line(45,20)(90,10)
    \SetWidth{1.}
    \DoublePhoton(0,50)(45,40){2}{7.5}{2}
    \DoublePhoton(45,40)(90,50){2}{7.5}{2}
    \DoublePhoton(45,20)(45,40){2}{3.5}{2}
    \Vertex(45,20){3}
    \Vertex(45,40){3}
    \Text(45,0)[b]{(b)}
  \end{axopicture}
  ~~~
  \begin{axopicture}(90,50)
    \SetWidth{1.5}
    \Line(0,10)(20,20)
    \Line(20,20)(70,20)
    \Line(70,20)(90,10)
    \SetWidth{1.}
    \DoublePhoton(0,50)(70,20){2}{8.5}{2}
    \DoublePhoton(20,20)(90,50){2}{8.5}{2}
    \Vertex(20,20){3}
    \Vertex(70,20){3}
    \Text(45,0)[b]{(c)}
  \end{axopicture}
  ~~~
  \begin{axopicture}(90,50)
    \SetWidth{1.5}
    \Line(0,10)(45,30)
    \Line(45,30)(90,10)
    \SetWidth{1.}
    \DoublePhoton(0,50)(45,30){2}{7.5}{2}
    \DoublePhoton(45,30)(90,50){2}{7.5}{2}
    \Vertex(45,30){3}
    \Text(45,0)[b]{(d)}
  \end{axopicture}
  \caption{scalar-graviton scattering at tree level.}
	\label{fig:TreeLevelScalarGraviton2}
\end{figure}

In order to study scalar-graviton scattering to one-loop order, we first
need to draw all the diagrams that can contribute to this process up to
one-loop. We start with the tree level diagrams as shown in
Fig.~\ref{fig:TreeLevelScalarGraviton2}. Then, we insert all possible one-loop
corrections, as shown in the next sections, into the tree level diagrams to get
all possible one-loop diagrams for scalar-graviton scattering. More
specifically, in this case we need to insert the propagator corrections, the triple graviton
vertex corrections, the scalar-scalar-graviton vertex corrections and the
scalar-scalar-graviton-graviton vertex corrections.
\vspace{5mm}

\begin{figure}[!h]
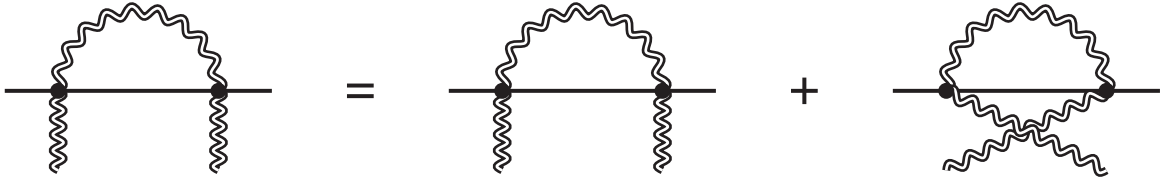

  \vspace{4mm}
  \centering
    \begin{axopicture}(100,90)(0,-40)
    \SetWidth{1.5}
    \Line(0,10)(100,10)
    \SetWidth{1.}
    \DoublePhotonArc(50,10)(30,0,180){2}{10}{2}
    \DoublePhoton(20,-20)(20,10){2}{5}{2}
    \DoublePhoton(80,-20)(80,10){2}{5}{2}
    \Vertex(20,10){3}
    \Vertex(80,10){3}
    \Text(50,-40)[b]{}
  \end{axopicture} \hspace{3mm}
  ~~~
  \begin{axopicture}(10,90)
    \SetWidth{1.5}
    \Line(0,52)(10,52)
    \Line(0,48)(10,48)
  \end{axopicture}\hspace{3mm}
  ~~~
  \begin{axopicture}(100,90)(0,-40)
    \SetWidth{1.5}
    \Line(0,10)(100,10)
    \SetWidth{1.}
    \DoublePhotonArc(50,10)(30,0,180){2}{10}{2}
    \DoublePhoton(20,-20)(20,10){2}{5}{2}
    \DoublePhoton(80,-20)(80,10){2}{5}{2}
    \Vertex(20,10){3}
    \Vertex(80,10){3}
    \Text(50,-40)[b]{}
  \end{axopicture}\hspace{3mm}
  ~~~
  \begin{axopicture}(10,90)
    \SetWidth{1.5}
    \Line(0,50)(10,50)
    \Line(5,55)(5,45)
  \end{axopicture}\hspace{3mm}
  ~~~
  \begin{axopicture}(100,90)(0,-40)
    \SetWidth{1.5}
    \Line(0,10)(100,10)
    \SetWidth{1.}
    \DoublePhotonArc(50,10)(30,0,180){2}{10}{2}
    \DoublePhoton(20,-20)(80,10){2}{8}{2}
    \DoublePhoton(80,-20)(20,10){2}{8}{2}
    \Vertex(20,10){3}
    \Vertex(80,10){3}
    \Text(50,-40)[b]{}
  \end{axopicture}
  \caption{Permutation of external graviton legs.}  \label{fig:Permutation}
\end{figure}

Moreover, we need to take
into account that there are permutations of external graviton legs whenever it
is possible. For example, the loop corrections to the scalar-scalar-graviton-graviton vertex has a permutation in external graviton legs as shown in Fig.~\ref{fig:Permutation}. In addition, when dealing with loops, it is important to remember the minus sign that arises from fermion loops.

The last point to describe before starting the calculations is how the graviton
propagator behaves in dimensional regularization at loop level. As starting point, we recall
the Lagrangian for the graviton propagator, Eq.~(\ref{eq:OurGravitonPropagator}), which can be written as
\begin{align}
  \mathcal{L}_{hh} =  \frac{1}{2} \partial_{\lambda} h_{\mu\nu}
  \partial^{\lambda} h^{\mu\nu} - \frac{1}{4} \partial_{\lambda}
  h_{\nu}^{\;\;\nu} \partial^{\lambda} h_{\mu}^{\;\;\mu} = \frac{1}{2} h_{\mu\nu} \partial^{\lambda} \partial_{\lambda} \Big (  I^{\mu\nu\alpha\beta} - \frac{1}{2} \eta^{\mu\nu} \eta^{\alpha\beta}     \Big ) h_{\alpha\beta} \, ,
\end{align}
where $\displaystyle  I^{\mu\nu\alpha\beta}=\frac{1}{2} (\eta^{\mu\alpha}\eta^{\nu\beta} + \eta^{\mu\beta} \eta^{\nu\alpha})$ is the identity tensor.

\noindent Then, we solve the Green's function equation in $d$ dimensions
\begin{align}
   \Big (  I^{\mu\nu\alpha\beta} - \frac{1}{2} \eta^{\mu\nu} \eta^{\alpha\beta}     \Big ) \partial^{\lambda} \partial_{\lambda} S_{\alpha\beta\gamma\delta} (x-y) = - I^{\mu\nu}_{\;\;\;\;\gamma\delta} \; \delta^{(d)} (x-y)  \, , \label{eq:GreenF}
\end{align}
to find the propagator. To solve this equation, we perform a Fourier transform which gives
\begin{align}
   \Big (  I^{\mu\nu\alpha\beta} - \frac{1}{2} \eta^{\mu\nu} \eta^{\alpha\beta}     \Big ) (-k^2)  S_{\alpha\beta\gamma\delta} (k) = - I^{\mu\nu}_{\;\;\;\;\gamma\delta}   \, . \label{eq:GreenFM}
\end{align}
To solve this equation, we use the following ansatz
\begin{align}
  S_{\alpha\beta\gamma\delta} (k) =  \frac{1}{k^2} \Big (a I_{\alpha\beta\gamma\delta} + b \eta_{\alpha\beta} \eta_{\gamma\delta} \Big) \, ,
\end{align}
which gives
\begin{align}
  a=1\, ,    \qquad\qquad\qquad\qquad        b=-\frac{1}{d-2} \, ,
\end{align}
where $\displaystyle \eta_{\mu\nu}\eta^{\mu\nu}= \delta_{\mu}^{\mu} = d = 4-2\epsilon $. 

\noindent Thus, the graviton propagator in momentum space in $d$ dimensions is
\begin{align}
S_{\mu\nu\alpha\beta} (k)  = \frac{i}{k^2} \Big (  \frac{1}{2}I_{\mu\nu\alpha\beta}-\frac{1}{d-2}\eta_{\mu\nu}
  \eta_{\alpha\beta} \Big ) = \frac{i}{k^2} P_{\mu\nu\alpha\beta} \, ,
\end{align}
where $\displaystyle P_{\mu\nu\alpha\beta}=\frac{1}{2}I_{\mu\nu\alpha\beta}-\frac{1}{d-2}\eta_{\mu\nu}
\eta_{\alpha\beta}$. 


\subsubsection{Self-Energy Corrections}
\begin{figure}[H]
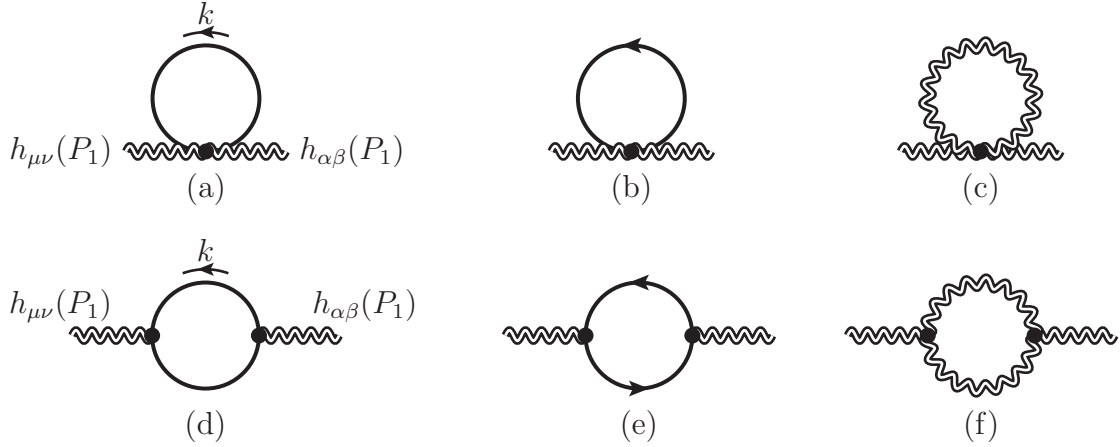

  \vspace{7mm}
  \centering
  \begin{axopicture}(60,60)(0,-10)
    \SetWidth{1.5}
    \Arc(30,30)(20,0,360)
    \SetWidth{1.}
    \Arc[arrow](30,30)(25,70,110)
    \DoublePhoton(0,10)(60,10){2}{10}{2}
    \Vertex(30,10){3}
    \Text(30,-10)[b]{(a)}
    \Text(30,58)[b]{$k$}
    \Text(-5,10)[r]{$ h_{\mu\nu} (P_1) $}
    \Text(65,10)[l]{$h_{\alpha\beta}(P_1)$}
  \end{axopicture} \hspace{28mm}
  ~~~
  \begin{axopicture}(60,60)(0,-10)
    \SetWidth{1.5}
    \Arc[arrow,arrowpos=0.25](30,30)(20,0,360)
    \SetWidth{1.}
    \DoublePhoton(0,10)(60,10){2}{10}{2}
    \Vertex(30,10){3}
    \Text(30,-10)[b]{(b)}
  \end{axopicture} \hspace{18mm}
  ~~~
  \begin{axopicture}(60,60)(0,-10)
    \SetWidth{1.}
    \DoublePhoton(0,10)(60,10){2}{10}{2}
    \DoublePhotonArc(30,30)(20,0,360){2}{18}{2}
    \Vertex(30,10){3}
    \Text(30,-10)[b]{(c)}
  \end{axopicture}\vspace{10mm}
  \\ 
  \begin{axopicture}(100,60)(0,-10)
    \SetWidth{1.5}
    \Arc(50,30)(20,0,360)
    \SetWidth{1.}
    \DoublePhoton(0,30)(30,30){2}{5}{2}
    \DoublePhoton(70,30)(100,30){2}{5}{2}
    \Arc[arrow](50,30)(25,70,110)
    \Vertex(30,30){3}
    \Vertex(70,30){3}
    \Text(50,-10)[b]{(d)}
    \Text(50,58)[b]{$k$}
    \Text(15,35)[rb]{$ h_{\mu\nu} (P_1) $}
    \Text(90,35)[lb]{$h_{\alpha\beta}(P_1)$}
  \end{axopicture}\hspace{15mm}
  ~~~
  \begin{axopicture}(100,60)(0,-10)
    \SetWidth{1.5}
    \Arc[arrow](50,30)(20,0,180)
    \Arc[arrow](50,30)(20,180,360)
    \SetWidth{1.}
    \DoublePhoton(0,30)(30,30){2}{5}{2}
    \DoublePhoton(70,30)(100,30){2}{5}{2}
    \Vertex(30,30){3}
    \Vertex(70,30){3}
    \Text(50,-10)[b]{(e)}
  \end{axopicture} \hspace{3mm}
  ~~~
  \begin{axopicture}(100,60)(0,-10)
    \SetWidth{1.}
    \DoublePhoton(0,30)(30,30){2}{5}{2}
    \DoublePhoton(70,30)(100,30){2}{5}{2}
    \DoublePhotonArc(50,30)(20,0,360){2}{18}{2}
    \Vertex(30,30){3}
    \Vertex(70,30){3}
    \Text(50,-10)[b]{(f)}
  \end{axopicture}\vspace{8mm}
  \caption{The graviton propagator at one-loop level with: one (a-c), two (d-f)
    propagators.}
	\label{fig:GravitonSelfEnergy}
\end{figure}

There are two types of propagators, scalar and graviton, in the tree level diagrams shown in
Fig.~\ref{fig:TreeLevelScalarGraviton2}, and each one has different loop
corrections. So, let us start with the graviton propagator which has six
one-loop diagrams as shown in Fig.~\ref{fig:GravitonSelfEnergy}.

\noindent In terms of the Feynman rules, the diagrams with one propagator in Fig.~\ref{fig:GravitonSelfEnergy}a-c can be written as follows:
\begin{alignat}{4}
  \mathcal{M}^{\text{(a)}}_{\mu\nu\alpha\beta} = & \frac{1}{2} \int \frac{d^4k}{(2\pi)^4}
  V_{\mu\nu\alpha\beta}^{\{\phi\phi h h\}} (k) \;\; && S_{\{\phi\phi\}}(k,m) \, ,
  && && \\[5mm]
  \mathcal{M}^{\text{(b)}}_{\mu\nu\alpha\beta} = &  \int \frac{d^4k}{(2\pi)^4}
  V_{\mu\nu\alpha\beta\rho\sigma}^{\{\bar{\chi}\chi h h\}} (P_1,k) \;\; && S_{\{\bar{\chi}\chi\}}^{\rho\sigma}(k)
  &&= 0 && \, ,  \\[5mm]
  \mathcal{M}^{\text{(c)}}_{\mu\nu\alpha\beta} = & \frac{1}{2} \int \frac{d^4k}{(2\pi)^4}
  V_{\mu\nu\alpha\beta\rho\sigma\gamma\delta}^{\{h h h h\}} (P_1,k) \;\; && S_{\{h h\}}^{\rho\sigma\gamma\delta}(k)
  &&= 0 && \, , \\ \nonumber
\end{alignat}
where $\frac{1}{2}$ is the symmetry factor. $\mathcal{M}^{\text{(b)}}_{\mu\nu\alpha\beta}$ and $
\mathcal{M}^{\text{(c)}}_{\mu\nu\alpha\beta}$  vanish according to the relation
Eq.~(\ref{eq:Tadpole}), which means that all massless tad-pole diagrams vanish.
Using the simplified rules, we get
\begin{align}
  \mathcal{M}^{\text{(a)}}_{\mu\nu\alpha\beta} = \frac{\kappa^2}{8} \int \frac{d^dk}{(2\pi)^d} \frac{1}{k^2-m^2} \big [    \eta_{\mu\nu} k_{\alpha} k_{\beta} - \eta_{\mu\alpha} k_{\nu} k_{\beta} -    \eta_{\mu\beta} k_{\nu} k_{\alpha} -
  \eta_{\nu\alpha} k_{\mu} k_{\beta} -    \eta_{\nu\beta} k_{\mu} k_{\alpha} +    \eta_{\alpha\beta} k_{\mu} k_{\nu} \big ] \, , \nonumber
\end{align}
which can be written in terms of scalar integrals as
\begin{align}
  \mathcal{M}^{\text{(a)}}_{\mu\nu\alpha\beta} = \kappa^2 \; \frac{m^2}{4 d} \; A_0(m) \; \big [ \eta_{\mu\nu}\eta_{\alpha\beta} -  \eta_{\mu\alpha}\eta_{\nu\beta} - \eta_{\mu\beta}\eta_{\alpha\nu}  \big ] \, .
\end{align}

On the other hand, if we use the standard Feynman rules to calculate the
amplitude $\mathcal{M}^{\text{(a)}}_{\mu\nu\alpha\beta}$, then the calculation is slightly more complicated because the scalar-scalar-graviton-graviton vertex $V_{\mu\nu\alpha\beta}^{\{\phi\phi h h\}}$ has six terms in
the standard rules Eq.~(\ref{eq:2phi1Hb}), compared to two terms in the simplified rules
Eq.~(\ref{eq:OurScalarScalarGravitonGravitonVertex}).

\noindent Next we consider the diagrams with two propagators in the loop as in
Fig.~\ref{fig:GravitonSelfEnergy}d-f, which can be written as:
\begin{alignat}{4}
  \mathcal{M}^{\text{(d)}}_{\mu\nu\alpha\beta} =  & \frac{1}{2} \int \frac{d^4k}{(2\pi)^4} V_{\mu\nu}^{\{\phi\phi  h\}}
  (k,Q_1,m) \;\; && S_{\{\phi\phi\}}(k,m) \;\; && S_{\{\phi\phi\}}(Q_1,m)
  \;\;  && V_{\alpha\beta}^{\{\phi\phi  h\}}(k,Q_1,m) \, , \nonumber\\[4mm]
  \mathcal{M}^{\text{(e)}}_{\mu\nu\alpha\beta} =  & \int \frac{d^4k}{(2\pi)^4}
  V_{\mu\nu\rho\gamma}^{ \{\bar{\chi}\chi  h\} } (k,Q_1) \;\; && S_{\{ \bar{\chi}\chi \}}^{\rho\sigma}(k) \;\; && S_{\{ \bar{\chi}\chi \}}^{\gamma\delta}(Q_1)
  \;\;  && V_{\sigma\delta\alpha\beta}^{\{\bar{\chi}\chi  h\}}(k,Q_1) \, , \nonumber\\[4mm]
  \mathcal{M}^{\text{(f)}}_{\mu\nu\alpha\beta} =  & \frac{1}{2} \int \frac{d^4k}{(2\pi)^4}
  V_{\mu\nu\rho\sigma\eta\lambda}^{\{h h  h\}}
  (k,Q_1) \;\; && S_{\{h h\}}^{\rho\sigma\gamma\delta}(k) \;\; && S_{\{h h\}}^{\eta\lambda\epsilon\kappa}(Q_1)
  \;\;  && V_{\gamma\delta\epsilon\kappa\alpha\beta}^{\{h h  h\}}(k,Q_1) \, , \nonumber
\end{alignat}
where $\frac{1}{2}$ is the symmetry factor and $ Q_1=k+P_1$.

Plugging in the Feynman rules will give lengthy expressions. Therefore, we only
show $\mathcal{M}^{\text{(f)}}_{\mu\nu\alpha\beta}$, which has two triple
graviton vertices, while the other amplitudes
$\mathcal{M}^{\text{(d)}}_{\mu\nu\alpha\beta},\mathcal{M}^{\text{(e)}}_{\mu\nu\alpha\beta}$
are approximately the same in the standard and simplified rules. Using the
simplified rules and doing the Passarino-Veltman reduction, the amplitude
$\mathcal{M}^{\text{(f)}}_{\mu\nu\alpha\beta}$ can be written in terms of
the scalar integral $B_0(0,0,P_1)$ as
\begin{align*}
  \mathcal{M}^{\text{(f)}}_{\mu\nu\alpha\beta} = &  
        \frac{\kappa^2 \; B_0(0,0,P_1)}{64 \; d^4 - 256 \; d^3 + 192 \; d^2 + 256 \; d - 256} \Big  [  \\
& \quad        + P_{1\mu} P_{1\nu} P_{1\alpha} P_{1\beta}  \big (\; d^6 - 2 \; d^4 - 116 \; d^3 + 
         312 \; d^2 + 144 \; d - 256 \big )  \\
& \quad         + \eta_{\mu\nu} \eta_{\alpha\beta} P_1^3  \big (\; d^6 - 5 \; d^5 + 31 \; d^3 + 6
          \; d^2 - 36 \; d - 8 \big ) \\
&  \quad  + P_1^2 \big ( \eta_{\mu\nu} P_{1\alpha} P_{1\beta}  +\eta_{\alpha\beta} P_{1\mu} P_{1\nu}  \big )  \big ( 64 - \; d^6 + 3 \; d^5 + 13 \; d^4 - 34 \; d^3 - 76 \; d^2 + 40 \; d  \big )  \Big ] \\[2mm]
& + \frac{\kappa^2 \; B_0(0,0,P_1)}{64 \; d^3 - 128 \; d^2 - 64 \; d + 128}    \Big  [  \\
& \quad   + P_1^3 \big ( \eta_{\mu\alpha} \eta_{\nu\beta} + \eta_{\mu\beta} \eta_{\nu\alpha} \big )  \big (3 \; d^3 - 24 \; d^2 - 8 \; d + 16 \big ) \\
& \quad  +  P_1^2 \big ( \eta_{\mu\alpha} P_{1\nu} P_{1\beta} + \eta_{\mu\beta} P_{1\nu} P_{1\alpha} + \eta_{\nu\alpha} P_{1\mu} P_{1\beta} + \eta_{\nu\beta} P_{1\mu} P_{1\alpha}   \big )  \big ( 32 \; d^2 - 7 \; d^3 + 20 \; d - 16 \big )  \Big ] \, .
\end{align*}

Now, if we instead use the standard Feynman rules to calculate the amplitude
$\mathcal{M}^{\text{(f)}}_{\mu\nu\alpha\beta}$, then the calculation is more complicated because the triple graviton vertex $V_{\mu\nu\rho\sigma\eta\lambda}^{\{h h
  h\}}$ has 40 terms in the standard rules Eq.~(\ref{eq:3Hb}), compared to only four
terms in the simplified rules Eq.~(\ref{eq:OurTripleGravitonVertex}). In addition, this amplitude has
two triple graviton vertices, so the number of Lagrangian terms involved from
the vertices in the standard rules is 1600, compared to 16 terms in the
simplified rules. Since these calculations are very lengthy, we only show the comparison between the running time in the FORM program, after doing Passarino-Veltman reduction and writing the amplitude
$\mathcal{M}^{\text{(f)}}_{\mu\nu\alpha\beta}$ in terms of scalar integrals:\vspace{1mm}
\begin{Verbatim}[gobble=2,frame=single,framesep=2mm,label=simplified way,labelposition=all,numbers=left]
  WTime =       0.58 sec   Generated terms =         10
              HH2a         Terms in output =         10
                           Bytes used      =       4024
\end{Verbatim}
\vspace{1mm}
\begin{Verbatim}[gobble=2,frame=single,framesep=2mm,label=standard way,labelposition=all,numbers=left]
  WTime =      26.38 sec   Generated terms =         10
              HH2b         Terms in output =         10
                           Bytes used      =       3960
\end{Verbatim}
As shown above, there is a large difference in the running time. In addition,
the running time will increase considerably for more complicated diagrams as we will
show in the next sections.

\begin{figure}[H]
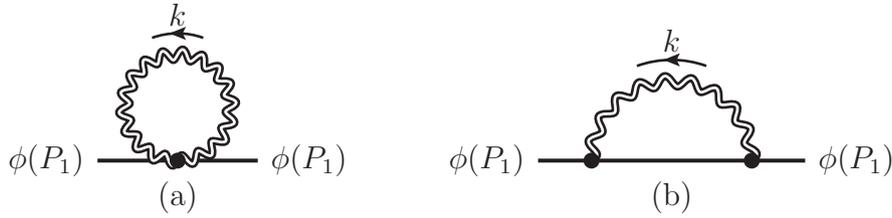

  \vspace{7mm}
  \centering
  \begin{axopicture}(60,60)(0,-10)
    \SetWidth{1.5}
    \Line(0,10)(60,10)
    \SetWidth{1.}
    \DoublePhotonArc(30,30)(20,0,360){2}{18}{2}
    \Arc[arrow](30,30)(28,70,110)
    \Text(30,61)[b]{$k$}
    \Text(-5,10)[r]{$\phi(P_1)$}
    \Text(65,10)[l]{$\phi(P_1)$}
    \Vertex(30,10){3}
    \Text(30,-10)[b]{(a)}
  \end{axopicture} \hspace{30mm}
  ~~~
  \begin{axopicture}(100,50)(0,-10)
    \SetWidth{1.5}
    \Line(0,10)(100,10)
    \SetWidth{1.}
    \DoublePhotonArc(50,10)(30,0,180){2}{10}{2}
    \Arc[arrow](50,10)(38,70,110)
    \Text(50,51)[b]{$k$}
    \Text(-5,10)[r]{$\phi(P_1)$}
    \Text(105,10)[l]{$\phi(P_1)$}
    \Vertex(20,10){3}
    \Vertex(80,10){3}
    \Text(50,-10)[b]{(b)}
  \end{axopicture}  
  \caption{The scalar propagator at one-loop level with: one (a), two (b) propagators.}
	\label{fig:ScalarSelfEnergy}
\end{figure}

Finally, we consider the scalar propagator which has just two diagrams that can
contribute as shown in Fig.~\ref{fig:ScalarSelfEnergy}. Again, the
massless tad-pole in Fig.~\ref{fig:ScalarSelfEnergy}a vanishes. So, the diagrams can be written as:
\begin{alignat}{4}
  \mathcal{M}^{\text{(a)}} = & \frac{1}{2} \int \frac{d^4k}{(2\pi)^4}
  V_{\mu\nu\alpha\beta}^{\{\phi\phi h h\}} (P_1) \;\; && S_{\{h h\}}^{\mu\nu\alpha\beta}(k)
  && = 0\, , &&  \\[3mm]
  \mathcal{M}^{\text{(b)}} =  & \int \frac{d^4k}{(2\pi)^4}
  V_{\mu\nu}^{\{\phi\phi  h\}} (P_1,Q_1,m) \;\; && S_{\{h
    h\}}^{\mu\nu\alpha\beta}(k) \;\; && S_{\{ \phi\phi \}}(Q_1) \;\;  &&
  V_{\alpha\beta}^{\{\phi\phi h\}}(P_1,Q_1,m) \, , \nonumber
\end{alignat}
where $\frac{1}{2}$ is the symmetry factor and $ Q_1=k+P_1 $.\\
Plugging the Feynman rules in $\mathcal{M}^{\text{(b)}}$ gives
\begin{align}
  \mathcal{M}^{\text{(b)}} =& \;\;\;\; \kappa^2\; \big [P_1^2+m^2 \big] \int \frac{d^dk}{(2\pi)^d} \frac{k \cdot P_1}{k^2((k+P_1)^2-m^2)}
  \nonumber  \\[3mm] &  + \kappa^2\; \frac{P_1^2}{2} \int \frac{d^dk}{(2\pi)^d} \frac{k^2}{k^2((k+P_1)^2-m^2)} 
\\[3mm] &  + \kappa^2 \; \Bigg [ \frac{-2 P_1^4 -4 m^2 P_1^2 + d \; P_1^4 + 2 d\; m^2 P_1^2 - d \; m^4}{2(d-2)} \Bigg ] \int \frac{d^dk}{(2\pi)^d} \frac{1}{k^2((k+P_1)^2-m^2)}  \, , \nonumber
\end{align}
which can be written in terms of scalar integrals as
\begin{align}
  \mathcal{M}^{\text{(b)}} = - \kappa^2 \; \frac{m^2}{2} \; A_0(m) + \kappa^2 \; \frac{m^2}{d-2} \; \big [ d\; P_1^2 -2 P_1^2 - m^2   \big ] B_0(0,m,P_1) \, .
\end{align}
\noindent Both the simplified and the standard rules give the same results, since
the scalar-scalar-graviton vertex $V_{\mu\nu}^{\{\phi\phi  h\}}$ and the
propagators are the same in both.

\subsubsection{Triple Graviton Vertex Corrections}
\begin{figure}[H]
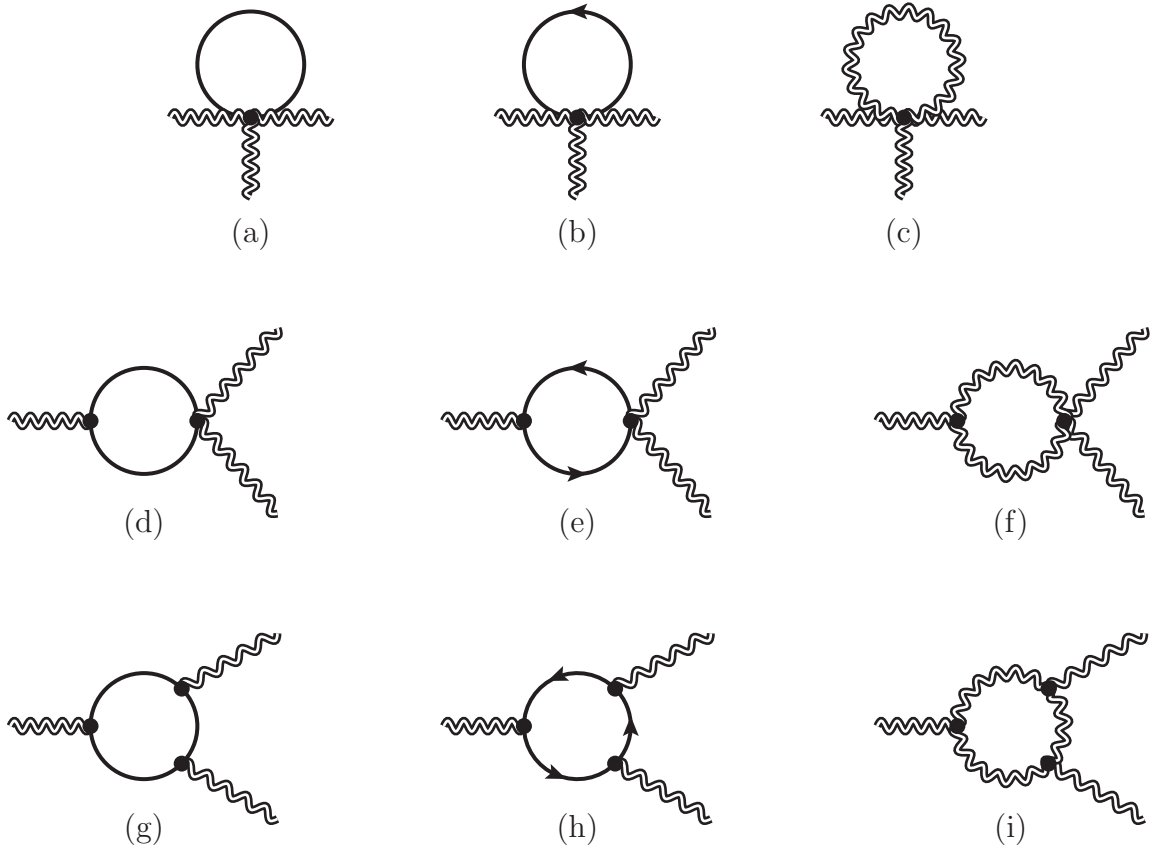

  \vspace{4mm}
	\centering\noindent
  \begin{axopicture}(60,90)(0,-40)
    \SetWidth{1.5}
    \Arc(30,30)(20,0,360)
    \SetWidth{1.}
    \DoublePhoton(0,10)(60,10){2}{10}{2}
    \DoublePhoton(30,10)(30,-20){2}{5}{2}
    \Vertex(30,10){3}
    \Text(30,-40)[b]{(a)}
  \end{axopicture} \hspace{15mm}
  ~~~
  \begin{axopicture}(60,90)(0,-40)
    \SetWidth{1.5}
    \Arc[arrow,arrowpos=0.25](30,30)(20,0,360)
    \SetWidth{1.}
    \DoublePhoton(0,10)(60,10){2}{10}{2}
    \DoublePhoton(30,10)(30,-20){2}{5}{2}
    \Vertex(30,10){3}
    \Text(30,-40)[b]{(b)}
  \end{axopicture} \hspace{15mm}
  ~~~
  \begin{axopicture}(60,90)(0,-40)
    \SetWidth{1.}
    \DoublePhoton(0,10)(60,10){2}{10}{2}
    \DoublePhotonArc(30,30)(20,0,360){2}{18}{2}
    \DoublePhoton(30,10)(30,-20){2}{5}{2}
    \Vertex(30,10){3}
    \Text(30,-40)[b]{(c)}
  \end{axopicture}
  \\[10mm]
  \begin{axopicture}(100,80)(0,-15)
    \SetWidth{1.5}
    \Arc(50,30)(20,0,360)
    \SetWidth{1.}
    \DoublePhoton(0,30)(30,30){2}{5}{2}
    \DoublePhoton(70,30)(100,65){2}{6}{2}
    \DoublePhoton(70,30)(100,-5){2}{6}{2}
    \Vertex(30,30){3}
    \Vertex(70,30){3}
    \Text(50,-15)[b]{(d)}
  \end{axopicture} \hspace{15mm}
  ~~~
  \begin{axopicture}(100,80)(0,-15)
    \SetWidth{1.5}
    \Arc[arrow](50,30)(20,0,180)
    \Arc[arrow](50,30)(20,180,360)
    \SetWidth{1.}
    \DoublePhoton(0,30)(30,30){2}{5}{2}
    \DoublePhoton(70,30)(100,65){2}{6}{2}
    \DoublePhoton(70,30)(100,-5){2}{6}{2}
    \Vertex(30,30){3}
    \Vertex(70,30){3}
    \Text(50,-15)[b]{(e)}
  \end{axopicture} \hspace{15mm}
  ~~~
  \begin{axopicture}(100,80)(0,-15)
    \SetWidth{1.}
    \DoublePhoton(0,30)(30,30){2}{5}{2}
    \DoublePhoton(70,30)(100,65){2}{6}{2}
    \DoublePhoton(70,30)(100,-5){2}{6}{2}
    \DoublePhotonArc(50,30)(20,0,360){2}{18}{2}
    \Vertex(30,30){3}
    \Vertex(70,30){3}
    \Text(50,-15)[b]{(f)}
  \end{axopicture}
  \\[12mm]
  \begin{axopicture}(100,80)(0,-15)
    \SetWidth{1.5}
    \Arc(50,30)(20,0,360)
    \SetWidth{1.}
    \DoublePhoton(0,30)(30,30){2}{5}{2}
    \DoublePhoton(64.1421,44.1421)(100,65){2}{5}{2}
    \DoublePhoton(64.1421,15.8579)(100,-5){2}{5}{2}
    \Vertex(30,30){3}
    \Vertex(64.1421,44.1421){3}
    \Vertex(64.1421,15.8579){3}
    \Text(50,-15)[b]{(g)}
  \end{axopicture} \hspace{15mm}
  ~~~
  \begin{axopicture}(100,80)(0,-15)
    \SetWidth{1.5}
    \Arc[arrow](50,30)(20,45,180)
    \Arc[arrow](50,30)(20,180,315)
    \Arc[arrow](50,30)(20,315,45)
    \SetWidth{1.}
    \DoublePhoton(0,30)(30,30){2}{5}{2}
    \DoublePhoton(64.1421,44.1421)(100,65){2}{5}{2}
    \DoublePhoton(64.1421,15.8579)(100,-5){2}{5}{2}
    \Vertex(30,30){3}
    \Vertex(64.1421,44.1421){3}
    \Vertex(64.1421,15.8579){3}
    \Text(50,-15)[b]{(h)}
  \end{axopicture} \hspace{15mm}
  ~~~
  \begin{axopicture}(100,80)(0,-15)
    \SetWidth{1.}
    \DoublePhoton(0,30)(30,30){2}{5}{2}
    \DoublePhoton(64.1421,44.1421)(100,65){2}{5}{2}
    \DoublePhoton(64.1421,15.8579)(100,-5){2}{5}{2}
    \DoublePhotonArc(50,30)(20,0,360){2}{18}{2}
    \Vertex(30,30){3}
    \Vertex(64.1421,44.1421){3}
    \Vertex(64.1421,15.8579){3}
    \Text(50,-15)[b]{(i)}
  \end{axopicture}
  \caption{Triple graviton vertex at one-loop level with: one (a-c), two (d-e), three (g-i) propagators.}
	\label{fig:TripleGravitonVertex}
\end{figure}

\begin{figure}[H]
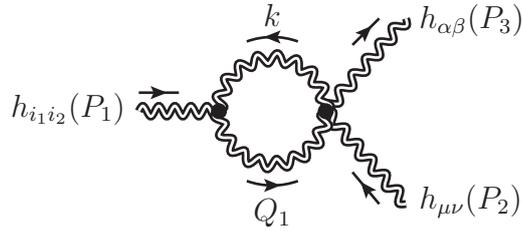

  \vspace{7mm}
  \centering  
  \begin{axopicture}(100,80)(0,-15)
    \SetWidth{1.}
    \Arc[arrow](50,30)(28,70,110)
    \Arc[arrow](50,30)(28,250,290)
    \DoublePhoton(0,30)(30,30){2}{5}{2}
    \DoublePhoton(70,30)(100,65){2}{6}{2}
    \DoublePhoton(70,30)(100,-5){2}{6}{2}
    \Line[arrow](80,55)(90,65)
    \Line[arrow](90,-5)(80,7)
    \Line[arrow](0,37)(15,37)
    \DoublePhotonArc(50,30)(20,0,360){2}{18}{2}
    \Vertex(30,30){3}
    \Vertex(70,30){3}
    \Text(50,62)[b]{$k$}
    \Text(50,-3)[t]{$Q_1$}
    \Text(-5,25)[rb]{$ h_{i_1 i_2} (P_1) $}
    \Text(105,58)[lb]{$h_{\alpha\beta}(P_3)$}
    \Text(105,-10)[lb]{$h_{\mu\nu}(P_2)$}
  \end{axopicture}
  \caption{Triple graviton vertex at one-loop level with two graviton propagators.}
  \label{fig:TripleExample}
\end{figure}

At tree level, the triple graviton vertex appears in the t-channel diagram as shown in
Fig.~\ref{fig:TreeLevelScalarGraviton2}b, and it has nine one-loop contributions:
three of them are tad-pole diagrams as in Fig.~\ref{fig:TripleGravitonVertex}a-c, three are bubble diagrams as in Fig.~\ref{fig:TripleGravitonVertex}d-f and three are triangle diagrams as in Fig.~\ref{fig:TripleGravitonVertex}g-i. However, in this section we only discuss the
bubble diagram with two graviton propagators, as shown in Fig.~\ref{fig:TripleExample}f in detail, while we only give overall comparisons for the other diagrams.

\noindent Using the momentum assignments in Fig.~\ref{fig:TripleExample}, the amplitude of this diagram can be written as
\begin{alignat}{4}
  \mathcal{M}_{\mu\nu\alpha\beta\, i_1 i_2} =  & \frac{1}{2} \int \frac{d^4k}{(2\pi)^4}
  V_{\mu\nu\rho\sigma\eta\lambda}^{\{h h  h\}}
  (k,Q_1) \;\; && S_{\{h h\}}^{\rho\sigma\gamma\delta}(k) \;\; && S_{\{h h\}}^{\eta\lambda\epsilon\kappa}(Q_1)
  \;\;  && V_{\gamma\delta\epsilon\kappa\alpha\beta\, i_1 i_2}^{\{h h h
    h\}}(k,Q_1,P_2) \, , \nonumber
\end{alignat}
where $\frac{1}{2}$ is the symmetry factor and $ Q_1=k+P_1$.

In this case, we note that the standard triple graviton vertex
$V_{\mu\nu\rho\sigma\eta\lambda}^{\{h h h\}}$ Eq.~(\ref{eq:3Hb}) has 40 terms and the standard quadruple graviton vertex $V_{\gamma\delta\epsilon\kappa\alpha\beta\, i_1 i_2}^{\{h h h h\}}$ Eq.~(\ref{eq:4Hb})
has 113 terms whereas the simplified ones
Eqs.~(\ref{eq:OurTripleGravitonVertex}, \ref{eq:OurQuadrupoleGravitonVertex}) have
only 4 and 12 terms respectively. Thus, for this diagram, the number of Lagrangian terms involved
from the vertices in the standard way is 4520, compared to 48 terms in
the simplified way. To compare the two, we again consider the running time in FORM, after doing
Passarino-Veltman reduction and writing the amplitude in terms of scalar integrals:\vspace{3mm}
\begin{Verbatim}[gobble=2,frame=single,framesep=2mm,label=simplified way,labelposition=all,numbers=left]
  WTime =      20.15 sec   Generated terms =         99
            V3HH2a         Terms in output =         99
                           Bytes used      =      32776
\end{Verbatim}
\vspace{3mm}
\begin{Verbatim}[gobble=2,frame=single,framesep=2mm,label=standard way,labelposition=all,numbers=left]
  WTime =     803.35 sec   Generated terms =         99
            V3HH2b         Terms in output =         99
                           Bytes used      =      34184
\end{Verbatim}
which shows that the running time in the standard way is about 40 times the
running time in the simplified way. Similarly, for the diagram in Fig~\ref{fig:TripleGravitonVertex}i the running time is about seven minutes in the simplified way while in the standard
way it is more than two hours.

Finally, we list, in Tab.~\ref{table:ComparisonTripleGraviton}, the number of
Lagrangian terms for all one-loop corrections to the triple graviton vertex in the standard and simplified ways.


\begin{table}[H]
  \caption{The number of Lagrangian terms when using the standard rules and the simplified ones for
    calculating the one-loop corrections to the triple graviton vertex as shown in
    Fig.~\ref{fig:TripleGravitonVertex}.}\smallskip
  \label{table:ComparisonTripleGraviton}
  \centering
  \begin{threeparttable}
  \begin{tabular}{c|C|C}
  \hline\hline  
  \text{Diagram\;\;} & \text{\;  The standard way\tnote{1}\;\;}&  \text{\;  The simplified way\tnote{1}} \\
  \hline
  (a)                &    $10$                                & $7$                                \\
  \hline
  (b)                & \text{The amplitude vanishes\tnote{2}} & \text{The amplitude vanishes\tnote{2}}      \\
  \hline
  (c)                & \text{The amplitude vanishes\tnote{2}}          & \text{The amplitude vanishes\tnote{2}}      \\
  \hline
  (d)                &    $18$                                & $6$                                \\
  \hline
  (e)                & \text{The amplitude vanishes}          & $319$                               \\
  \hline
  (f)                &    $4 520$                              & $48$                               \\
  \hline
  (g)                &    $27$                                & $27$                                 \\
  \hline
  (h)                &    $512$                               & $1 331$                               \\
  \hline
  (i)                &    $64 000$                             & $64$                                 \\
  \hline
  \end{tabular}
  \begin{tablenotes}
  \item[1] \scriptsize{Since the propagators are the same in the standard and
      simplified rules, we only consider the terms of the vertices in all our comparisons of the numbers of the Lagrangian terms.}
  \item[2] \scriptsize{all massless tad-pole diagrams vanish according to the relation Eq.~(\ref{eq:Tadpole}).}
  \end{tablenotes}
\end{threeparttable}
\end{table}
\normalsize

\subsubsection{Scalar-Scalar-Graviton Vertex Corrections}
\begin{figure}[H]
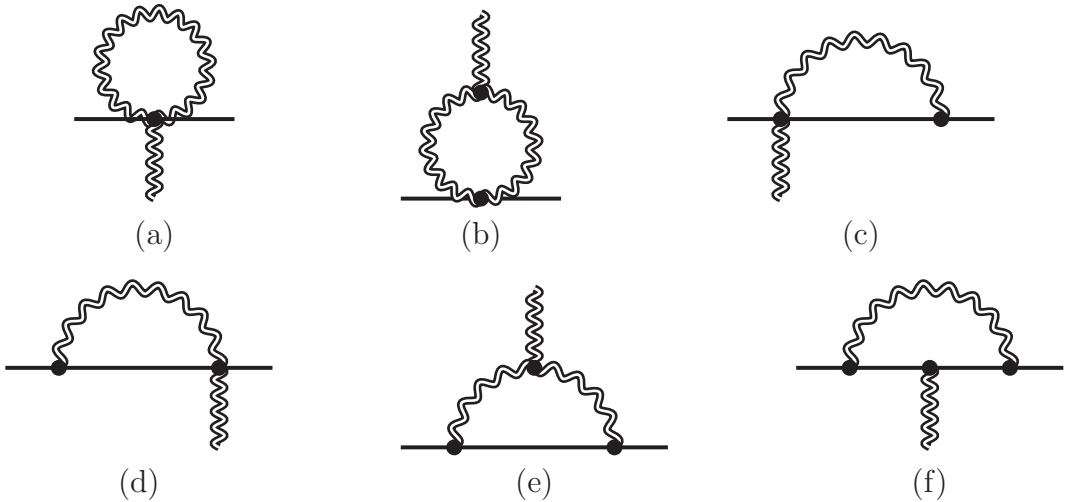

  \vspace{3mm}
	\centering
  \begin{axopicture}(60,90)(0,-40)
    \SetWidth{1.}
    \DoublePhotonArc(30,30)(20,0,360){2}{18}{2}
    \DoublePhoton(30,10)(30,-20){2}{5}{2}
    \SetWidth{1.5}
    \Line(0,10)(60,10)
    \Vertex(30,10){3}
    \Text(30,-40)[b]{(a)}
  \end{axopicture} \hspace{15mm}
  ~~~
  \begin{axopicture}(60,90)(0,-10)
    \SetWidth{1.5}
    \Line(0,10)(60,10)
    \SetWidth{1.}
    \DoublePhoton(30,50)(30,80){2}{5}{2}
    \DoublePhotonArc(30,30)(20,0,360){2}{18}{2}
    \Vertex(30,10){3}
    \Vertex(30,50){3}
    \Text(30,-10)[b]{(b)}
  \end{axopicture} \hspace{15mm}
  ~~~
  \begin{axopicture}(100,90)(0,-40)
    \SetWidth{1.5}
    \Line(0,10)(100,10)
    \SetWidth{1.}
    \DoublePhoton(20,10)(20,-20){2}{5}{2}
    \DoublePhotonArc(50,10)(30,0,180){2}{10}{2}
    \Vertex(20,10){3}
    \Vertex(80,10){3}
    \Text(50,-40)[b]{(c)}
  \end{axopicture} 
  \\[1mm]
  \begin{axopicture}(100,90)(0,-40)
    \SetWidth{1.5}
    \Line(0,10)(100,10)
    \SetWidth{1.}
    \DoublePhoton(80,10)(80,-20){2}{5}{2}
    \DoublePhotonArc(50,10)(30,0,180){2}{10}{2}
    \Vertex(20,10){3}
    \Vertex(80,10){3}
    \Text(50,-40)[b]{(d)}
  \end{axopicture} \hspace{10mm}
  ~~~
  \begin{axopicture}(100,90)(0,-10)
    \SetWidth{1.5}
    \Line(0,10)(100,10)
    \SetWidth{1.}
    \DoublePhoton(50,40)(50,70){2}{5}{2}
    \DoublePhotonArc(50,10)(30,0,180){2}{10}{2}
    \Vertex(20,10){3}
    \Vertex(80,10){3}
    \Vertex(50,40){3}
    \Text(50,-10)[b]{(e)}
  \end{axopicture} \hspace{10mm}
  ~~~
  \begin{axopicture}(100,90)(0,-40)
    \SetWidth{1.5}
    \Line(0,10)(100,10)
    \SetWidth{1.}
    \DoublePhoton(50,10)(50,-20){2}{5}{2}
    \DoublePhotonArc(50,10)(30,0,180){2}{10}{2}
    \Vertex(20,10){3}
    \Vertex(80,10){3}
    \Vertex(50,10){3}
    \Text(50,-40)[b]{(f)}
  \end{axopicture}
  \caption{Scalar-Scalar-Graviton vertex at one-loop level with: one (a), two (b-d), three (e-f) propagators.}
	\label{fig:ScalarScalarGravitonVertex}
\end{figure}

At tree level, the scalar-scalar-graviton vertex appears twice in the s
and u-channel diagrams and once in the t-channel as shown in
Fig.~\ref{fig:TreeLevelScalarGraviton2}, and it has six one-loop diagrams as
shown in Fig.~\ref{fig:ScalarScalarGravitonVertex}: the tad-pole diagram in
Fig.~\ref{fig:ScalarScalarGravitonVertex}a, the three bubble diagrams in
Fig.~\ref{fig:ScalarScalarGravitonVertex}b-d and the two triangle
diagrams in Fig.~\ref{fig:ScalarScalarGravitonVertex}e-f.

Again, we list in Tab.~\ref{table:ComparisonScalarScalarGraviton} the
comparison of the number of Lagrangian terms using the standard and simplified
rules to show the usefulness of the simplified rules and how the number of Lagrangian terms can be reduced by a factor 30 for some of the diagrams in Fig.~\ref{fig:ScalarScalarGravitonVertex}.

\begin{table}[H]
  \caption{The number of Lagrangian terms when using the standard rules and the simplified ones for
    calculating the one-loop corrections to the scalar-scalar-graviton vertex as
    shown in Fig.~\ref{fig:ScalarScalarGravitonVertex}.}\smallskip
  \label{table:ComparisonScalarScalarGraviton}
  \centering
  \begin{threeparttable}
  \begin{tabular}{c|C|C}
  \hline\hline  
  \text{Diagram\;\;} & \text{\; The standard way\;\;}&  \text{\; The simplified way} \\
  \hline
  (a)                & \text{The amplitude vanishes\tnote{1}}          & \text{The amplitude vanishes\tnote{1}}      \\
  \hline
  (b)                &  $240$                                 & $8$                                \\
  \hline
  (c)                &    $18$                                & $6$                                \\
  \hline
  (d)                &    $18$                                & $6$                               \\
  \hline
  (e)                &    $360$                               & $36$                                \\
  \hline
  (f)                &    $27$                                & $27$                                \\
  \hline
  \end{tabular}
 \begin{tablenotes}
  \item[1] \scriptsize{all massless tad-pole diagrams vanish according to the relation Eq.~(\ref{eq:Tadpole}).}
  \end{tablenotes}
\end{threeparttable}
\end{table}

\subsubsection{Scalar-Scalar-Graviton-Graviton Vertex Corrections}
\begin{figure}[H]
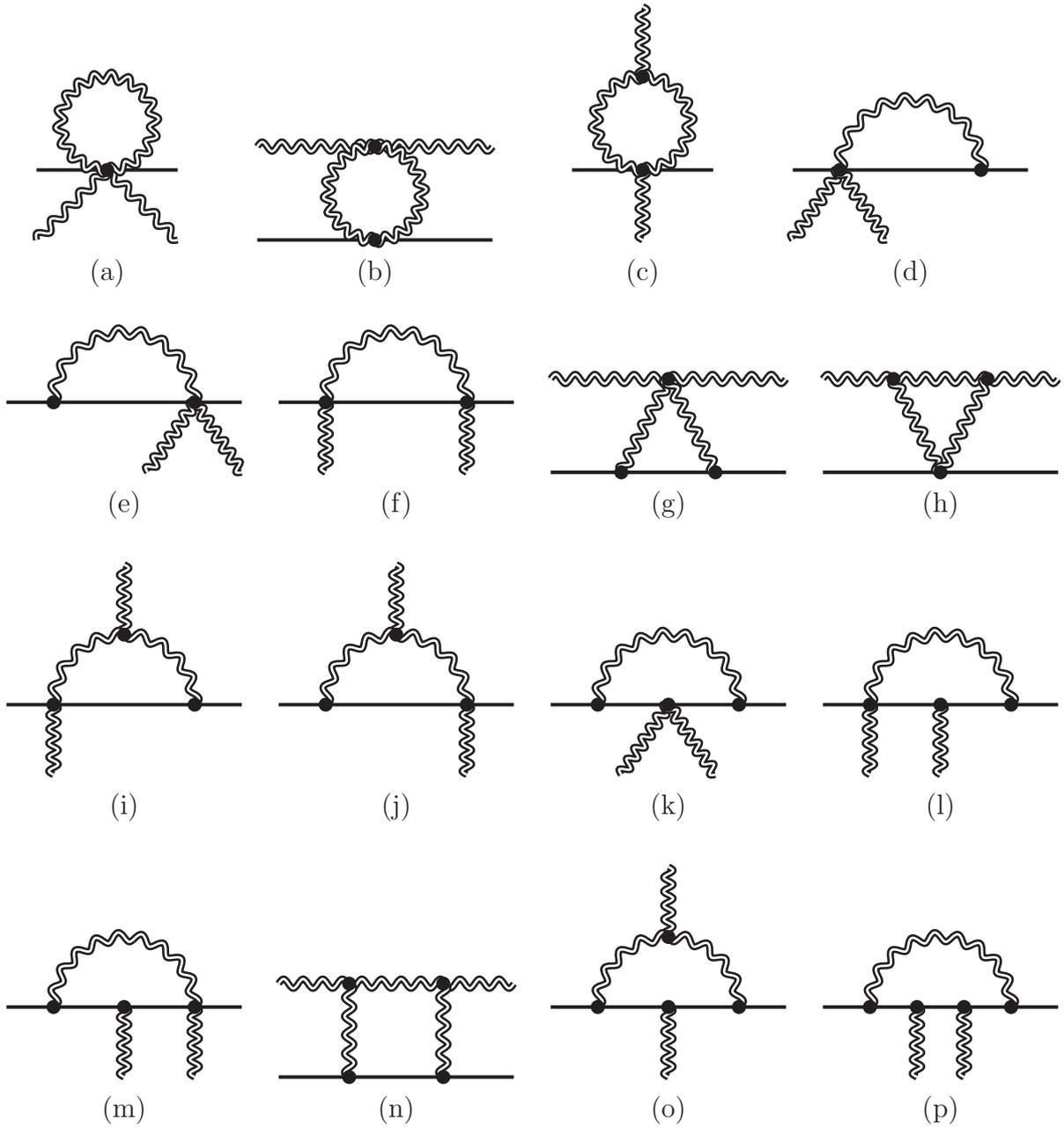

	\centering
  \vspace{3mm}
  \begin{axopicture}(60,90)(0,-40)
    \SetWidth{1.5}
    \Line(0,10)(60,10)
    \SetWidth{1.}
    \DoublePhoton(0,-20)(30,10){2}{5}{2}
    \DoublePhoton(30,10)(60,-20){2}{5}{2}
    \DoublePhotonArc(30,30)(20,0,360){2}{18}{2}
    \Vertex(30,10){3}
    \Text(30,-40)[b]{(a)}
  \end{axopicture} \hspace{5mm}
  ~~~
  \begin{axopicture}(100,60)(0,-10)
    \SetWidth{1.5}
    \Line(0,10)(100,10)
    \SetWidth{1.}
    \DoublePhoton(0,50)(100,50){2}{12}{2}
    \DoublePhotonArc(50,30)(20,0,360){2}{20}{2}
    \Vertex(50,10){3}
    \Vertex(50,50){3}
    \Text(50,-10)[b]{(b)}
  \end{axopicture} \hspace{5mm}
  ~~~
  \begin{axopicture}(60,120)(0,-40)
    \SetWidth{1.5}
    \Line(0,10)(60,10)
    \SetWidth{1.}
    \DoublePhoton(30,50)(30,80){2}{5}{2}
    \DoublePhoton(30,10)(30,-20){2}{5}{2}
    \DoublePhotonArc(30,30)(20,0,360){2}{18}{2}
    \Vertex(30,10){3}
    \Vertex(30,50){3}
    \Text(30,-40)[b]{(c)}
  \end{axopicture} \hspace{5mm}
  ~~~
  \begin{axopicture}(100,90)(0,-40)
    \SetWidth{1.5}
    \Line(0,10)(100,10)
    \SetWidth{1.}
    \DoublePhotonArc(50,10)(30,0,180){2}{10}{2}
    \DoublePhoton(0,-20)(20,10){2}{6.5}{2}
    \DoublePhoton(40,-20)(20,10){2}{6.5}{2}
    \Vertex(20,10){3}
    \Vertex(80,10){3}
    \Text(50,-40)[b]{(d)}
  \end{axopicture}
  \\[3mm]
  \begin{axopicture}(100,90)(0,-40)
    \SetWidth{1.5}
    \Line(0,10)(100,10)
    \SetWidth{1.}
    \DoublePhotonArc(50,10)(30,0,180){2}{10}{2}
    \DoublePhoton(100,-20)(80,10){2}{6.5}{2}
    \DoublePhoton(60,-20)(80,10){2}{6.5}{2}
    \Vertex(20,10){3}
    \Vertex(80,10){3}
    \Text(50,-40)[b]{(e)}
  \end{axopicture}
  ~~~
  \begin{axopicture}(100,90)(0,-40)
    \SetWidth{1.5}
    \Line(0,10)(100,10)
    \SetWidth{1.}
    \DoublePhotonArc(50,10)(30,0,180){2}{10}{2}
    \DoublePhoton(20,-20)(20,10){2}{5}{2}
    \DoublePhoton(80,-20)(80,10){2}{5}{2}
    \Vertex(20,10){3}
    \Vertex(80,10){3}
    \Text(50,-40)[b]{(f)}
  \end{axopicture}
  ~~~
  \begin{axopicture}(100,60)(0,-10)
    \SetWidth{1.5}
    \Line(0,10)(100,10)
    \SetWidth{1.}
    \DoublePhoton(0,50)(100,50){2}{12}{2}
    \DoublePhoton(30,10)(50,50){2}{6.5}{2}
    \DoublePhoton(70,10)(50,50){2}{6.5}{2}
    \Vertex(30,10){3}
    \Vertex(70,10){3}
    \Vertex(50,50){3}
    \Text(50,-10)[b]{(g)}
  \end{axopicture} 
  ~~~
  \begin{axopicture}(100,60)(0,-10)
    \SetWidth{1.5}
    \Line(0,10)(100,10)
    \SetWidth{1.}
    \DoublePhoton(0,50)(100,50){2}{12}{2}
    \DoublePhoton(50,10)(30,50){2}{6.5}{2}
    \DoublePhoton(50,10)(70,50){2}{6.5}{2}
    \Vertex(30,50){3}
    \Vertex(70,50){3}
    \Vertex(50,10){3}
    \Text(50,-10)[b]{(h)}
  \end{axopicture} 
  \\[3mm]
  \begin{axopicture}(100,120)(0,-40)
    \SetWidth{1.5}
    \Line(0,10)(100,10)
    \SetWidth{1.}
    \DoublePhoton(50,40)(50,70){2}{5}{2}
    \DoublePhotonArc(50,10)(30,0,180){2}{10}{2}
    \DoublePhoton(20,10)(20,-20){2}{5}{2}
    \Vertex(20,10){3}
    \Vertex(80,10){3}
    \Vertex(50,40){3}
    \Text(50,-40)[b]{(i)}
  \end{axopicture}
  ~~~
  \begin{axopicture}(100,120)(0,-40)
    \SetWidth{1.5}
    \Line(0,10)(100,10)
    \SetWidth{1.}
    \DoublePhoton(50,40)(50,70){2}{5}{2}
    \DoublePhotonArc(50,10)(30,0,180){2}{10}{2}
    \DoublePhoton(80,10)(80,-20){2}{5}{2}
    \Vertex(20,10){3}
    \Vertex(80,10){3}
    \Vertex(50,40){3}
    \Text(50,-40)[b]{(j)}
  \end{axopicture} 
  ~~~
  \begin{axopicture}(100,90)(0,-40)
    \SetWidth{1.5}
    \Line(0,10)(100,10)
    \SetWidth{1.}
    \DoublePhotonArc(50,10)(30,0,180){2}{10}{2}
    \DoublePhoton(30,-20)(50,10){2}{6.5}{2}
    \DoublePhoton(70,-20)(50,10){2}{6.5}{2}
    \Vertex(20,10){3}
    \Vertex(80,10){3}
    \Vertex(50,10){3}
    \Text(50,-40)[b]{(k)}
  \end{axopicture} 
  ~~~
  \begin{axopicture}(100,90)(0,-40)
    \SetWidth{1.5}
    \Line(0,10)(100,10)
    \SetWidth{1.}
    \DoublePhotonArc(50,10)(30,0,180){2}{10}{2}
    \DoublePhoton(20,-20)(20,10){2}{5}{2}
    \DoublePhoton(50,-20)(50,10){2}{5}{2}
    \Vertex(20,10){3}
    \Vertex(80,10){3}
    \Vertex(50,10){3}
    \Text(50,-40)[b]{(l)}
  \end{axopicture}
  \\[3mm]
  \begin{axopicture}(100,90)(0,-40)
    \SetWidth{1.5}
    \Line(0,10)(100,10)
    \SetWidth{1.}
    \DoublePhotonArc(50,10)(30,0,180){2}{10}{2}
    \DoublePhoton(80,-20)(80,10){2}{5}{2}
    \DoublePhoton(50,-20)(50,10){2}{5}{2}
    \Vertex(20,10){3}
    \Vertex(80,10){3}
    \Vertex(50,10){3}
    \Text(50,-40)[b]{(m)}
  \end{axopicture}
  ~~~
  \begin{axopicture}(100,60)(0,-10)
    \SetWidth{1.5}
    \Line(0,10)(100,10)
    \SetWidth{1.}
    \DoublePhoton(0,50)(100,50){2}{12}{2}
    \DoublePhoton(30,10)(30,50){2}{5}{2}
    \DoublePhoton(70,10)(70,50){2}{5}{2}
    \Vertex(30,10){3}
    \Vertex(70,10){3}
    \Vertex(30,50){3}
    \Vertex(70,50){3}
    \Text(50,-10)[b]{(n)}
  \end{axopicture} 
  ~~~
  \begin{axopicture}(100,120)(0,-40)
    \SetWidth{1.5}
    \Line(0,10)(100,10)
    \SetWidth{1.}
    \DoublePhoton(50,40)(50,70){2}{5}{2}
    \DoublePhotonArc(50,10)(30,0,180){2}{10}{2}
    \DoublePhoton(50,10)(50,-20){2}{5}{2}
    \Vertex(20,10){3}
    \Vertex(80,10){3}
    \Vertex(50,40){3}
    \Vertex(50,10){3}
    \Text(50,-40)[b]{(o)}
  \end{axopicture}
  ~~~
  \begin{axopicture}(100,90)(0,-40)
    \SetWidth{1.5}
    \Line(0,10)(100,10)
    \SetWidth{1.}
    \DoublePhotonArc(50,10)(30,0,180){2}{10}{2}
    \DoublePhoton(40,-20)(40,10){2}{5}{2}
    \DoublePhoton(60,-20)(60,10){2}{5}{2}
    \Vertex(20,10){3}
    \Vertex(80,10){3}
    \Vertex(40,10){3}
    \Vertex(60,10){3}
    \Text(50,-40)[b]{(p)}
  \end{axopicture}
  \caption{Scalar-Scalar-Graviton-Graviton vertex at one-loop level with: one
    (a), two (b-f), three (g-m), four (n-p) propagators.}
  \label{fig:ScalarScalarGravitonGraviton}
\end{figure}

The scalar-scalar-graviton-graviton vertex also appears once at tree level in the last diagram in Fig.~\ref{fig:TreeLevelScalarGraviton2}. However, it has 16 one-loop diagrams as
shown in Fig.~\ref{fig:ScalarScalarGravitonGraviton}: the tad-pole diagram in
Fig.~\ref{fig:ScalarScalarGravitonGraviton}a, the five bubble diagrams in
Fig.~\ref{fig:ScalarScalarGravitonGraviton}b-f, the seven triangle diagrams in
Fig.~\ref{fig:ScalarScalarGravitonGraviton}g-m and the three box diagrams in Fig.~\ref{fig:ScalarScalarGravitonGraviton}n-p.

Again, since the aim of this section is to show the usefulness of the simplified
rules, we only discuss the statistics of the results for the
diagram in Fig.~\ref{fig:ScalarScalarGravitonGraviton}h. For this
diagram, the number of Lagrangian terms involved from the vertices in the
standard way is 9 600, compared to 32 terms in the simplified way. To compare the two, we again consider the
running time in FORM, after doing Passarino-Veltman reduction and writing the amplitude
in terms of scalar integrals:\vspace{1mm}
\begin{Verbatim}[gobble=2,frame=single,framesep=2mm,label=simplified way,labelposition=all,numbers=left]
  WTime =      91.16 sec   Generated terms =      21830
         V2phi2H3a         Terms in output =        116
                           Bytes used      =    1786232
\end{Verbatim}
\vspace{1mm}
\begin{Verbatim}[gobble=2,frame=single,framesep=2mm,label=standard way,labelposition=all,numbers=left]
  WTime =    3940.32 sec   Generated terms =      34572
         V2phi2H3b         Terms in output =        116
                           Bytes used      =    1823344
\end{Verbatim}
which shows that the running time in the simplified way is less than two minutes
while in the standard way it is more than one hour.

Finally, we list, in Tab.~\ref{table:ComparisonScalarScalarGravitonGraviton},
the number of Lagrangian terms for all one-loop corrections to the triple graviton vertex in the standard and simplified ways.

\begin{table}[H]
  \caption{The number of Lagrangian terms when using the standard rules and the simplified ones for
    calculating the one-loop corrections to the scalar-scalar-graviton-graviton
    vertex as shown in Fig.~\ref{fig:ScalarScalarGravitonGraviton}.}\smallskip
  \label{table:ComparisonScalarScalarGravitonGraviton}
  \centering
  \begin{threeparttable}
  \begin{tabular}{c|C|C}
  \hline\hline  
  \text{Diagram\;\;} & \text{\; The standard way\;\;}&  \text{\; The simplified way} \\
  \hline
  (a)                & \text{The amplitude vanishes\tnote{1}}          & \text{The amplitude vanishes\tnote{1}}      \\
  \hline
  (b)                &    $678$                                & $24$                                \\
  \hline
  (c)                & $400$          & $28$      \\
  \hline
  (d)                &    $30$                                & $21$                                \\
  \hline
  (e)                & $30$          & $21$                               \\
  \hline
  (f)                &    $36$                              & $4$                                \\
  \hline
  (g)                &    $1 017$                                & $108$                                \\
  \hline
  (h)                & $9 600$          & $32$      \\
  \hline
  (i)                 & $720$          & $24$      \\
  \hline
  (j)                &    $720$                                & $24$                                \\
  \hline
  (k)                & $54$          & $18$                               \\
  \hline
  (l)                &    $54$                              & $18$                                \\
  \hline
  (m)                &    $54$                                & $18$                                \\
  \hline
  (n)                & $14 400$          & $144$      \\
  \hline
  (o)                & $1 080$          & $108$      \\
  \hline
  (p)                &    $81$                                & $81$                                \\
  \hline
  \end{tabular}
 \begin{tablenotes}
  \item[1] \scriptsize{all massless tad-pole diagrams vanish according to the relation Eq.~(\ref{eq:Tadpole}).}
  \end{tablenotes}
\end{threeparttable}
\end{table}
\normalsize

\section{Conclusions \label{se:conclusion}}
\setcounter{equation}{0}
In this work we have shown how it is possible to simplify the Feynman rules for gravity by
using the freedom of choosing the gauge, adding total derivative terms and
redefining the fields, which can change the form of Lagrangian without changing the information
that it contains. In particular, the triple graviton and quadruple graviton
vertices were reduced from 40 to 4 terms and from 113 to 12 terms respectively.

In order to check our simplified Feynman rules, we have compared the
resulting amplitudes from the simplified rules with the resulting amplitudes from
the standard rules for scalar-graviton and graviton-graviton scattering at tree
level, and we indeed found that the resulting amplitudes are in agreement. In
addition, our results at tree level also agree with the results of M. T.
Grisaru, P. Van Nieuwenhuizen and C. C. Wu \cite{TreeLevelResults} as well as J.
F. Donoghue and T. Torma \cite{DonoghueTreeLevel}.

Besides tree level amplitudes, we have also calculated some one-loop diagrams for
scalar-graviton scattering, and we have shown how the calculations become simpler by
using the simplified Feynman rules.
In particular, for those diagrams that have triple or quadruple graviton
vertices. Moreover, we have shown how the running time in the FORM program can
be considerably reduced, up to $40$ times faster for some diagrams, by using the
simplified Feynman rules to calculate the amplitudes.


These simplified rules can also be used to simplify the calculations of more complicated diagrams
such as the quadruple graviton vertex at one-loop level in graviton-graviton
scattering with four propagators. In the latter example, the number of Lagrangian
terms involved from the vertices in the standard way is 2 560 000 terms, compared to 256 terms in the simplified way.
In addition, our aim was to simplify the lowest order vertices. However, this
technique can also be used to simplify higher order vertices.
Finally, this thesis may open the door to finding more freedoms and tools to further manipulate the Lagrangian in order to reach even simpler Feynman rules for gravity.

\vspace{9mm}
\begin{figure}[H]
  \centering
  \includegraphics[scale=0.1]{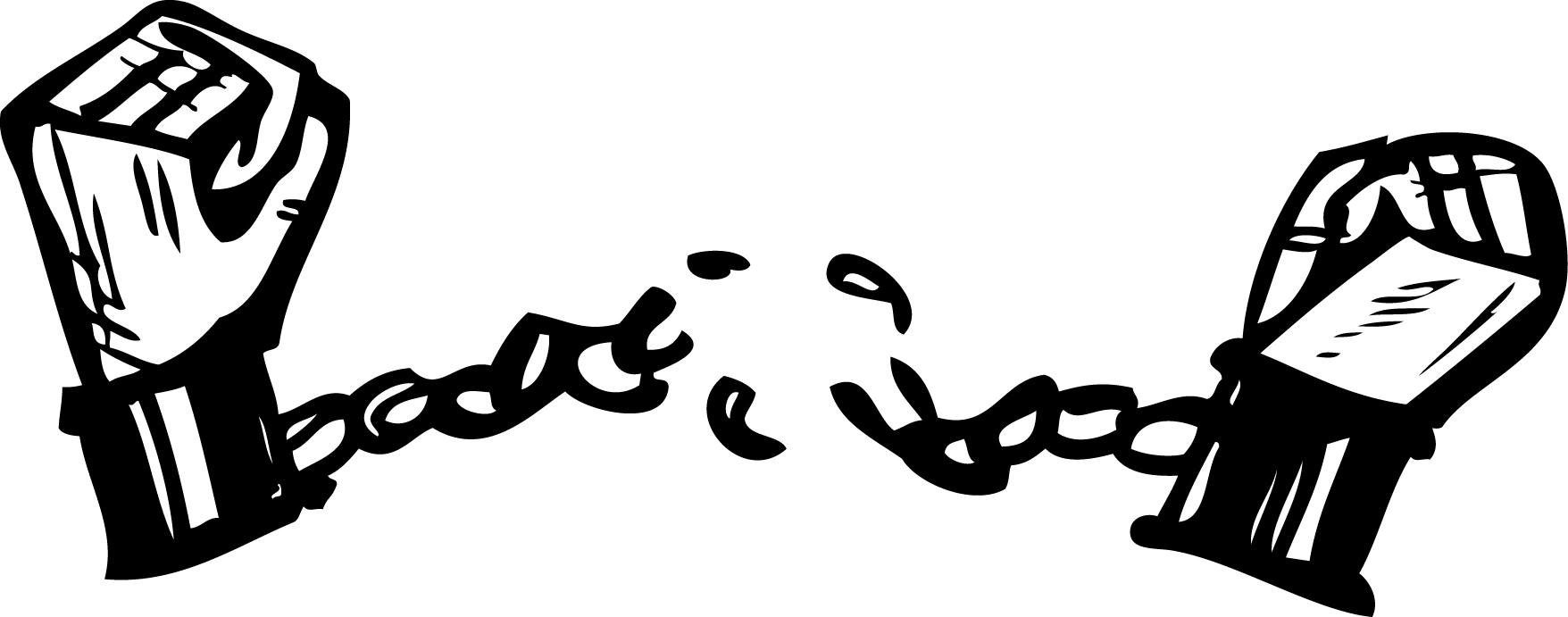}  
\end{figure}


\clearpage
\appendix

\section{Parameters}\label{Ap:Parameters}
\setcounter{equation}{0}
Table \ref{table:removeHtwoderiv}-\ref{table:simplestghost} in this appendix
shows the values of the parameters that were chosen to obtain the simplified Feynman rules as shown in Sec.~\ref{se:FeynmanRules}.
\vspace{4mm}
\begin{table}[H]
  \caption{The parameters of the total derivative Lagrangian for the
    gravitational field Eq.~(\ref{eq:LagrangianTD}) that remove all second order derivative terms
    for the propagator $S_{hh}$ and the vertices $V_{hhh}$, $V_{hhhh}$.}\smallskip
\label{table:removeHtwoderiv}
\centering
\begin{tabular}{L|CC|CC|CC}
\hline\hline  
  \text{Propagator/Vertex} & \text{Parameter} & \text{Value} &  \text{Parameter} & \text{Value} &  \text{Parameter} & \text{Value}  \\
  \hline
               & a_1   & $-2$     & a_2  &  $2$      &        &          \\
  \hline                                                                
  S_{hh}       & a_3   & $-1$     & a_4  &  $2$      &  a_5   & $1$      \\
               & a_6   & $-1$     & a_7  &  $-3$     &  a_8   & $2$      \\
  \hline                                                                
  V_{hhh}      & a_9   & $-1/4$   & a_{10}& $1/2$    & a_{11} & $1$      \\
               & a_{12}& $-2$     & a_{13}& $1/4$    & a_{14} & $-1/2$   \\
               & a_{15}& $-1/2$   & a_{16}& $1$      & a_{17} & $-3/2$   \\
               & a_{18}& $3$      & a_{19}& $1$      & a_{20} & $-2$     \\
               & a_{21}& $-2$     & a_{22}& $2$      &        &          \\
  \hline
  V_{hhhh}     & a_{23}& $-1/24$  & a_{24}& $ 1/4 $  & a_{25} & $ -1/3$  \\
               & a_{26}& $1/24 $  & a_{27}& $ -1/4$  & a_{28} & $ 1/3 $  \\
               & a_{29}& $1/4  $  & a_{30}& $ -1/2$  & a_{31} & $ -1/8$  \\
               & a_{32}& $1/4  $  & a_{33}& $ 1/4 $  & a_{34} & $ -1/2$  \\
               & a_{35}& $-1   $  & a_{36}& $ 2   $  & a_{37} & $ -3/8$  \\
               & a_{38}& $ 3/4 $  & a_{39}& $ 1   $  & a_{40} & $-2   $  \\
               & a_{41}& $ -1  $  & a_{42}& $ 2   $  & a_{43} & $3/2  $  \\
               & a_{44}& $ -2  $  & a_{45}& $ -3  $  & a_{46} & $ 2   $  \\
               & a_{47}& $ 1/2 $  & a_{48}& $ -1  $  & a_{49} & $-1   $  \\
               & a_{50}& $ 2   $  &       &          &        &          \\
  \hline
\end{tabular}
\end{table}
\vfill
\begin{table}[H]
  \caption{The parameters of the gravitational field redefinition
    Eq.~(\ref{eq:RedefinedField}) that reduce the number of Lagrangian terms for the
    vertices $V_{hhh}$ and $V_{hhhh}$.}\smallskip
  \label{table:fieldredefinitionparameters}
  \centering
  \begin{tabular}{L|CC|CC|CC}
  \hline\hline  
  \text{Vertex} & \text{Parameter} & \text{Value} &  \text{Parameter} & \text{Value} &  \text{Parameter} & \text{Value}  \\
  \hline
  V_{hhh}      & c_1   & $ 1/2 $  &   c_2  & $ -1/4$   &        &          \\
  \hline
  V_{hhhh}     & c_{3} & $3/32$   &   c_{4}& $ 0$      &  c_{5} & $ -1/8$  \\
               & c_{6} & $ 1/4$   &        &           &        &          \\   
  \hline
  \end{tabular}
\end{table}

\begin{table}[H]
  \caption{The parameters of the gauge condition Eq.~(\ref{eq:Gauge}), where the values of
    $b_1,b_2$ ensure the same de Donder propagator $S_{hh}$
    Eq.~(\ref{eq:2Hb}) as in the standard gauge Eq.~(\ref{eq:GaugeDonder}), and the other
    $b$'s parameters reduce the number of Largrangian terms for the vertices $V_{hhh}$ and $V_{hhhh}$.}\smallskip
  \label{table:setdeDondergauge}
  \centering
  \begin{tabular}{L|CC|CC|CC}
  \hline\hline  
  \text{Propagator/Vertex} & \text{Parameter} & \text{Value} &  \text{Parameter} & \text{Value} &  \text{Parameter} & \text{Value}  \\
  \hline                                                                
  S_{hh}       & b_1   & $1 $     &   b_2  & $-1/2$    &        &         \\
  \hline
  V_{hhh}      & b_3   & $ -1/8$  &   b_4  & $ 1/2$    &  b_5   &  $1/4$  \\
               & b_6   & $ -1/2$  &   b_7  & $ -1/2$   &  b_8   &  $1/2$  \\ 
  \hline
  V_{hhhh}     & b_{9 }& $ -1/64$ & b_{10} & $ 1/16$   & b_{11} &  $ 1/8$  \\
               & b_{12}& $ -1/2$  & b_{13} & $ 1/32$   & b_{14 }&  $ -1/8$ \\    
               & b_{15}& $ -1/8$  & b_{16} & $ 1/4$    & b_{17 }&  $ -1/8$ \\
               & b_{18}& $ 1/4$   & b_{19} & $ 3/8$    & b_{20 }&  $ -1/4$ \\
               & b_{21}& $ 1/8$   & b_{22} & $ -1/4$   &        &          \\
  \hline
  \end{tabular}
\end{table}
\vfill
\vspace{7mm}
\begin{table}[H]
  \caption{The parameters of the total derivative Lagrangian for the scalar field
    Eq.~(\ref{eq:LagrangianTDScalar}) that remove all second order derivative terms
    for the propagator $S_{\phi\phi}$ and the vertices $V_{\phi\phi h}$,
  $V_{\phi\phi hh}$, $V_{\phi\phi hhh}$.}\smallskip
  \label{table:removephitwoderiv}
  \centering
  \begin{tabular}{L|CC}
    \hline\hline
    \text{Propagator/Vertex} & \text{Parameter} & \text{Value}  \\
    \hline
     S_{\phi\phi} & d_1 & $0$    \\
    \hline
     V_{\phi\phi h} & d_2,\cdots,d_{5} & $0$    \\
    \hline
     V_{\phi\phi h h} & d_6,\cdots,d_{14} & $0$    \\
    \hline
     V_{\phi\phi h h h} & d_{15},\cdots,d_{22} & $0$  \\
    \hline
  \end{tabular}
\end{table}
\vfill
\vspace{7mm}
\begin{table}[H]
  \caption{The parameters of the scalar field redefinition
    Eq.~(\ref{eq:RedefinedFieldScalar}) that reduce the number of Lagrangian terms
    for the vertices $V_{\phi\phi h}$, $V_{\phi\phi hh}$ and $V_{\phi\phi hhh}$.}\smallskip
  \label{table:simplestphi}
  \centering
  \begin{tabular}{L|CC|CC|CC}
    \hline\hline
    \text{Vertex} & \text{Parameter} & \text{Value} &  \text{Parameter} & \text{Value} &  \text{Parameter} & \text{Value}  \\
    \hline
    V_{\phi\phi h} & e_1 & $0$ &  &  &  &   \\
    \hline
    V_{\phi\phi h h} & e_2 & $0$ & e_3  & $0$ &  &  \\
    \hline
    V_{\phi\phi h h h} & e_4 & $-1/384$ & e_5  & $0$ & e_6  & $-1/48$  \\
    \hline
  \end{tabular}
\end{table}
\vfill
\begin{table}[H]
  \caption{The parameters of the total derivative Lagrangian for the ghost and
    antighost fields Eq.~(\ref{eq:LagrangianTDGhost}) that remove all second order derivative terms
    for the propagator $S_{\bar{\chi}\chi}$ and the vertices $V_{\bar{\chi}\chi h}$,
  $V_{\bar{\chi}\chi hh}$, $V_{\bar{\chi}\chi hhh}$.}\smallskip
  \label{table:removeghosttwoderiv}
  \centering
  \begin{tabular}{L|CC|CC|CC}
    \hline\hline
    \text{Propagator/Vertex} & \text{Parameter} & \text{Value} &  \text{Parameter} & \text{Value} &  \text{Parameter} & \text{Value}  \\
    \hline
    S_{\bar{\chi}\chi} & h_1 & $-1$ & & & & \\
    \hline
    V_{\bar{\chi}\chi h} &  h_2,\cdots,h_{10}  & $0$  & h_{11}  & $1/2$  & h_{12}  & $-1/2$ \\
    & h_{13} & $-1/2$ & h_{14}  & $-1/4$  & h_{15}  & $-1/2$    \\
    \hline
    V_{\bar{\chi}\chi h h} &
    h_{20} & $ 0     $ & h_{21} & $ -1/8  $ & h_{22} & $ 1/4   $ \\
    & h_{23} & $ -1/32 $ & h_{24} & $ 1/8   $ & h_{25} & $ -1/8  $ \\
    & h_{26} & $ 1/8   $ & h_{27} & $ 1/8   $ & h_{28} & $ -1/4  $ \\
    & h_{29} & $ 0     $ & h_{30} & $ 0     $ & h_{31} & $ -1/8  $ \\
    & h_{32} & $ 1/8   $ & h_{33} & $ -1/2  $ & h_{34} & $ 1/4   $ \\
    & h_{35} & $ 1/4   $ & h_{36} & $ 0 $ & &  \\
    \hline
  \end{tabular}
\end{table}

\vspace{15mm}
\begin{table}[H]
  \caption{The parameters of the ghost and antighost field redefinition
    Eqs.~(\ref{eq:RedefinedFieldGhost}, \ref{eq:RedefinedFieldAntighost}) that reduce
    the number of Lagrangian terms for the vertices $V_{\bar{\chi}\chi h}$ and $V_{\bar{\chi}\chi hh}$.}\smallskip
  \label{table:simplestghost}
  \centering
  \begin{tabular}{L|CC|CC|CC}
    \hline\hline
    \text{Vertex} & \text{Parameter} & \text{Value} &  \text{Parameter} & \text{Value} &  \text{Parameter} & \text{Value}  \\
    \hline
    V_{\bar{\chi}\chi h} & f_1 & $1/4$ &  f_2  & $0$ & g_1,g_2 & $0$ \\
    \hline
    V_{\bar{\chi}\chi h h} &  f_3  & $-1/32$  & f_4  & $1/8$  & f_5 & $1/8$\\
    & f_6 & $-1/8$ &  g_3,\cdots,g_6 & $0$     &      &       \\
    \hline
  \end{tabular}
\end{table}
\vfill

\clearpage
\section{Standard Feynman Rules}\label{Ap:FeynmanRules}
\setcounter{equation}{0}
For completeness we give here the standard Feynman rules that we derived using the weak gravitational field
expansion Eq.~(\ref{eq:WeakFieldExpansion}) and de Donder gauge
Eq.~(\ref{eq:GaugeDonder}):\vspace{3mm}

\begin{itemize}
\item The scalar propagator:
\begin{flalign*}
  \mathcal{L}_{\phi\phi} = &
   \frac{1}{2} \Big (\partial^{\mu} \phi \partial_{\mu} \phi -  \phi^2 m^2\Big)
   \, . \numberthis\label{eq:2phib}&&
\end{flalign*}
\item The scalar-scalar-graviton vertex:
\begin{flalign*}
  \mathcal{L}_{\phi\phi h} = &
 \; \frac{\kappa}{4} \Big ( -  \phi^2  h_{\mu}^{\;\;\mu} m^2
  +  \partial^{\mu} \phi \partial_{\mu} \phi h_{\nu}^{\;\; \nu}
  - 2\, \partial_{\mu} \phi \partial_{\nu} \phi h^{\mu \nu} \Big )  \, . \numberthis\label{eq:2phi1Hb}&&
\end{flalign*}
\item The scalar-scalar-graviton-graviton vertex: 
\begin{flalign*}
  \mathcal{L}_{\phi\phi h h} = 
\;  \frac{\kappa^2}{8} \Big ( &  - \frac{1}{2} \phi^2  h_{\mu}^{\;\; \mu} h_{\nu}^{\;\; \nu} m^2
  +  \phi^2  h^{\mu \nu} h_{\mu \nu} m^2
  + \frac{1}{2} \partial^{\mu} \phi \partial_{\mu} \phi h_{\nu}^{\;\; \nu} h_{\alpha}^{\;\; \alpha}
  -  \partial^{\mu} \phi \partial_{\mu} \phi h^{\nu \alpha} h_{\nu\alpha} \\ &
  - 2\, \partial_{\mu} \phi \partial_{\nu} \phi h^{\mu \nu} h_{\alpha}^{\;\; \alpha}
  + 4\, \partial^{\mu} \phi \partial_{\nu} \phi h_{\mu \alpha} h^{\nu
    \alpha}  \Big )  \, . \numberthis\label{eq:2phi2Hb}&&
\end{flalign*}
\item The scalar-scalar-graviton-graviton-graviton vertex:
\begin{flalign*}     
  \mathcal{L}_{\phi\phi h h h} = 
  \; \frac{\kappa^3}{16} \Big ( & - \frac{1}{6} h_{\mu}^{\;\; \mu} h_{\nu }^{\;\;\nu} h_{\alpha}^{\;\; \alpha} \phi^2 m^2
  + \frac{1}{6} h_{\mu}^{\;\; \mu} h_{\nu}^{\;\; \nu} h_{\alpha}^{\;\; \alpha} \partial^{\beta} \phi \partial_{\beta} \phi
  -  h_{\mu}^{\;\; \mu} h_{\nu}^{\;\; \nu} h^{\alpha \beta} \partial_{\alpha} \phi
  \partial_{\beta} \phi \\ &
  +  h_{\mu}^{\;\; \mu} h^{\nu \alpha} h_{\nu \alpha} \phi^2  m^2 
  -  h_{\mu }^{\;\;\mu} h^{\nu \alpha} h_{\nu \alpha} \partial^{\beta} \phi \partial_{\beta} \phi
  + 4\, h_{\mu}^{\;\; \mu} h^{\nu \alpha} h_{\alpha \beta} \partial_{\nu} \phi
  \partial^{\beta} \phi \\ &
  + 2\, h^{\mu \nu} h_{\mu \nu} h^{\alpha \beta} \partial_{\alpha} \phi \partial_{\beta} \phi
  - \frac{4}{3} h^{\mu \nu} h_{\mu}^{\;\; \alpha} h_{\nu \alpha} \phi^2  m^2
  + \frac{4}{3} h^{\mu \nu} h_{\mu }^{\;\;\alpha} h_{\nu \alpha} \partial^{\beta} \phi
  \partial_{\beta} \phi \\ &
  - 8\, h^{\mu \nu} h_{\nu \alpha} h^{\alpha \beta} \partial_{\mu} \phi
  \partial_{\beta} \phi \Big )  \, . \numberthis\label{eq:2phi3Hb}&&
\end{flalign*}
\item The graviton propagator:
\begin{flalign*}
  \mathcal{L}_{hh} = &
     \; \frac{1}{2} \Big ( - \frac{1}{2} \partial^{\mu} h_{\nu}^{\;\;\nu} \partial_{\mu} h_{\alpha}^{\;\;\alpha}
     + \partial^{\mu} h^{\nu\alpha} \partial_{\mu} h_{\nu\alpha}\Big )  \, . \numberthis\label{eq:2Hb}&&
\end{flalign*}
\item The triple graviton vertex:
\begin{flalign*}     
\mathcal{L}_{hhh} = 
\; \kappa \Big ( &  - \frac{1}{4} h_{\mu}^{\;\;\mu} h_{\nu}^{\;\;\nu}\partial^{\alpha}\partial^{\beta} h_{\alpha\beta}
       + \frac{1}{4} h_{\mu}^{\;\;\mu} h_{\nu}^{\;\;\nu} \partial^{\alpha}\partial_{\alpha} h_{\beta}^{\;\;\beta}
       - h_{\mu}^{\;\;\mu} \partial^{\nu} h_{\nu\alpha} \partial^{\alpha}
       h_{\beta}^{\;\;\beta} 
       + h_{\mu}^{\;\;\mu} \partial_{\nu}h^{\nu\alpha} \partial^{\beta}h_{\alpha\beta} \\ &
       - h_{\mu}^{\;\;\mu} \partial^{\nu}\partial_{\nu}h^{\alpha\beta} h_{\alpha\beta}
       - h_{\mu}^{\;\;\mu} h^{\nu\alpha} \partial_{\nu}\partial_{\alpha} h_{\beta}^{\;\;\beta}
       + h_{\mu}^{\;\;\mu} h^{\nu\alpha} \partial_{\nu}\partial^{\beta}h_{\alpha\beta}
       + h_{\mu}^{\;\;\mu} h^{\nu\alpha} \partial_{\alpha}\partial^{\beta}h_{\nu\beta} \\ &
       + \frac{1}{4} h_{\mu}^{\;\;\mu} \partial^{\nu} h_{\alpha}^{\;\;\alpha} \partial_{\nu} h_{\beta}^{\;\;\beta}
       - \frac{3}{4} h_{\mu}^{\;\;\mu} \partial^{\nu} h^{\alpha\beta} \partial_{\nu} h_{\alpha\beta}
       + \frac{1}{2} h_{\mu}^{\;\;\mu} \partial^{\nu}h^{\alpha\beta} \partial_{\alpha}h_{\nu\beta}
       + \partial^{\mu} h_{\mu\nu} h^{\nu\alpha} \partial_{\alpha}
       h_{\beta}^{\;\;\beta}  &&
\end{flalign*}
\begin{flalign*}     
\qquad   & - \frac{3}{2} \partial_{\mu} h^{\mu\nu} h_{\nu\alpha} \partial_{\beta} h^{\alpha\beta}
       + \partial_{\mu} h^{\mu\nu} \partial_{\nu} h_{\alpha\beta} h^{\alpha\beta}
       - 2 \partial_{\mu} h^{\mu\nu} \partial_{\alpha} h_{\nu\beta} h^{\alpha\beta}
       - \frac{1}{2} \partial^{\mu}\partial_{\mu}h_{\nu}^{\;\;\nu} h^{\alpha\beta}
       h_{\alpha\beta} \\ &
       + 2 \partial^{\mu}\partial_{\mu} h^{\nu\alpha} h_{\nu\beta} h_{\alpha}^{\;\;\beta}
       + \frac{1}{2} h^{\mu\nu} h_{\mu\nu} \partial_{\alpha}\partial_{\beta}h^{\alpha\beta}
       + h^{\mu\nu} \partial_{\mu} h_{\nu\alpha} \partial^{\alpha} h_{\beta}^{\;\;\beta}
       - h^{\mu\nu} \partial_{\mu} h_{\nu\alpha} \partial_{\beta}
       h^{\alpha\beta} \\ &
       + 2 h^{\mu\nu} \partial_{\mu}\partial_{\nu}h^{\alpha\beta} h_{\alpha\beta}
       + \frac{1}{2} h^{\mu\nu} h_{\mu\alpha} \partial_{\nu}\partial^{\alpha} h_{\beta}^{\;\;\beta}
       - h^{\mu\nu} h_{\mu\alpha} \partial_{\nu}\partial_{\beta}h^{\alpha\beta}
       - h^{\mu\nu} \partial_{\mu}\partial_{\alpha} h_{\nu\beta} h^{\alpha\beta} \\ &
       - \frac{1}{2} h^{\mu\nu} \partial_{\mu} h_{\alpha}^{\;\;\alpha} \partial_{\nu} h_{\beta}^{\;\;\beta}
       + \frac{1}{2} h^{\mu\nu} \partial_{\mu} h_{\alpha}^{\;\;\alpha} \partial^{\beta} h_{\nu\beta}
       - 2 h^{\mu\nu} \partial_{\mu} \partial_{\alpha} h^{\alpha\beta} h_{\nu\beta}
       + \frac{3}{2} h^{\mu\nu} \partial_{\mu} h^{\alpha\beta} \partial_{\nu}
       h_{\alpha\beta} \\ &
       - 2 h^{\mu\nu} \partial_{\mu} h^{\alpha\beta} \partial_{\alpha} h_{\nu\beta}
       + \frac{3}{2} h^{\mu\nu} \partial_{\mu}\partial^{\alpha}h_{\beta}^{\;\;\beta} h_{\nu\alpha}
       + h^{\mu\nu} \partial_{\nu} h_{\mu\alpha} \partial^{\alpha} h_{\beta}^{\;\;\beta}
       - h^{\mu\nu} \partial_{\nu} h_{\mu\alpha} \partial_{\beta}
       h^{\alpha\beta} \\ &
       - h^{\mu\nu} h_{\nu\alpha} \partial^{\alpha}\partial^{\beta}h_{\mu\beta}
       - h^{\mu\nu} \partial_{\nu}\partial_{\alpha} h_{\mu\beta} h^{\alpha\beta}
       + \frac{1}{2} h^{\mu\nu} \partial_{\nu} h_{\alpha}^{\;\;\alpha} \partial^{\beta} h_{\mu\beta}
       - h^{\mu\nu} \partial^{\alpha} h_{\mu\nu} \partial_{\alpha}
       h_{\beta}^{\;\;\beta} \\ &
       + h^{\mu\nu} \partial_{\alpha}h_{\mu\nu} \partial_{\beta}h^{\alpha\beta}
       - \frac{1}{2} h^{\mu\nu} \partial^{\alpha} h_{\mu\alpha} \partial^{\beta} h_{\nu\beta}
       + 3 h^{\mu\nu} \partial^{\alpha} h_{\mu}^{\;\;\beta} \partial_{\alpha} h_{\nu\beta}
       - h^{\mu\nu} \partial^{\alpha} h_{\mu\beta} \partial^{\beta}
       h_{\nu\alpha}   \Big )    \, . \numberthis\label{eq:3Hb}&&
\end{flalign*}
\item The quadruple graviton vertex:
\begin{flalign*}
  \mathcal{L}_{hhhh} = 
  \; \kappa^2 \Big ( & - 2\, h_{\mu \mu} h_{\nu \nu} h_{\alpha \alpha} \partial_{\beta} \partial_{\gamma} h_{\beta \gamma}
  + 2\, h_{\mu \mu} h_{\nu \nu} h_{\alpha \alpha} \partial_{\beta} \partial_{\beta} h_{\gamma \gamma}
  - \frac{1}{4} h_{\mu \mu} h_{\nu \nu} h_{\alpha \beta} \partial_{\alpha} \partial_{\beta} h_{\gamma \gamma}
\\ &  + \frac{1}{4} h_{\mu \mu} h_{\nu \nu} h_{\alpha \beta} \partial_{\alpha} \partial_{\gamma} h_{\beta \gamma}
  + \frac{1}{4} h_{\mu \mu} h_{\nu \nu} h_{\alpha \beta} \partial_{\beta} \partial_{\gamma} h_{\alpha \gamma}
  - \frac{1}{4} h_{\mu \mu} h_{\nu \nu} \partial_{\alpha} h_{\alpha \beta} \partial_{\beta} h_{\gamma \gamma}
\\ &  + \frac{1}{4} h_{\mu \mu} h_{\nu \nu} \partial_{\alpha} h_{\alpha \beta} \partial_{\gamma} h_{\beta \gamma}
  + \frac{1}{16} h_{\mu \mu} h_{\nu \nu} \partial_{\alpha} h_{\beta \beta} \partial_{\alpha} h_{\gamma \gamma}
  - \frac{3}{16} h_{\mu \mu} h_{\nu \nu} \partial_{\alpha} h_{\beta \gamma} \partial_{\alpha} h_{\beta \gamma}
\\ &  + \frac{1}{8} h_{\mu \mu} h_{\nu \nu} \partial_{\alpha} h_{\beta \gamma} \partial_{\beta} h_{\alpha \gamma}
  - \frac{1}{4} h_{\mu \mu} h_{\nu \nu} \partial_{\alpha} \partial_{\alpha} h_{\beta \gamma} h_{\beta \gamma}
  + \frac{1}{4} h_{\mu \mu} h_{\nu \alpha} h_{\nu \alpha} \partial_{\beta} \partial_{\gamma} h_{\beta \gamma}
\\ &  + \frac{1}{4} h_{\mu \mu} h_{\nu \alpha} h_{\nu \beta} \partial_{\alpha} \partial_{\beta} h_{\gamma \gamma}
  - \frac{1}{2} h_{\mu \mu} h_{\nu \alpha} h_{\nu \beta} \partial_{\alpha} \partial_{\gamma} h_{\beta \gamma}
  - \frac{1}{2} h_{\mu \mu} h_{\nu \alpha} h_{\alpha \beta} \partial_{\beta} \partial_{\gamma} h_{\nu \gamma}
\\ &  + \frac{1}{2} h_{\mu \mu} h_{\nu \alpha} \partial_{\nu} h_{\alpha \beta} \partial_{\beta} h_{\gamma \gamma}
  - \frac{1}{2} h_{\mu \mu} h_{\nu \alpha} \partial_{\nu} h_{\alpha \beta} \partial_{\gamma} h_{\beta \gamma}
  - \frac{1}{4} h_{\mu \mu} h_{\nu \alpha} \partial_{\nu} h_{\beta \beta} \partial_{\alpha} h_{\gamma \gamma}
\\ &  + \frac{1}{4} h_{\mu \mu} h_{\nu \alpha} \partial_{\nu} h_{\beta \beta} \partial_{\gamma} h_{\alpha \gamma}
  + \frac{3}{4} h_{\mu \mu} h_{\nu \alpha} \partial_{\nu} h_{\beta \gamma} \partial_{\alpha} h_{\beta \gamma}
  - h_{\mu \mu} h_{\nu \alpha} \partial_{\nu} h_{\beta \gamma} \partial_{\beta} h_{\alpha \gamma}
\\ &  + h_{\mu \mu} h_{\nu \alpha} \partial_{\nu} \partial_{\alpha} h_{\beta \gamma} h_{\beta \gamma}
  - \frac{1}{2} h_{\mu \mu} h_{\nu \alpha} \partial_{\nu} \partial_{\beta} h_{\alpha \gamma} h_{\beta \gamma}
  - h_{\mu \mu} h_{\nu \alpha} \partial_{\nu} \partial_{\beta} h_{\beta \gamma} h_{\alpha \gamma}
\\ &  + \frac{3}{4} h_{\mu \mu} h_{\nu \alpha} \partial_{\nu} \partial_{\beta} h_{\gamma \gamma} h_{\alpha \beta}
  + \frac{1}{2} h_{\mu \mu} h_{\nu \alpha} \partial_{\alpha} h_{\nu \beta} \partial_{\beta} h_{\gamma \gamma}
  - \frac{1}{2} h_{\mu \mu} h_{\nu \alpha} \partial_{\alpha} h_{\nu \beta} \partial_{\gamma} h_{\beta \gamma}
\\ &  + \frac{1}{4} h_{\mu \mu} h_{\nu \alpha} \partial_{\alpha} h_{\beta \beta} \partial_{\gamma} h_{\nu \gamma}
  - \frac{1}{2} h_{\mu \mu} h_{\nu \alpha} \partial_{\alpha} \partial_{\beta} h_{\nu \gamma} h_{\beta \gamma}
  - \frac{1}{2} h_{\mu \mu} h_{\nu \alpha} \partial_{\beta} h_{\nu \alpha} \partial_{\beta} h_{\gamma \gamma}
\\ &  + \frac{1}{2} h_{\mu \mu} h_{\nu \alpha} \partial_{\beta} h_{\nu \alpha} \partial_{\gamma} h_{\beta \gamma}
  - \frac{1}{4} h_{\mu \mu} h_{\nu \alpha} \partial_{\beta} h_{\nu \beta} \partial_{\gamma} h_{\alpha \gamma}
  + \frac{3}{2} h_{\mu \mu} h_{\nu \alpha} \partial_{\beta} h_{\nu \gamma} \partial_{\beta} h_{\alpha \gamma}
\\ &  - \frac{1}{2} h_{\mu \mu} h_{\nu \alpha} \partial_{\beta} h_{\nu \gamma} \partial_{\gamma} h_{\alpha \beta}
  + \frac{1}{2} h_{\mu \mu} \partial_{\nu} h_{\nu \alpha} h_{\alpha \beta} \partial_{\beta} h_{\gamma \gamma}
  - \frac{3}{4} h_{\mu \mu} \partial_{\nu} h_{\nu \alpha} h_{\alpha \beta}
  \partial_{\gamma} h_{\beta \gamma} &&
\end{flalign*}
\begin{flalign*}
  \qquad\qquad\quad  &  + \frac{1}{2} h_{\mu \mu} \partial_{\nu} h_{\nu \alpha} \partial_{\alpha} h_{\beta \gamma} h_{\beta \gamma}
  - h_{\mu \mu} \partial_{\nu} h_{\nu \alpha} \partial_{\beta} h_{\alpha \gamma} h_{\beta \gamma}
  - \frac{1}{4} h_{\mu \mu} \partial_{\nu} \partial_{\nu} h_{\alpha \alpha} h_{\beta \gamma} h_{\beta \gamma}
\\ &  + h_{\mu \mu} \partial_{\nu} \partial_{\nu} h_{\alpha \beta} h_{\alpha \gamma} h_{\beta \gamma}
  + \frac{1}{2} h_{\mu \nu} h_{\mu \nu} h_{\alpha \beta} \partial_{\alpha} \partial_{\beta} h_{\gamma \gamma}
  - \frac{1}{2} h_{\mu \nu} h_{\mu \nu} h_{\alpha \beta} \partial_{\alpha} \partial_{\gamma} h_{\beta \gamma}
\\ &  - \frac{1}{2} h_{\mu \nu} h_{\mu \nu} h_{\alpha \beta} \partial_{\beta} \partial_{\gamma} h_{\alpha \gamma}
  - \frac{1}{8} h_{\mu \nu} h_{\mu \nu} \partial_{\alpha} h_{\beta \beta} \partial_{\alpha} h_{\gamma \gamma}
  + \frac{3}{8} h_{\mu \nu} h_{\mu \nu} \partial_{\alpha} h_{\beta \gamma} \partial_{\alpha} h_{\beta \gamma}
\\ &  - \frac{1}{4} h_{\mu \nu} h_{\mu \nu} \partial_{\alpha} h_{\beta \gamma} \partial_{\beta} h_{\alpha \gamma}
  - \frac{1}{3} h_{\mu \nu} h_{\mu \alpha} h_{\nu \alpha} \partial_{\beta} \partial_{\gamma} h_{\beta \gamma}
  - h_{\mu \nu} h_{\mu \alpha} h_{\nu \beta} \partial_{\alpha} \partial_{\beta} h_{\gamma \gamma}
\\ &  + \frac{3}{4} h_{\mu \nu} h_{\mu \alpha} h_{\nu \beta} \partial_{\alpha} \partial_{\gamma} h_{\beta \gamma}
  + \frac{1}{4} h_{\mu \nu} h_{\mu \alpha} h_{\nu \beta} \partial_{\alpha} \partial_{\gamma} h_{\gamma \beta}
  + h_{\mu \nu} h_{\mu \alpha} h_{\nu \beta} \partial_{\beta} \partial_{\gamma} h_{\alpha \gamma}
\\ &  - \frac{1}{2} h_{\mu \nu} h_{\mu \alpha} \partial_{\nu} h_{\alpha \beta} \partial_{\beta} h_{\gamma \gamma}
  + \frac{1}{8} h_{\mu \nu} h_{\mu \alpha} \partial_{\nu} h_{\beta \beta} \partial_{\alpha} h_{\gamma \gamma}
  - \frac{3}{2} h_{\mu \nu} h_{\mu \alpha} \partial_{\nu} h_{\beta \gamma} \partial_{\alpha} h_{\beta \gamma}
\\ &  + \frac{1}{4} h_{\mu \nu} h_{\mu \alpha} \partial_{\nu} h_{\beta \gamma} \partial_{\beta} h_{\alpha \gamma}
  + \frac{1}{4} h_{\mu \nu} h_{\mu \alpha} \partial_{\nu} h_{\beta \gamma} \partial_{\gamma} h_{\alpha \beta}
  - \frac{1}{2} h_{\mu \nu} h_{\mu \alpha} \partial_{\nu} \partial_{\alpha} h_{\beta \gamma} h_{\beta \gamma}
\\ &  + h_{\mu \nu} h_{\mu \alpha} \partial_{\nu} \partial_{\beta} h_{\alpha \gamma} h_{\beta \gamma}
  + \frac{1}{4} h_{\mu \nu} h_{\mu \alpha} \partial_{\beta} h_{\nu \alpha} \partial_{\beta} h_{\gamma \gamma}
  + \frac{1}{8} h_{\mu \nu} h_{\mu \alpha} \partial_{\beta} h_{\nu \beta} \partial_{\gamma} h_{\alpha \gamma}
\\ &  - 3 h_{\mu \nu} h_{\mu \alpha} \partial_{\beta} h_{\nu \gamma} \partial_{\beta} h_{\alpha \gamma}
  + \frac{1}{4} h_{\mu \nu} h_{\mu \alpha} \partial_{\beta} h_{\nu \gamma} \partial_{\gamma} h_{\alpha \beta}
  + \frac{3}{4} h_{\mu \nu} h_{\nu \alpha} h_{\alpha \beta} \partial_{\beta}
  \partial_{\gamma} h_{\mu \gamma}
  \\ &- \frac{1}{2} h_{\mu \nu} h_{\nu \alpha} \partial_{\alpha} h_{\mu \beta} \partial_{\beta} h_{\gamma \gamma}
  + \frac{1}{4} h_{\mu \nu} h_{\nu \alpha} \partial_{\alpha} h_{\beta \gamma} \partial_{\beta} h_{\mu \gamma}
  + \frac{1}{4} h_{\mu \nu} h_{\nu \alpha} \partial_{\alpha} h_{\beta \gamma}
  \partial_{\gamma} h_{\mu \beta}
  \\ & + \frac{3}{4} h_{\mu \nu} h_{\nu \alpha} \partial_{\alpha} \partial_{\beta} h_{\mu \gamma} h_{\beta \gamma}
  + \frac{3}{4} h_{\mu \nu} h_{\nu \alpha} \partial_{\beta} h_{\mu \alpha} \partial_{\beta} h_{\gamma \gamma}
  + \frac{3}{8} h_{\mu \nu} h_{\nu \alpha} \partial_{\beta} h_{\mu \beta} \partial_{\gamma} h_{\alpha \gamma}
\\ &  + \frac{3}{4} h_{\mu \nu} h_{\nu \alpha} \partial_{\beta} h_{\mu \gamma} \partial_{\gamma} h_{\alpha \beta}
  - \frac{5}{4} h_{\mu \nu} \partial_{\mu} h_{\nu \alpha} h_{\alpha \beta} \partial_{\beta} h_{\gamma \gamma}
  + h_{\mu \nu} \partial_{\mu} h_{\nu \alpha} h_{\alpha \beta} \partial_{\gamma} h_{\beta \gamma}
\\ &  - h_{\mu \nu} \partial_{\mu} h_{\nu \alpha} \partial_{\alpha} h_{\beta \gamma} h_{\beta \gamma}
  + 2 h_{\mu \nu} \partial_{\mu} h_{\nu \alpha} \partial_{\beta} h_{\alpha \gamma} h_{\beta \gamma}
  + \frac{3}{8} h_{\mu \nu} \partial_{\mu} h_{\alpha \alpha} h_{\nu \beta} \partial_{\beta} h_{\gamma \gamma}
\\ &  + \frac{1}{2} h_{\mu \nu} \partial_{\mu} h_{\alpha \alpha} \partial_{\nu} h_{\beta \gamma} h_{\beta \gamma}
  - \frac{1}{4} h_{\mu \nu} \partial_{\mu} h_{\alpha \alpha} \partial_{\beta} h_{\nu \gamma} h_{\beta \gamma}
  - \frac{1}{2} h_{\mu \nu} \partial_{\mu} h_{\alpha \beta} h_{\nu \alpha} \partial_{\beta} h_{\gamma \gamma}
\\ &  - \frac{1}{2} h_{\mu \nu} \partial_{\mu} h_{\alpha \beta} h_{\nu \beta} \partial_{\alpha} h_{\gamma \gamma}
  + \frac{1}{2} h_{\mu \nu} \partial_{\mu} h_{\alpha \beta} h_{\nu \gamma} \partial_{\alpha} h_{\beta \gamma}
  + \frac{1}{2} h_{\mu \nu} \partial_{\mu} h_{\alpha \beta} h_{\nu \gamma} \partial_{\beta} h_{\alpha \gamma}
\\ &  + \frac{1}{2} h_{\mu \nu} \partial_{\mu} h_{\alpha \beta} h_{\alpha \gamma} \partial_{\beta} h_{\nu \gamma}
  + h_{\mu \nu} \partial_{\mu} h_{\alpha \beta} h_{\alpha \gamma} \partial_{\gamma} h_{\nu \beta}
  - \frac{3}{2} h_{\mu \nu} \partial_{\mu} h_{\alpha \beta} \partial_{\nu}
  h_{\alpha \gamma} h_{\beta \gamma}
  \\ & - \frac{3}{2} h_{\mu \nu} \partial_{\mu} h_{\alpha \beta} \partial_{\nu} h_{\beta \gamma} h_{\alpha \gamma}
  + \frac{1}{2} h_{\mu \nu} \partial_{\mu} h_{\alpha \beta} \partial_{\nu} h_{\gamma \gamma} h_{\alpha \beta}
  + \frac{1}{2} h_{\mu \nu} \partial_{\mu} h_{\alpha \beta} \partial_{\alpha} h_{\nu \gamma} h_{\beta \gamma}
\\ &  - 2 h_{\mu \nu} \partial_{\mu} \partial_{\nu} h_{\alpha \beta} h_{\alpha \gamma} h_{\beta \gamma}
  + \frac{3}{4} h_{\mu \nu} \partial_{\mu} \partial_{\alpha} h_{\nu \beta} h_{\alpha \gamma} h_{\beta \gamma}
  + \frac{5}{4} h_{\mu \nu} \partial_{\mu} \partial_{\alpha} h_{\alpha \beta} h_{\nu \gamma} h_{\beta \gamma}
\\ &  - h_{\mu \nu} \partial_{\mu} \partial_{\alpha} h_{\beta \beta} h_{\nu \gamma} h_{\alpha \gamma}
  - \frac{3}{2} h_{\mu \nu} \partial_{\mu} \partial_{\alpha} h_{\beta \gamma} h_{\nu \alpha} h_{\beta \gamma}
  + \frac{5}{8} h_{\mu \nu} \partial_{\mu} \partial_{\alpha} h_{\beta \gamma} h_{\nu \beta} h_{\alpha \gamma}
\\ &  + \frac{5}{8} h_{\mu \nu} \partial_{\mu} \partial_{\alpha} h_{\beta \gamma} h_{\nu \gamma} h_{\alpha \beta}
  - \frac{1}{2} h_{\mu \nu} \partial_{\nu} h_{\mu \alpha} h_{\alpha \beta} \partial_{\beta} h_{\gamma \gamma}
  + h_{\mu \nu} \partial_{\nu} h_{\mu \alpha} h_{\alpha \beta} \partial_{\gamma} h_{\beta \gamma}
\\ &  - h_{\mu \nu} \partial_{\nu} h_{\mu \alpha} \partial_{\alpha} h_{\beta \gamma} h_{\beta \gamma}
  + \frac{1}{2} h_{\mu \nu} \partial_{\nu} h_{\alpha \beta} h_{\alpha \gamma} \partial_{\beta} h_{\mu \gamma}
  + \frac{1}{2} h_{\mu \nu} \partial_{\nu} h_{\alpha \beta} \partial_{\alpha}
  h_{\mu \gamma} h_{\beta \gamma} \\ &
  + \frac{1}{4} h_{\mu \nu} \partial_{\nu} \partial_{\alpha} h_{\mu \beta} h_{\alpha \gamma} h_{\beta \gamma}
  + \frac{1}{2} h_{\mu \nu} \partial_{\alpha} h_{\mu \nu} \partial_{\alpha} h_{\beta \gamma} h_{\beta \gamma}
  - \frac{3}{2} h_{\mu \nu} \partial_{\alpha} h_{\mu \beta} \partial_{\alpha}
  h_{\nu \gamma} h_{\beta \gamma} &&
\end{flalign*}
\begin{flalign*}
  \qquad\qquad\quad  &  - 2 \partial_{\mu} h_{\mu \nu} h_{\nu \alpha} h_{\alpha \beta} \partial_{\beta} h_{\gamma \gamma}
  + \frac{3}{2} \partial_{\mu} h_{\mu \nu} h_{\nu \alpha} h_{\alpha \beta} \partial_{\gamma} h_{\beta \gamma}
  - 2 \partial_{\mu} h_{\mu \nu} h_{\nu \alpha} \partial_{\alpha} h_{\beta \gamma} h_{\beta \gamma}
\\ &  + 2 \partial_{\mu} h_{\mu \nu} h_{\nu \alpha} \partial_{\beta} h_{\alpha \gamma} h_{\beta \gamma}
  + \frac{1}{2} \partial_{\mu} h_{\mu \nu} \partial_{\nu} h_{\alpha \alpha} h_{\beta \gamma} h_{\beta \gamma}
  - 2 \partial_{\mu} h_{\mu \nu} \partial_{\nu} h_{\alpha \beta} h_{\alpha \gamma} h_{\beta \gamma}
\\ &  - \frac{1}{2} \partial_{\mu} h_{\mu \nu} \partial_{\alpha} h_{\nu \alpha} h_{\beta \gamma} h_{\beta \gamma}
  + 4 \partial_{\mu} h_{\mu \nu} \partial_{\alpha} h_{\nu \beta} h_{\alpha \gamma} h_{\beta \gamma}
  + \frac{1}{3} \partial_{\mu} \partial_{\mu} h_{\nu \nu} h_{\alpha \beta} h_{\alpha \gamma} h_{\beta \gamma}
\\ &  + \frac{1}{2} \partial_{\mu} \partial_{\mu} h_{\nu \alpha} h_{\nu \alpha} h_{\beta \gamma} h_{\beta \gamma}
  - 2 \partial_{\mu} \partial_{\mu} h_{\nu \alpha} h_{\nu \beta} h_{\alpha
    \gamma} h_{\beta \gamma}  \Big )    \, . \numberthis\label{eq:4Hb}&&
\end{flalign*}      
\item The ghost propagator:
\begin{flalign*}
  \mathcal{L}_{\bar{\chi}\chi} = &
        \bar{\chi}^{\mu} \partial^{\nu}\partial_{\nu} \chi_{\mu}   \, . \numberthis\label{eq:2ghb}&&
\end{flalign*}      
\item The ghost-ghost-graviton vertex:
\begin{flalign*}
  \mathcal{L}_{\bar{\chi}\chi h} = 
\; \kappa \Big ( & - \frac{1}{2} \bar{\chi}^{\mu} \partial_{\mu} \chi^{\nu} \partial_{\nu}  h_{\alpha}^{\;\;\alpha}
       + \bar{\chi}^{\mu} \partial_{\mu} \chi^{\nu} \partial^{\alpha} h_{\nu\alpha}
  - \frac{1}{2} \bar{\chi}^{\mu} \chi^{\nu} \partial_{\mu}\partial_{\nu}h_{\alpha}^{\;\;\alpha}
    + \bar{\chi}^{\mu} \chi^{\nu} \partial_{\nu}\partial^{\alpha}h_{\mu\alpha} \nonumber\\
      & + \bar{\chi}^{\mu} \partial^{\nu}\partial_{\nu}\chi^{\alpha} h_{\mu\alpha}
       - \bar{\chi}^{\mu} \partial^{\nu}\chi^{\alpha} \partial_{\mu}h_{\nu\alpha}
       + \bar{\chi}^{\mu} \partial^{\nu}\chi^{\alpha} \partial_{\nu}h_{\mu\alpha}
       + \bar{\chi}^{\mu} \partial^{\nu}\chi^{\alpha}
       \partial_{\alpha}h_{\mu\nu}  \Big )  \, . \numberthis\label{eq:2gh1Hb}&&
\end{flalign*}
\item The ghost-ghost-graviton-graviton vertex vanishes.
\end{itemize}

\clearpage
\section{Kinematics}\label{Ap:Kinematics}
\setcounter{equation}{0}

\subsection[Scalar-Graviton Scattering]{Scalar-Graviton Scattering ($\phi(p_1) h_{\mu\nu}(p_2) \rightarrow \phi(p_3) h_{\alpha\beta} (p_4)$):}
There are four diagrams for this process as shown in
Fig.~\ref{fig:TreeLevelScalarGraviton}, three of them (a, b, c) representing s,
t and u-channels respectively. According to the conventions that we use in this
thesis, the Mandelstam variables are given by
\begin{align}
  s&=(p_1+p_2)^2=(p_3+p_4)^2 \, , \nonumber\\
  t&=(p_1-p_3)^2=(p_2-p_4)^2 \, , \nonumber \\
  u&=(p_1-p_4)^2=(p_2-p_3)^2  \, , \nonumber\\
  s+t+u&=\sum_i M_i^2=2m^2 \, , \label{eq:MandelstamVariables}
\end{align}
where $m$ is the mass of scalar field $\phi$.


\indent In the \acrshort{cm} frame, with the incoming particles along the z-axis, the momenta and the polarization vector can be chosen as:
\begin{align}
  p_1 &=(E,0,0,-k) \, , \nonumber\\
  p_2&=(k,0,0,k)   \, , \label{eq:IncomingScalarGraviton}\\
  \epsilon_{\mu}^{\pm 1}(p_2) &=
                                \begin{pmatrix}
                                  0,\frac{1}{\sqrt{2}},\pm \frac{ i}{\sqrt{2}},0
                                \end{pmatrix} \, . \nonumber
\end{align}
Choosing the outgoing particles to be in the yz-plane, the momenta and the complex conjugate polarization vector are:
\begin{align}
  p_3 &=(E,0,-k \sin(\theta),-k \cos(\theta)) \, , \nonumber \\
  p_4&=(k,0,k \sin(\theta),k \cos(\theta)) \, ,  \label{eq:OutgoingScalarGraviton}\\
  \epsilon_{\alpha}^{* \pm 1}(p_4) &= \Big (0,\frac{1}{\sqrt{2}},\frac{\mp i \cos(\theta) }{\sqrt{2}},\frac{\pm i \sin(\theta) }{\sqrt{2}} \Big) \, , \nonumber
\end{align}
where $\theta$ is the scattering angle and $\epsilon_{\alpha}^{\pm 1}(p_4) = R^{\;\;\mu}_{\alpha}(\hat{p}_4) \; \epsilon_{\mu}^{\pm 1}(p_2)$ where $R^{\;\;\mu}_{\alpha}(\hat{p}_4)$ is a rotation.
It is also possible to write the graviton polarization directly as a tensor:
\begin{align}
    \epsilon_{\mu\nu}^{\pm 2}(p_2) &=
            \begin{pmatrix}
              0 & 0 & 0 & 0 \\
              0 & \frac{1}{2} & \pm \frac{i}{2} & 0 \\
              0 & \pm \frac{i}{2} & -\frac{1}{2} & 0 \\
              0 & 0 & 0 & 0 \\                 
            \end{pmatrix} \, , \\[5mm]
    \epsilon_{\alpha\beta}^{* \pm 2}(p_4) &=
            \begin{pmatrix}
              0 & 0 & 0 & 0 \\
              0 & \frac{1}{2} & \mp \frac{i}{2}\cos(\theta) & \pm \frac{i}{2}\sin(\theta)  \\
              0 & \mp \frac{i}{2}\cos(\theta) & -\frac{1}{2}\cos^2(\theta) & \frac{1}{2}\sin(\theta)\cos(\theta) \\
              0 & \pm \frac{i}{2}\sin(\theta) &  \frac{1}{2}\sin(\theta)\cos(\theta) & -\frac{1}{2}\sin^2(\theta) \\[3mm]                 
            \end{pmatrix} \, ,
\end{align}
where $\epsilon_{\alpha\beta}^{\pm 2}(p_4) = R^{\;\;\mu}_{\alpha}(\hat{p}_4) \;
R^{\;\;\nu}_{\beta}(\hat{p}_4) \; \epsilon_{\mu\nu}^{\pm 2}(p_2)$ where again
$R^{\;\;\mu}_{\alpha}(\hat{p}_4), \, R^{\;\;\nu}_{\beta}(\hat{p}_4)$ are
rotation matrices.
Finally, the following useful relations are valid in the \acrshort{cm} frame:
\begin{align}
  p_1^{\mu} \epsilon_{\mu}^{\pm 1} (p_2)&=0 \, , \\
  p_3^{\mu} \epsilon_{\mu}^{\pm 1} (p_4)&=0 \, .
\end{align}

\subsection[Graviton-Graviton Scattering]{Graviton-Graviton Scattering ($h^{\mu\nu}(p_1) h^{\alpha\beta}(p_2) \rightarrow h^{\gamma\delta}(p_3) h^{\lambda\rho}(p_4) $):}
There are also four diagrams for this process as shown in
Fig.~\ref{fig:TreeLevelGravitonGraviton}, but the relation between Mandelstam
variables for this process is now
\begin{align}
  s+t+u&=0 \, .
\end{align}


In the \acrshort{cm} frame, with the incoming particles, along the z-axis, the momenta and the
polarization vectors can be chosen as:
\begin{align}
  p_1 &=(k,0,0,k) \, , \nonumber\\
  p_2&=(k,0,0,-k) \, , \nonumber\\
  \epsilon_{\mu}^{\pm 1}(p_1) &=
    \begin{pmatrix}
      0,\frac{1}{\sqrt{2}},\pm \frac{ i}{\sqrt{2}},0
    \end{pmatrix} \, ,   \label{eq:IncomingGravitonGraviton}\\
  \epsilon_{\alpha}^{\pm 1}(p_2) &=
    \begin{pmatrix}
      0,\frac{1}{\sqrt{2}},\mp \frac{ i}{\sqrt{2}},0
    \end{pmatrix} \, . \nonumber
\end{align}
Choosing the outgoing particles to be in the yz-plane, the momenta and the complex
conjugate polarization vectors are:
\begin{align}
  p_3 &=(k,0,k \sin(\theta),k \cos(\theta)) \, , \nonumber\\
  p_4&=(k,0,-k \sin(\theta),-k \cos(\theta)) \, , \nonumber\\
  \epsilon_{\gamma}^{* \pm 1}(p_3) &= \Big (0,\frac{1}{\sqrt{2}},\frac{\mp i \cos(\theta) }{\sqrt{2}},\frac{\pm i \sin(\theta) }{\sqrt{2}} \Big) \, ,   \label{eq:OutgoingGravitonGraviton}\\
  \epsilon_{\lambda}^{* \pm 1}(p_4) &= \Big (0,\frac{1}{\sqrt{2}},\frac{\pm i \cos(\theta) }{\sqrt{2}},\frac{\mp i \sin(\theta) }{\sqrt{2}} \Big) \, , \nonumber
\end{align}
where $\theta$ is the scattering angle.

\noindent Finally, the following useful relations are valid in the \acrshort{cm} frame:
\begin{align}
  p_1^{\mu} \epsilon_{\mu}^{\pm 1} (p_2)&=0 \, , \nonumber\\
  p_3^{\mu} \epsilon_{\mu}^{\pm 1} (p_4)&=0 \, , \\
  p_4^{\mu} \epsilon_{\mu}^{\pm 1} (p_1)= - p_3^{\mu} \epsilon_{\mu}^{\pm 1} (p_1)  \qquad &, \qquad   p_4^{\mu} \epsilon_{\mu}^{\pm 1} (p_2)= - p_3^{\mu} \epsilon_{\mu}^{\pm 1} (p_2) \, , \nonumber\\
  p_2^{\mu} \epsilon_{\mu}^{\pm 1} (p_3)= - p_1^{\mu} \epsilon_{\mu}^{\pm 1} (p_3)   \qquad &, \qquad   p_2^{\mu} \epsilon_{\mu}^{\pm 1} (p_4)= - p_1^{\mu} \epsilon_{\mu}^{\pm 1} (p_4) \, . \nonumber
\end{align}

\clearpage

\section{Dimensional Regularization of Scalar Integrals}\label{Ap:DimensionalRegularization}
\setcounter{equation}{0}
To illustrate how the scalar integrals are calculated in dimensional
regularization, let us take as an example the scalar integral $B_0$
Eq.~(\ref{eq:BZero}) and then follow the standard procedure of dimensional regularization \cite{Peskin}.
First, we move the loop integral from Minkowski space to Euclidean space by
replacing the zeroth component of momentum in Minkowski space \(k_0\) by the
imaginary fourth component in Euclidean space $ik_4$. As a result of this replacement $k_0 \rightarrow ik_4$:
\begin{align}
  k^2 = k_0^2 - \vec{k}^2 \;\; &\rightarrow \;\; - k^2_E = - k_4^2 - \vec{k}^2 \, , \\
  d^4 k \;\;&\rightarrow \;\; i d^4 k_E\, .  
\end{align}
Then, the scalar integral $B_0$ for $m_0=m_1=m$ and $q_1=p_1$ becomes
\begin{align}
   B_0(p_1,m,m) =&  \int \frac{i d^d k_E}{(2\pi)^d}
         \frac{1}{(-k^2_E-{m}^2+i\epsilon)(-(k+p_1)^2_E-{m}^2+i\epsilon)} \nonumber \\[5mm]
   =&  \int \frac{i d^d k_E}{(2\pi)^d}
\frac{1}{(k^2_E+{m}^2-i\epsilon)((k+p_1)^2_E+{m}^2-i\epsilon)} \, .
\end{align}
Second, since the integration over the fourth component of the momentum goes along the
imaginary axis $ik_4$, we need to perform a Wick rotation to move to the
integration along the real axis, making sure that the contour does not cross the poles of $\frac{1}{(k^2_E+{m}^2-i\epsilon)((k+p_1)^2_E+{m}^2-i\epsilon)} $.
Third, in order to transform the product of several brackets in the denominator into a single bracket, we use the Feynman parameterization
\begin{align}
\frac{1}{A_1^{\alpha_1} A_2^{\alpha_2} \cdots A_n^{\alpha_n}} &= \frac{\Gamma(\alpha_1 + \alpha_2 + \cdots + \alpha_n)}{\Gamma(\alpha_1) \Gamma(\alpha_2) \cdots \Gamma(\alpha_n) } \\[5mm]
& \qquad  \int dx_1 dx_2 \cdots dx_n \frac{\delta(1-x_1 - x_2 - \cdots - x_n ) x_1^{\alpha_1 - 1} x_2^{\alpha_2 - 1} \cdots x_n^{\alpha_n - 1}} {[A_1 x_1 + A_2 x_2 + \cdots + A_n x_n]^{\alpha_1 + \alpha_2 + \cdots + \alpha_n}} \nonumber \, ,
\end{align}
where $\Gamma$ is the gamma function.

\noindent In our case, \(\alpha_1=\alpha_2=1, n=2\), and the integral becomes
\begin{align}
  B_0 & = \frac{i}{(2 \pi)^d} \int d^d k_E \; \frac{1}{k^2_E+ m^2-i\epsilon} \; \frac{1}{(k+p_1)^2_E + m^2-i\epsilon} \nonumber \\
&  = \frac{i}{(2 \pi)^d} \int d^d k_E \; \frac{\Gamma(2)}{\Gamma(1) \Gamma(1)}  \int_{0}^{1} \frac{dx_1 dx_2 \; \delta(1-x_1-x_2)}{[[k^2_E+ m^2-i\epsilon] x_1 + [(k+p_1)^2_E + m^2-i\epsilon] x_2]^2} \nonumber \\
      &  =  \frac{i}{(2 \pi)^d} \int d^d k_E \; \int_{0}^{1} \frac{dx}{[k^2_E + 2 (k \cdot p_{1})_E\, x + p^2_{1E} \, x + m^2-i\epsilon]^2} \, ,
\end{align}
where $\Gamma(1)=\Gamma(2)=1$, and from the second line to the
third we used the delta function to do the integration over $x_1$ and finally renamed $x_2$ to $x$.

\noindent Fourth, to complete the square of $k_E$ in the denominator, we need to shift
the integration variable $k_E \rightarrow k_E - p_{1E} \, x$  to absorb the term $(k \cdot p_{1})_E$ and obtain
\begin{align}
  B_0  = \frac{i}{(2 \pi)^d}  \; \int_{0}^{1} dx \int  \frac{d^d k_E}{[k^2_E + p^2_{1E}\, x (1-x) + m^2-i\epsilon]^2}  = \frac{i}{(2 \pi)^d}  \; \int_{0}^{1} dx \int  \frac{d^d k_E}{[k^2_E + M^2-i\epsilon]^2}  \nonumber \, ,
\end{align}
where $ M^2 = p^2_{1E} \, x (1-x) + m^2$  .

\noindent Fifth, we perform the momentum integral in Euclidean space, using
spherical coordinates as follows
\begin{align}
  B_0 & = \frac{i}{(2 \pi)^d}  \; \int_{0}^{1} dx \; \Omega_d  \int_{0}^{\infty} d k_E \;  k_E \;   \frac{ (k^2_E)^{d/2 -1}}{[k^2_E + M^2]^2} \nonumber \\[2mm]
      &  = \frac{i}{(2 \pi)^d}  \; \int_{0}^{1} dx \; \Omega_d  \int_{0}^{\infty} \frac{d k^2_E}{2} \;  \frac{ (k^2_E)^{d/2 -1}}{[k^2_E + M^2]^2} \nonumber \\[2mm]
      & = \frac{i}{(2 \pi)^d}  \; \int_{0}^{1} dx \; \frac{\Omega_d}{2} \; (M^2)^{d/2 -2} \int_{0}^{\infty}  \frac{d x \; x^{d/2 -1}}{ [x+1]^2} \nonumber  \\[2mm]
      & = \frac{i}{(4 \pi)^{d/2}}  \; \int_{0}^{1} dx  \; (M^2)^{d/2 -2} \; \Gamma(2 - d/2) \label{eq:IntegralPerformed} \, ,
\end{align}
where $\Omega_d = \frac{2 \pi^{d/2}}{\Gamma(d/2)}$ follows from the
angular integration, and from the second line to the third $k^2_E \rightarrow
k^2_E\, M^2$ is used, and then $k^2_E \rightarrow x$. However, we note that the
gamma function $\Gamma(2-d/2)$ still diverges in the last result if we are in 4-dimensions.

\noindent Sixth, we transform back into Minkowski space by replacing again
\(d = 4 - 2 \epsilon\) and doing an expansion in $\epsilon$ using the following relation \cite{Peskin}:
  \begin{equation}
    \frac{\Gamma(2 - d/2)}{(4 \pi)^{d/2}} \; \bigg ( \frac{1}{M^2} \bigg )^{2-d/2} = \frac{1}{16 \pi^2} \Bigg ( \frac{1}{\epsilon} - \gamma  + \log(4 \pi) - \log(M^2) + \mathcal{O}(\epsilon) \Bigg )    \, .
  \end{equation}
\noindent Finally, we get
\begin{equation}
  \label{eq:BZeroFinal}
  B_0(p_1,m,m) = \frac{i}{16 \pi^2}  \; \Bigg ( \frac{1}{\epsilon} - \gamma  + \log(4 \pi) - \int_{0}^{1} dx \log[ m^2-p^2_1 x (1-x)]  \Bigg ) \, , 
\end{equation}
where $ M^2 = p^2_{1E} \, x (1-x) + m^2= - p^2_{1} \, x (1-x) +
m^2$. The integral still blows up for \(\epsilon \rightarrow 0\) but now the
divergent part is separated from the finite terms. Similarly, we can calculate the other scalar integrals.

\clearpage

\section{Passarino-Veltman Reduction of Tensor Integrals}\label{Ap:PassarinoVeltman}
\setcounter{equation}{0}
In the Passarino-Veltman method, the tensorial integrals can be written in terms of
scalar functions as listed below. However, we only include the cases that are
relevant to the scalar-graviton scattering to one-loop order (i.e., $A$ and $D$
integrals up to four indices, $B$ integrals up to five indices and $C$ integrals
up to six indices in the numerator). In addition, all scalar
functions $C,D$ are symmetric under $i,j,k,l,m,n$ indices in our notation below (e.g., $C_{21}=C_{12}$), and $i\epsilon$ was dropped for simplicity:
\begin{alignat*}{3}
  &A  && = \int \frac{d^dk}{(2\pi)^d} \frac{1}{(k^2 - m^2_0)} = A_0(m_0)
  \numberthis  \label{eq:AIntegrals}  \, , \\[4mm]
  &A^{\mu}  && = \int \frac{d^dk}{(2\pi)^d} \frac{k^{\mu}}{(k^2 - m^2_0)} = 0 \numberthis \, ,   \\[4mm]
 &A^{\mu\nu}  && = \int \frac{d^dk}{(2\pi)^d} \frac{k^{\mu}k^{\nu}}{(k^2 -
   m^2_0)} = \eta^{\mu\nu} A_{00}(m_0) \numberthis  \, ,  \\[4mm]
 &A^{\mu\nu\alpha}  && = \int \frac{d^dk}{(2\pi)^d}
 \frac{k^{\mu}k^{\nu}k^{\alpha}}{(k^2 - m^2_0)} = 0 \numberthis  \, ,  \\[4mm]
 &A^{\mu\nu\alpha\beta}  && = \int \frac{d^dk}{(2\pi)^d} \frac{k^{\mu}k^{\nu}k^{\alpha}k^{\beta}}{(k^2 - m^2_0)} = \Big [ \eta^{\mu \nu} \eta^{\alpha
  \beta}+\eta^{\mu \alpha} \eta^{\nu \beta}+\eta^{\mu \beta} \eta^{\nu \alpha}
\Big ] A_{0000}(m_0) \numberthis \label{eq:AAIntegrals} \, , \\[4mm]
&B  && = \int \frac{d^dk}{(2\pi)^d} \frac{1}{(k^2 - m^2_0) ((k + q_1)^2- m^2_1)}
=  B_0(p_1,m_0,m_1) \numberthis \label{eq:BIntegrals}  \, , \\[4mm]
& B^{\mu}  && = \int \frac{d^dk}{(2\pi)^d} \frac{k^{\mu}}{(k^2 - m^2_0)((k + q_1)^2- m^2_1)} = p_{1}^{\mu} \,
B_{1}(p_1,m_0,m_1) \numberthis  \, ,  \\[4mm]
& B^{\mu\nu} && = \int \frac{d^dk}{(2\pi)^d} \frac{k^{\mu} k^{\nu}}{(k^2-
  m^2_0)((k + q_1)^2- m^2_1)} \numberthis \\
& && = \Big [ p_{1}^{\mu} \, p_{1}^{\nu} \Big ] B_{11}(p_1,m_0, m_1)+\eta^{\mu \nu} B_{00}(p_1,m_0,m_1)  \, ,  \\[4mm]  
& B^{\mu\nu\alpha} && = \int \frac{d^dk}{(2\pi)^d} \frac{k^{\mu} k^{\nu}
  k^{\alpha}}{(k^2- m^2_0) ((k + q_1)^2- m^2_1)} \numberthis \\
& && = \Big [ p_{1}^{\mu}\, p_{1}^{\nu}\, p_{1}^{\alpha} \Big]\, B_{111}(p_1,m_0,m_1) + \Big  [\eta^{\mu \nu}
p_{1}^{\alpha}+\eta^{\mu \alpha} p_{1}^{\nu}+\eta^{\nu \alpha} p_{1}^{\mu} \Big] B_{001}(p_1,m_0,m_1)  \, , \\[4mm]
& B^{\mu\nu\alpha\beta}  && = \int \frac{d^dk}{(2\pi)^d} \frac{k^{\mu} k^{\nu}
  k^{\alpha} k^{\beta}}{(k^2 - m^2_0)((k + q_1)^2 - m^2_1)} \numberthis \label{eq:BBIntegrals} \\
\end{alignat*}
\begin{alignat*}{3}
& && =\;\;\; \Big [ p_{1}^{\mu} p_{1}^{\nu} p_{1}^{\alpha} p_{1}^{\beta}\Big ]
B_{1111}(p_1,m_0,m_1) + \Big [ \eta^{\mu \nu} \eta^{\alpha
  \beta}+\eta^{\mu \alpha} \eta^{\nu \beta} +\eta^{\mu \beta} \eta^{\nu \alpha}
\Big ] B_{0000}(p_1,m_0,m_1)  \\
& && \quad  + \Big [ \eta^{\mu \nu} p_{1}^{\alpha} p_{1}^{\beta} +
\eta^{\mu \alpha} p_{1}^{\nu} p_{1}^{\beta}  + \eta^{\nu \alpha} p_{1}^{\mu} p_{1}^{\beta}
 + \eta^{\beta \nu} p_{1}^{\alpha}
p_{1}^{\mu} +\eta^{\beta \alpha} p_{1}^{\nu} p_{1}^{\mu}+\eta^{\mu \beta} p_{1}^{\nu}
p_{1}^{\alpha} \Big ] B_{0011}(p_1,m_0,m_1)  \, ,  \\[5mm]
& B^{\mu\nu\alpha\beta\rho}&& = \int \frac{d^dk}{(2\pi)^d} \frac{k^{\mu} k^{\nu}
  k^{\alpha} k^{\beta} k^{\rho}}{(k^2 - m^2_0)((k + q_1)^2 - m^2_1)} \numberthis \\
& && = \;  p^{\mu}_{1} p^{\nu}_{1} p^{\alpha}_{1}
p^{\beta}_{1} p^{\rho}_{1}
\; B_{11111}(p_1,m_0,m_1) \\
& && \quad + \Big [
\eta^{\mu \nu} p^{\alpha}_1 p^{\beta}_1p^{\rho}_1 + \eta^{\mu \alpha} p^{\nu}_1 p^{\beta}_1 p^{\rho}_1 +
\eta^{\nu \alpha} p^{\mu}_1 p^{\beta}_1 p^{\rho}_1 + \eta^{\beta \nu} p^{\alpha}_1p^{\mu}_1 p^{\rho}_1 + \eta^{\beta \alpha} p^{\nu}_1 p^{\mu}_1 p^{\rho}_1 +\eta^{\mu \beta} p^{\nu}_1
p^{\alpha}_1 p^{\rho}_1 \\
& && \qquad\qquad\;\; + \eta^{\rho\mu} p^{\alpha}_1 p^{\beta}_1p^{\nu}_1 +
\eta^{\rho \nu} p^{\alpha}_1 p^{\beta}_1p^{\mu}_1 + \eta^{\rho \alpha} p^{\nu}_1 p^{\beta}_1 p^{\mu}_1
+\eta^{\rho \beta} p^{\nu}_1p^{\alpha}_1 p^{\mu}_1 \Big ] B_{00111}(p_1,m_0,m_1) \\
& && \quad +  \Big [ \eta^{\mu \nu} \eta^{\alpha
  \beta} p^{\rho}_1 +\eta^{\mu \alpha} \eta^{\nu \beta} p^{\rho}_1+\eta^{\mu
  \beta} \eta^{\nu \alpha} p^{\rho}_1 +
\eta^{\rho \nu} \eta^{\alpha\beta} p^{\mu}_1 +\eta^{\rho \alpha} \eta^{\nu
  \beta} p^{\mu}_1+\eta^{\rho\beta} \eta^{\nu \alpha} p^{\mu}_1 \\
& && \qquad\qquad + \eta^{\mu \rho} \eta^{\alpha\beta} p^{\nu}_1 +\eta^{\mu \alpha} \eta^{\rho \beta} p^{\nu}_1+\eta^{\mu\beta} \eta^{\rho \alpha} p^{\nu}_1 +
\eta^{\mu\nu} \eta^{\rho\beta} p^{\alpha}_1 +\eta^{\mu \rho} \eta^{\nu \beta}
p^{\alpha}_1+\eta^{\mu\beta} \eta^{\nu \rho} p^{\alpha}_1  \\
& && \qquad\qquad + \eta^{\mu \nu} \eta^{\alpha\rho} p^{\beta}_1 +\eta^{\mu \alpha} \eta^{\nu \rho} p^{\beta}_1+\eta^{\mu\rho} \eta^{\nu \alpha} p^{\beta}_1
\Big ] B_{00001}(p_1,m_0,m_1)  \, , \\[5mm]
&C \quad &&= \int \frac{d^dk}{(2\pi)^d} \frac{1}{(k^2 - m^2_0) ((k + q_1)^2-
  m^2_1)((k + q_2)^2- m^2_2)} \numberthis \label{eq:CIntegrals} \\
& && =  C_0(p_1,p_2,m_0,m_1,m_2)  \, ,  \\[5mm]
&\mathcal{C}^{\mu} &&= \int \frac{d^dk}{(2\pi)^d} \frac{k^{\mu}}{(k^2 - m^2_0)((k +
  q_1)^2- m^2_1)((k + q_2)^2- m^2_2)} \numberthis \\
& && = \sum_{i=1}^2 p_{i}^{\mu} \, C_i(p_1,p_2,m_0,m_1,m_2)  \, , \\[5mm]
&C^{\mu\nu} &&= \int \frac{d^dk}{(2\pi)^d} \frac{k^{\mu} k^{\nu}}{(k^2-
  m^2_0)((k + q_1)^2- m^2_1)((k + q_2)^2- m^2_2)}  \numberthis \\
& &&= \sum_{i,j=1}^2   p_{i}^{\mu} \, p_{j}^{\nu} \, C_{ij}(p_1,p_2,m_0,m_1,m_2) + \eta^{\mu \nu} C_{00}(p_1,p_2,m_0,m_1,m_2)  \, , \\[5mm]
& C^{\mu\nu\alpha}&& = \int \frac{d^dk}{(2\pi)^d} \frac{k^{\mu} k^{\nu}
  k^{\alpha}}{(k^2- m^2_0) ((k + q_1)^2- m^2_1)((k + q_2)^2- m^2_2)} \numberthis  \\
& &&= \; \sum_{i,j,k=1}^2  p^{\mu}_{i} \, p^{\nu}_{j} \, p^{\alpha}_{k} \;
C_{ijk}(p_1,p_2,m_0,m_1,m_2)  \\
& && \quad + \sum_{i=1}^2  \Big  [\eta^{\mu \nu} p^{\alpha}_{i}+\eta^{\mu \alpha}
p^{\nu}_{i} +\eta^{\nu \alpha} p^{\mu}_i \Big] C_{00i}(p_1,p_2,m_0,m_1,m_2) \, , 
\end{alignat*}

\begin{alignat*}{3}
& C^{\mu\nu\alpha\beta}&& = \int \frac{d^dk}{(2\pi)^d} \frac{k^{\mu} k^{\nu}
  k^{\alpha} k^{\beta}}{(k^2 - m^2_0)((k + q_1)^2 - m^2_1)((k + q_2)^2 - m^2_2)} \numberthis
\label{eq:CCIntegrals}\\
& && = \; \sum_{i,j,k,l=1}^2  p^{\mu}_{i} p^{\nu}_{j} p^{\alpha}_{k} p^{\beta}_{l}
\; C_{ijkl}(p_1,p_2,m_0,m_1,m_2) \\
& && \quad + \sum_{i,j=1}^2 \Big [ \eta^{\mu \nu} p^{\alpha}_i p^{\beta}_j + \eta^{\mu \alpha} p^{\nu}_i p^{\beta}_j +
\eta^{\nu \alpha} p^{\mu}_i p^{\beta}_j + \eta^{\beta \nu} p^{\alpha}_i
p^{\mu}_j + \eta^{\beta \alpha} p^{\nu}_i p^{\mu}_j \\
& && \qquad\qquad\qquad\qquad\qquad\qquad\qquad +\eta^{\mu \beta} p^{\nu}_i p^{\alpha}_j \Big ] C_{00ij}(p_1,p_2,m_0,m_1,m_2) \\
& && \quad + \Big [ \eta^{\mu \nu} \eta^{\alpha
  \beta}  +\eta^{\mu \alpha} \eta^{\nu \beta}+\eta^{\mu \beta} \eta^{\nu \alpha}
\Big ] C_{0000}(p_1,p_2,m_0,m_1,m_2)  \, ,  \\[4mm]
& C^{\mu\nu\alpha\beta\rho}&& = \int \frac{d^dk}{(2\pi)^d} \frac{k^{\mu} k^{\nu}
  k^{\alpha} k^{\beta} k^{\rho}}{(k^2 - m^2_0)((k + q_1)^2 - m^2_1)((k + q_2)^2 - m^2_2)}\\
& && = \; \sum_{i,j,k,l,m=1}^2  p^{\mu}_{i} p^{\nu}_{j} p^{\alpha}_{k}
p^{\beta}_{l} p^{\rho}_{m}
\; C_{ijklm}(p_1,p_2,m_0,m_1,m_2) \\
& && \quad + \sum_{i,j,k=1}^2 \Big [
\eta^{\mu \nu} p^{\alpha}_i p^{\beta}_jp^{\rho}_k + \eta^{\mu \alpha} p^{\nu}_i p^{\beta}_j p^{\rho}_k +
\eta^{\nu \alpha} p^{\mu}_i p^{\beta}_j p^{\rho}_k + \eta^{\beta \nu} p^{\alpha}_ip^{\mu}_j p^{\rho}_k + \eta^{\beta \alpha} p^{\nu}_i p^{\mu}_j p^{\rho}_k +\eta^{\mu \beta} p^{\nu}_i
p^{\alpha}_j p^{\rho}_k \\
& && \qquad\qquad\;\; + \eta^{\rho\mu} p^{\alpha}_i p^{\beta}_jp^{\nu}_k +
\eta^{\rho \nu} p^{\alpha}_i p^{\beta}_jp^{\mu}_k + \eta^{\rho \alpha} p^{\nu}_i p^{\beta}_j p^{\mu}_k
+\eta^{\rho \beta} p^{\nu}_ip^{\alpha}_j p^{\mu}_k \Big ] C_{00ijk}(p_1,p_2,m_0,m_1,m_2) \\
& && \quad + \sum_{i=1}^2 \Big [ \eta^{\mu \nu} \eta^{\alpha
  \beta} p^{\rho}_i +\eta^{\mu \alpha} \eta^{\nu \beta} p^{\rho}_i+\eta^{\mu
  \beta} \eta^{\nu \alpha} p^{\rho}_i +
\eta^{\rho \nu} \eta^{\alpha\beta} p^{\mu}_i +\eta^{\rho \alpha} \eta^{\nu
  \beta} p^{\mu}_i+\eta^{\rho\beta} \eta^{\nu \alpha} p^{\mu}_i \\
& && \qquad\qquad + \eta^{\mu \rho} \eta^{\alpha\beta} p^{\nu}_i +\eta^{\mu \alpha} \eta^{\rho \beta} p^{\nu}_i+\eta^{\mu\beta} \eta^{\rho \alpha} p^{\nu}_i +
\eta^{\mu\nu} \eta^{\rho\beta} p^{\alpha}_i +\eta^{\mu \rho} \eta^{\nu \beta}
p^{\alpha}_i+\eta^{\mu\beta} \eta^{\nu \rho} p^{\alpha}_i  \\
& && \qquad\qquad + \eta^{\mu \nu} \eta^{\alpha\rho} p^{\beta}_i +\eta^{\mu \alpha} \eta^{\nu \rho} p^{\beta}_i+\eta^{\mu\rho} \eta^{\nu \alpha} p^{\beta}_i
\Big ] C_{0000i}(p_1,p_2,m_0,m_1,m_2)  \, , \\[4mm]
& C^{\mu\nu\alpha\beta\rho\sigma}&& = \int \frac{d^dk}{(2\pi)^d} \frac{k^{\mu} k^{\nu}
  k^{\alpha} k^{\beta} k^{\rho} k^{\sigma}}{(k^2 - m^2_0)((k + q_1)^2 - m^2_1)((k + q_2)^2 - m^2_2)}\\
& && = \; \sum_{i,j,k,l,m,n=1}^2  p^{\mu}_{i} p^{\nu}_{j} p^{\alpha}_{k}
p^{\beta}_{l} p^{\rho}_{m} p^{\sigma}_{n}  \; C_{ijklmn}(p_1,p_2,m_0,m_1,m_2) \\
& && \quad + \sum_{i,j,k,l=1}^2 \Big [
\eta^{\mu \nu} p^{\alpha}_i p^{\beta}_jp^{\rho}_k p^{\sigma}_{l} + \eta^{\mu \alpha} p^{\nu}_i p^{\beta}_j p^{\rho}_kp^{\sigma}_{l} +\eta^{\nu \alpha} p^{\mu}_i p^{\beta}_j p^{\rho}_k p^{\sigma}_{l} + \eta^{\beta
  \nu} p^{\alpha}_ip^{\mu}_j p^{\rho}_k p^{\sigma}_{l} + \eta^{\beta \alpha}
p^{\nu}_i p^{\mu}_j p^{\rho}_k p^{\sigma}_{l} \\
& && \qquad\qquad\;\; +\eta^{\mu \beta} p^{\nu}_i p^{\alpha}_j p^{\rho}_k p^{\sigma}_{l}
 + \eta^{\rho\mu} p^{\alpha}_i p^{\beta}_jp^{\nu}_kp^{\sigma}_{l} + \eta^{\rho \nu} p^{\alpha}_i p^{\beta}_jp^{\mu}_k
p^{\sigma}_{l} + \eta^{\rho \alpha} p^{\nu}_i p^{\beta}_j p^{\mu}_k
p^{\sigma}_{l}+\eta^{\rho \beta} p^{\nu}_ip^{\alpha}_j p^{\mu}_k p^{\sigma}_{l}\\
& && \qquad\qquad\;\;+ \eta^{\sigma \mu} p^{\alpha}_i p^{\beta}_jp^{\rho}_k p^{\nu}_{l} +  \eta^{\sigma
  \nu} p^{\alpha}_i p^{\beta}_jp^{\rho}_k p^{\mu}_{l} +  \eta^{\sigma \alpha}
p^{\nu}_i p^{\beta}_jp^{\rho}_k p^{\mu}_{l} +  \eta^{\sigma \beta} p^{\alpha}_i
p^{\nu}_jp^{\rho}_k p^{\mu}_{l} \\
& && \qquad\qquad\;\; + \eta^{\sigma \rho} p^{\alpha}_i p^{\beta}_jp^{\nu}_k p^{\mu}_{l} 
\Big ] C_{00ijkl}(p_1,p_2,m_0,m_1,m_2) \\
\end{alignat*}
\begin{alignat*}{3}
& && \quad + \sum_{i,j=1}^2 \Big [
\eta^{\mu \nu} \eta^{\alpha\beta} p^{\rho}_i p^{\sigma}_{j} +\eta^{\mu \alpha}
\eta^{\nu \beta} p^{\rho}_ip^{\sigma}_{j}+\eta^{\mu\beta} \eta^{\nu \alpha}
p^{\rho}_i p^{\sigma}_{j}
+\eta^{\rho \nu} \eta^{\alpha\beta} p^{\mu}_i p^{\sigma}_{j} +\eta^{\rho \alpha} \eta^{\nu
  \beta} p^{\mu}_ip^{\sigma}_{j} \\
& && \qquad\qquad\;\; +\eta^{\rho\beta} \eta^{\nu \alpha}
p^{\mu}_ip^{\sigma}_{j}  + \eta^{\mu \rho} \eta^{\alpha\beta} p^{\nu}_i
p^{\sigma}_{j} +\eta^{\mu \alpha} \eta^{\rho \beta} p^{\nu}_i p^{\sigma}_{j} +\eta^{\mu\beta}
\eta^{\rho \alpha} p^{\nu}_ip^{\sigma}_{j} + \eta^{\mu\nu} \eta^{\rho\beta}
p^{\alpha}_i p^{\sigma}_{j} \\
& && \qquad\qquad\;\;+\eta^{\mu \rho} \eta^{\nu \beta}
p^{\alpha}_i p^{\sigma}_{j}+\eta^{\mu\beta} \eta^{\nu \rho} p^{\alpha}_i
p^{\sigma}_{j}  + \eta^{\mu \nu} \eta^{\alpha\rho} p^{\beta}_ip^{\sigma}_{j}
+\eta^{\mu \alpha} \eta^{\nu \rho} p^{\beta}_ip^{\sigma}_{j}+\eta^{\mu\rho}
\eta^{\nu \alpha} p^{\beta}_ip^{\sigma}_{j} \\
& && \qquad\qquad\;\; + \eta^{\sigma \nu} \eta^{\alpha\beta} p^{\rho}_i p^{\mu}_{j} +\eta^{\sigma \alpha}\eta^{\nu \beta} p^{\rho}_ip^{\mu}_{j}+\eta^{\sigma\beta} \eta^{\nu \alpha}p^{\rho}_i p^{\mu}_{j}
+ \eta^{\mu \sigma} \eta^{\alpha\beta} p^{\rho}_i p^{\nu}_{j} +\eta^{\mu \alpha}\eta^{\sigma \beta} p^{\rho}_ip^{\nu}_{j}\\
& && \qquad\qquad\;\;+\eta^{\mu\beta} \eta^{\sigma \alpha}p^{\rho}_i p^{\nu}_{j}
+ \eta^{\mu \nu} \eta^{\sigma\beta} p^{\rho}_i p^{\alpha}_{j} +\eta^{\mu \sigma}\eta^{\nu \beta} p^{\rho}_ip^{\alpha}_{j}+\eta^{\mu\beta} \eta^{\nu \sigma}p^{\rho}_i p^{\alpha}_{j}
+ \eta^{\mu \nu} \eta^{\alpha\sigma} p^{\rho}_i p^{\beta}_{j} \\
& && \qquad\qquad\;\;+\eta^{\mu \alpha}\eta^{\nu \sigma} p^{\rho}_ip^{\beta}_{j}+\eta^{\mu\sigma} \eta^{\nu \alpha}p^{\rho}_i p^{\beta}_{j}
+ \eta^{\sigma \nu} \eta^{\alpha\rho} p^{\beta}_i p^{\mu}_{j} +\eta^{\sigma \alpha}\eta^{\nu \rho} p^{\beta}_ip^{\mu}_{j}+\eta^{\sigma\rho} \eta^{\nu \alpha}p^{\beta}_i p^{\mu}_{j}\\
& && \qquad\qquad\;\; + \eta^{\mu \sigma} \eta^{\alpha\rho} p^{\beta}_i p^{\nu}_{j} +\eta^{\mu \alpha}\eta^{\sigma \rho} p^{\beta}_ip^{\nu}_{j}+\eta^{\mu\rho} \eta^{\sigma \alpha}p^{\beta}_i p^{\nu}_{j}
+ \eta^{\mu \nu} \eta^{\sigma\rho} p^{\beta}_i p^{\alpha}_{j} +\eta^{\mu \sigma}\eta^{\nu \rho} p^{\beta}_ip^{\alpha}_{j}\\
& && \qquad\qquad\;\;+\eta^{\mu\rho} \eta^{\nu \sigma}p^{\beta}_i p^{\alpha}_{j}
+ \eta^{\sigma \nu} \eta^{\beta\rho} p^{\alpha}_i p^{\mu}_{j} +\eta^{\sigma \beta}\eta^{\nu \rho} p^{\alpha}_ip^{\mu}_{j}+\eta^{\sigma\rho} \eta^{\nu \beta}p^{\alpha}_i p^{\mu}_{j}
+ \eta^{\mu \sigma} \eta^{\beta\rho} p^{\alpha}_i p^{\nu}_{j} \\
& && \qquad\qquad\;\;+\eta^{\mu \beta}\eta^{\sigma \rho} p^{\alpha}_ip^{\nu}_{j}+\eta^{\mu\rho} \eta^{\sigma \beta}p^{\alpha}_i p^{\nu}_{j}
+ \eta^{\sigma \alpha} \eta^{\beta\rho} p^{\nu}_i p^{\mu}_{j} +\eta^{\sigma \beta}\eta^{\alpha \rho} p^{\nu}_ip^{\mu}_{j}\\
& && \qquad\qquad\;\;+\eta^{\sigma\rho} \eta^{\alpha \beta}p^{\nu}_i p^{\mu}_{j}
\Big ] C_{0000ij}(p_1,p_2,m_0,m_1,m_2) \\
& && \quad + \Big [
\eta^{\mu \nu} \eta^{\alpha\beta} \eta^{\rho\sigma}  +\eta^{\mu \alpha} \eta^{\nu \beta}\eta^{\rho\sigma}+\eta^{\mu \beta} \eta^{\nu \alpha}\eta^{\rho\sigma}
+\eta^{\sigma \nu} \eta^{\alpha\beta} \eta^{\rho\mu}  +\eta^{\sigma \alpha} \eta^{\nu \beta}\eta^{\rho\mu}+\eta^{\sigma \beta} \eta^{\nu \alpha}\eta^{\rho\mu}\\
& && \qquad\qquad +\eta^{\mu \sigma} \eta^{\alpha\beta} \eta^{\rho\nu}  +\eta^{\mu \alpha} \eta^{\sigma \beta}\eta^{\rho\nu}+\eta^{\mu \beta} \eta^{\sigma \alpha}\eta^{\rho\nu}
+\eta^{\mu \nu} \eta^{\sigma\beta} \eta^{\rho\alpha}  +\eta^{\mu \sigma} \eta^{\nu \beta}\eta^{\rho\alpha}+\eta^{\mu \beta} \eta^{\nu \sigma}\eta^{\rho\alpha}\\
& && \qquad\qquad +\eta^{\mu \nu} \eta^{\alpha\sigma} \eta^{\rho\beta}  +\eta^{\mu \alpha} \eta^{\nu \sigma}\eta^{\rho\beta}+\eta^{\mu \sigma} \eta^{\nu \alpha}\eta^{\rho\beta}
\Big ] C_{000000}(p_1,p_2,m_0,m_1,m_2)  \, ,  \\[4mm]
&D  &&= \int \frac{d^dk}{(2\pi)^d} \frac{1}{(k^2 - m^2_0) \big((k + q_1)^2-
  m^2_1 \big) \big((k + q_2)^2- m^2_2\big) \big((k + q_3)^2- m^2_3 \big)} \numberthis  \label{eq:DIntegrals}  \\
& && = D_0(p_1,p_2,p_3,m_0,m_1,m_2,m_3)  \, , \\[4mm]
&D^{\mu}  &&= \int \frac{d^dk}{(2\pi)^d} \frac{k^{\mu}}{(k^2 - m^2_0) \big((k + q_1)^2-
  m^2_1 \big) \big((k + q_2)^2- m^2_2\big) \big((k + q_3)^2- m^2_3 \big)} \numberthis \\
& && = \sum_{i=1}^3 p_{i}^{\mu} \; D_i(p_1,p_2,p_3,m_0,m_1,m_2,m_3)  \, , \\[4mm]
&D^{\mu\nu}  &&= \int \frac{d^dk}{(2\pi)^d} \frac{k^{\mu}k^{\nu}}{(k^2 - m^2_0) \big((k + q_1)^2-
  m^2_1 \big) \big((k + q_2)^2- m^2_2\big) \big((k + q_3)^2- m^2_3 \big)} \numberthis \\
& && = \eta^{\mu\nu} D_{00}(p_1,p_2,p_3,m_0,m_1,m_2,m_3) + \sum_{i,j=1}^3 p_{i}^{\mu} p_{j}^{\nu} \; D_{ij}(p_1,p_2,p_3,m_0,m_1,m_2,m_3)  \, , \\[4mm]
&D^{\mu\nu\alpha}  &&= \int \frac{d^dk}{(2\pi)^d} \frac{k^{\mu}k^{\nu}k^{\alpha}}{(k^2 - m^2_0) \big((k + q_1)^2-
  m^2_1 \big) \big((k + q_2)^2- m^2_2\big) \big((k + q_3)^2- m^2_3 \big)} \numberthis \\
& && = \;\;\;\;\; \sum_{i=1}^3 \Big [ \eta^{\mu\nu} p_i^{\alpha} +  \eta^{\mu\alpha} p_i^{\nu} + \eta^{\nu\alpha} p_i^{\mu} \Big ] D_{00i}(p_1,p_2,p_3,m_0,m_1,m_2,m_3) \\
\end{alignat*}
\begin{alignat*}{3}
& && \quad + \sum_{i,j,k=1}^3 p_{i}^{\mu} p_{j}^{\nu} p_{j}^{\alpha} \; D_{ijk}(p_1,p_2,p_3,m_0,m_1,m_2,m_3) \, ,  \\[4mm]
&D^{\mu\nu\alpha\beta}  &&= \int \frac{d^dk}{(2\pi)^d}
\frac{k^{\mu}k^{\nu}k^{\alpha}k^{\beta}}{(k^2 - m^2_0) \big((k + q_1)^2-  m^2_1
  \big) \big((k + q_2)^2- m^2_2\big) \big((k + q_3)^2- m^2_3 \big)} \numberthis \label{eq:DDIntegrals}  \\
& && = \;\;\; \Big [
\eta^{\mu\nu}\eta^{\alpha\beta}+\eta^{\mu\alpha}\eta^{\nu\beta}+\eta^{\mu\beta}\eta^{\alpha\nu}
\Big ]  D_{0000}(p_1,p_2,p_3,m_0,m_1,m_2,m_3)  \\
& && \quad + \sum_{i,j=1}^3 \Big [ \eta^{\mu\nu} p_i^{\alpha} p_j^{\beta}  +
\eta^{\mu\alpha} p_i^{\nu} p_j^{\beta} + \eta^{\nu\alpha} p_i^{\mu} p_j^{\beta}
+ \eta^{\mu\beta} p_i^{\alpha} p_j^{\nu}  +
\eta^{\nu\beta} p_i^{\alpha} p_j^{\mu} \\
& && \qquad\qquad\qquad\qquad\qquad\qquad\qquad + \eta^{\alpha\beta} p_i^{\mu} p_j^{\nu} \Big ] D_{00ij}(p_1,p_2,p_3,m_0,m_1,m_2,m_3) \\
& && \quad + \sum_{i,j,k,l=1}^3 p_{i}^{\mu} p_{j}^{\nu} p_{k}^{\alpha}p_{l}^{\beta} \; D_{ijkl}(p_1,p_2,p_3,m_0,m_1,m_2,m_3)  \, .
\end{alignat*}
Also, we use the following reduction formulas
\begin{align}
  k.p_i &= \frac{1}{2} \Big [((k+q_i)^2 - m_i^2) - ((k+q_{i-1})^2 - m_{i-1}^2) + m_i^2 - m_{i-1}^2 - q_i^2 + q_{i-1}^2 \Big ] \, , \label{eq:red1} \\
  p_{i}.Q_{i} &= \frac{1}{2} \Big [ (Q_i^2-m_i^2) - (Q_{i-1}^2 -m_{i-1}^2) +m_i^2-m_{i-1}^2 + p_i^2   \Big] \, , \\
  p_{i}.Q_{i-1} &= \frac{1}{2} \Big [ (Q_i^2-m_i^2) - (Q_{i-1}^2 -m_{i-1}^2) +m_i^2-m_{i-1}^2 - p_i^2   \Big] \, , \\
  p_{i}.Q_{i-2} &= \frac{1}{2} \Big [ (Q_i^2-m_i^2) - (Q_{i-1}^2 -m_{i-1}^2) +m_i^2-m_{i-1}^2 - p_i^2  -2 p_{i-1}.p_i \Big] \, , \\
  p_i.k &= \frac{1}{2} \Big [ (Q_i^2-m_i^2) - (Q_{i-1}^2 -m_{i-1}^2) +m_i^2-m_{i-1}^2 - 2 q_{i}.p_i + p_i^2  \Big] \, , \\
  p_i.k &= \frac{1}{2} \Big [ (Q_i^{\prime\prime \, 2}-m_i^2) - (Q_{i-1}^{\prime \, 2} -m_{i-1}^2) +m_i^2-m_{i-1}^2 - p_{i}^2 -2 p_i.p_{i-1} \Big] \, ,  \\
  p_i.k &= \frac{1}{2} \Big [ (Q_i^{\prime \, 2}-m_i^2) - (k^2 -m_0^2) +m_i^2-m_0^2 - p_{i}^2  \Big] \, , \label{eq:red2}
\end{align}
where the relations between the momenta are:
\begin{align*}
  Q_i &=k+p_1+\cdots+p_i=k+q_i  \, , \\
  q_i &= p_1+\cdots +p_i \, , \\
  Q_i^{\prime} &=k+p_i \, , \\
  Q_i^{\prime\prime}&=k+p_{i-1}+p_i  \, .
\end{align*}

Finally, we follow the Passarino-Veltman method as explained in
Sec.~\ref{se:PassarinoVeltman} using the reduction formulas
Eqs.~(\ref{eq:red1}$-$\ref{eq:red2}) and other integral tools such as change of
variables. As a result, we write all above scalar functions
$A_{00},A_{0000},B_1,B_{00},\ldots$ in terms of the
scalar functions $A_0, B_0, C_0$ and $D_0$.

\clearpage
\section{The FORM Program}\label{Ap:formprogram}
\setcounter{equation}{0}
In this appendix, we show short pieces of our code to illustrate as much
as possible how we perform the calculations in the FORM program. At the same
time, we ensure that all the indices are contracted in a proper way in the code.
However, let us start by introducing some notations in Tab.~\ref{ta:notations}.
\begin{table}[H]
  \caption{Some notations that we use in our code.}\smallskip
  \label{ta:notations}
  \centering
  \begin{threeparttable}
  \begin{tabular}{c|c|c}
    \hline\hline
    $h_{\mu\nu}=\;$\texttt{H(mu,nu)}  &  $\partial_{\alpha} h_{\mu\nu}=\;$\texttt{H(al,mu,nu)}  &  $\partial_{\alpha}\partial_{\beta}h_{\mu\nu}=\;$\texttt{H(al,be,mu,nu)} \\
    \hline
    $g_{\mu\nu}=\;$\texttt{GL(mu,nu)}  &  $\partial_{\alpha} g_{\mu\nu}=\;$\texttt{GL(al,mu,nu)}  &  $\partial_{\alpha}\partial_{\beta}g_{\mu\nu}=\;$\texttt{GL(al,be,mu,nu)} \\
    \hline
    $g^{\mu\nu}=\;$\texttt{GU(mu,nu)}  &  $\partial_{\alpha} g^{\mu\nu}=\;$\texttt{GU(al,mu,nu)}  &  $\partial_{\alpha}\partial_{\beta}g^{\mu\nu}=\;$\texttt{GU(al,be,mu,nu)} \\
    \hline
    $\phi=\;$\texttt{phi}  &  $\partial_{\alpha} \phi=\;$\texttt{phi(al)}  &  $\partial_{\alpha}\partial_{\beta}\phi=\;$\texttt{phi(al,be)} \\
    \hline
   $\partial_{\alpha}\Gamma_{\nu\beta}^{\mu}=\;$\texttt{Gamma(al,mu,nu,be)} & $\delta_{\mu\nu}=\;$\texttt{d\_(mu,nu)}  &  \texttt{MAXH}\tnote{1}$\;\;=\;4$ \\
    \hline
  \end{tabular}
  \begin{tablenotes}
  \item[1] \scriptsize{Since our calculations are up to the quadruple graviton vertex.}
  \end{tablenotes}
\end{threeparttable}
\end{table}

As an example, the square root of the determinant of metric $\sqrt{-g}=\;$\texttt{SQRTMG}
can be calculated from the relation Eq.~(\ref{eq:determinant}) as:
\begin{Verbatim}[gobble=2,frame=single,framesep=2mm,label=Code,labelposition=all,numbers=left]
  Local SQRTMG = 1+sum_(n,1,`MAXH',invfac_(n)/2^n*trlng^n);
  #do i=1,`MAXH'
      id,once trlng = -sum_(n,1,`MAXH'
                     ,sign_(n)/n*epsh^n*kappa^n*HH(n,i`i'x1,i`i'x1));
      id epsh^{`MAXH'+1} = 0;
  #enddo
  #do i=1,`MAXH'
  #do j=2,`MAXH'
      id,once HH(n?,mu?,nu?) = H(mu,i`i'x`j')*HH(n-1,i`i'x`j',nu);
      id HH(0,mu?,nu?) = d_(mu,nu);
  #enddo
  #enddo
\end{Verbatim}

In addition, the field redefinition can be done by replacing each field and its derivatives with
the proper expansion. The code below shows the gravitational field redefinition Eq.~(\ref{eq:RedefinedField}), where the other fields were treated in the same way:
\begin{Verbatim}[gobble=2,frame=single,framesep=2mm,label=Code,labelposition=all,numbers=left]
  #do i=1,`MAXH'
  multiply aa;
  id,once aa*H(?a,mu?,nu?) =  a1*Der(?a,H(mu,nu))
   +kappa*epsh*(a3*Der(?a,H(mu,i{`i'+1}x1),
     H(i{`i'+1}x1,nu))+a2*Der(?a,H(mu,nu),H(i{`i'+1}x1,i{`i'+1}x1)))
   +kappa^2*epsh^2*(a4*Der(?a,H(mu,nu),H(i{`i'+1}x1,i{`i'+1}x1),
     H(i{`i'+1}x2,i{`i'+1}x2))+a5*Der(?a,H(mu,nu),H(i{`i'+1}x1,i{`i'+1}x2),
     H(i{`i'+1}x1,i{`i'+1}x2))+a6*Der(?a,H(mu,i{`i'+1}x1),H(nu,i{`i'+1}x1),
     H(i{`i'+1}x2,i{`i'+1}x2))+a7*Der(?a,H(mu,i{`i'+1}x1),H(nu,i{`i'+1}x2),
     H(i{`i'+1}x1,i{`i'+1}x2)))
   +kappa^3*epsh^3*(a8*Der(?a,H(mu,nu),H(i{`i'+1}x1,i{`i'+1}x1),
     H(i{`i'+1}x2,i{`i'+1}x2),H(i{`i'+1}x3,i{`i'+1}x3))+a9*Der(?a,H(mu,nu),
     H(i{`i'+1}x1,i{`i'+1}x1),H(i{`i'+1}x2,i{`i'+1}x3),
     H(i{`i'+1}x2,i{`i'+1}x3))+a10*Der(?a,H(mu,nu),H(i{`i'+1}x1,i{`i'+1}x2),
     H(i{`i'+1}x2,i{`i'+1}x3),H(i{`i'+1}x3,i{`i'+1}x1))+a11*Der(?a,
     H(mu,i{`i'+1}x1),H(nu,i{`i'+1}x1),H(i{`i'+1}x2,i{`i'+1}x2),
     H(i{`i'+1}x3,i{`i'+1}x3))+a12*Der(?a,H(mu,i{`i'+1}x1),H(nu,i{`i'+1}x1),
     H(i{`i'+1}x2,i{`i'+1}x3),H(i{`i'+1}x2,i{`i'+1}x3))+a13*Der(?a,
     H(mu,i{`i'+1}x1),H(nu,i{`i'+1}x2),H(i{`i'+1}x1,i{`i'+1}x2),
     H(i{`i'+1}x3,i{`i'+1}x3))+a14*Der(?a,H(mu,i{`i'+1}x1),H(nu,i{`i'+1}x2),
     H(i{`i'+1}x1,i{`i'+1}x3),H(i{`i'+1}x2,i{`i'+1}x3)));
  id aa = 1;
  id epsh^`MAXH' = 0;
  .sort
  #enddo
\end{Verbatim}
Taking a derivative in our notation can also be achieved in FORM with the following procedure:
\begin{Verbatim}[gobble=2,frame=single,framesep=2mm,label=Code,labelposition=all,numbers=left]
  repeat;
    id Deriv(H?fields(?a),?b) = H(?a)*Deriv(?b);
    id Deriv(mu?,H?fields(?a),?b) = H(mu,?a)*Deriv(?b)
                                   +H(?a)*Deriv(mu,?b);
    id Deriv(mu?,nu?,H?fields(?a),?b) = H(mu,nu,?a)*Deriv(?b)
                                       +H(mu,?a)*Deriv(nu,?b)
                                       +H(nu,?a)*Deriv(mu,?b)
                                       +H(?a)*Deriv(mu,nu,?b);
    id Deriv(mu?) = 0;
    id Deriv(mu?,nu?) = 0;
    id Deriv = 1;
  endrepeat;
\end{Verbatim}
Finally, the amplitude of a Feynman diagram can be calculated in FORM as shown
below for the s-channel diagram in scalar-graviton scattering:\vspace{2mm}
\begin{Verbatim}[gobble=2,frame=single,framesep=2mm,label=Code,labelposition=all,numbers=left]
  * S-channel in scalar-graviton scattering: 
  *
  * p2-> x             x p4->
  * mu,nu x           x al,be
  *        x         x
  *  aaaaaaaXaaaaaaaYaaaaaaaa
  *   p1->    P1->     p3->
  *
  *   X=p1x,i1x Y=p2x,i2x  

  #include vertexprocedures.hf
  #include symbols.hf
  #include setexternal.hf  
  .sort
  L Vleft =
  #call vertex2phi1H`VERTEXTYPE'
   ;
  #call setmom(p1x,3)
  #call takederiv
  print +s;
  .sort
  skip;
  L Vright =
  #call vertex2phi1H`VERTEXTYPE'
    ;
  #call relabelindexij(1,2)
  #call setmom(p2x,3)
  #call takederiv
  print +s;
  .sort
  #call pickoutphiin(p1,p1x,p1xp1x,2,Vleft)
  #call pickoutphiout(p3,p2x,p2xp2x,2,Vright)
  #call pickoutHin(mu,nu,p2,p1x,p1xp1x,2,Vleft)
  #call pickoutHout(al,be,p4,p2x,p2xp2x,1,Vright)
  .sort
  drop Vleft,Vright;
  G PPHH1`VERTEXTYPE' = i_^3*Vleft*Vright;
  #call connectphi(P1,p1x,p1xp1x,1,p2x,p2xp2x,1,PPHH1`VERTEXTYPE')
  print +s;
  .store
  save save/PPHH1`VERTEXTYPE'.sav PPHH1`VERTEXTYPE';
  .end
\end{Verbatim}
Where we start the code by defining the vertices and importing their
corresponding expressions. In the example above, the left vertex defined in
lines 15 and 16 is a scalar-scalar-graviton vertex, with its
corresponding expression given by Eq.~(\ref{eq:OurphiphiH}). Then in lines 18,
19, 31 and 33, we convert this vertex to momentum space. After that in line 37, we
put all together and multiply by ($i$) for each vertex and propagator. Finally
in line 38, we connect the two vertices by calling the propagator procedure, which also contains the \texttt{pickoutphiin} and \texttt{pickoutphiout} procedures.

\clearpage
\centering
\printglossary[type=\acronymtype] 


\cleardoublepage
\phantomsection
\addcontentsline{toc}{section}{References}

\centering


\begin{thebibliography}{}

\end{thebibliography}


\begin{thebibliography}{1}
\vspace{2mm}

\bibitem{Peskin}
  M.~E. Peskin and D.~V. Schroeder, {\em {An Introduction to quantum field
      theory}}.
  \newblock Reading, USA: Addison-Wesley, 1995.

\bibitem{GR}
  B.~Schutz, {\em A First Course in General Relativity}.
  \newblock Cambridge University Press, 2~ed., 2009.

\bibitem{GRasQFTDonoghue}
J.~F. {Donoghue}, M.~M. {Ivanov}, and A.~{Shkerin}, ``{EPFL Lectures on General
  Relativity as a Quantum Field Theory},'' {\em ArXiv e-prints} 1702.00319, Feb. 2017.

\bibitem{HigherSpin}
S.~Weinberg and E.~Witten, ``{Limits on Massless Particles},'' {\em Phys.
  Lett.}, vol.~96B, pp.~59--62, 1980.

\bibitem{EFT}
J.~F. Donoghue, ``{General relativity as an effective field theory: The leading
  quantum corrections},'' {\em Phys. Rev.}, vol.~D50, pp.~3874--3888, 1994.

\bibitem{GhostVertex}
M.~T. Grisaru, P.~van Nieuwenhuizen, and C.~C. Wu, ``Background-field method
  versus normal field theory in explicit examples: One-loop divergences in the
  $s$ matrix and green's functions for yang-mills and gravitational fields,''
  {\em Phys. Rev. D}, vol.~12, pp.~3203--3213, Nov 1975.

\bibitem{TreeLevelResults}
M.~T. Grisaru, P.~van Nieuwenhuizen, and C.~C. Wu, ``Gravitational born
  amplitudes and kinematical constraints,'' {\em Phys. Rev. D}, vol.~12,
  pp.~397--403, Jul 1975.
  
\bibitem{DonoghueTreeLevel}
  J.~F. Donoghue and T.~Torma, ``{Infrared behavior of graviton-graviton
    scattering},'' {\em Phys. Rev.}, vol.~D60, p.~024003, 1999.
  
\bibitem{SameWork}
J.~D. Gonçalves, T.~de~Paula~Netto, and I.~L. Shapiro, ``{Gauge and
  parametrization ambiguity in quantum gravity},'' {\em Phys. Rev.}, vol.~D97,
  no.~2, p.~026015, 2018.

\bibitem{EquivalenceTheorem}
S.~Kamefuchi, L.~O'Raifeartaigh, and A.~Salam, ``Change of variables and
  equivalence theorems in quantum field theories,'' {\em Nuclear Physics},
  vol.~28, no.~4, pp.~529 -- 549, 1961.

\bibitem{HelicityAmplitudes}
  S.~Weinberg, ``Photons and gravitons in perturbation theory: Derivation of
  maxwell's and einstein's equations,'' {\em Phys. Rev.}, vol.~138,
  pp.~B988--B1002, May 1965.
  
\bibitem{MasslessTadpole}
G.~Leibbrandt, ``Introduction to the technique of dimensional regularization,''
  {\em Rev. Mod. Phys.}, vol.~47, pp.~849--876, Oct 1975.

\bibitem{PassarinoVeltman}
  G.~Passarino and M.~J.~G. Veltman, ``{One Loop Corrections for e+ e-
    Annihilation Into mu+ mu- in the Weinberg Model},'' {\em Nucl. Phys.},
  vol.~B160, pp.~151--207, 1979.  

\bibitem{FORM}
B.~Ruijl, T.~Ueda, and J.~Vermaseren, ``{FORM version 4.2},'' {\em {ArXiv
  e-prints}} 1707.06453, 2017.

\bibitem{FORMGitHub}
  J.~Vermaseren, ``{FORM 4.2,'' ``\it{GitHub repository}}''
  \url{https://github.com/vermaseren/form}, 2018.  
  
\end{thebibliography}

%
%




\vspace{85mm}
\ornamento

\end{document}